\documentclass{elsart}   
\usepackage[intlimits]{amsmath}
\usepackage{graphicx}
\usepackage{amsfonts}
\usepackage{amssymb}
\usepackage{dsfont}

\newcommand{\be}{\begin{equation}}
\newcommand{\ee}{\end{equation}}
\newcommand{\beqa}{\begin{eqnarray}}
\newcommand{\eeqa}{\end{eqnarray}}

\def\Slash#1{\setbox0=\hbox{$#1$} 
\dimen0=\wd0 
\setbox1=\hbox{/} \dimen1=\wd1 
\ifdim\dimen0>\dimen1 
\rlap{\hbox to \dimen0{\hfil/\hfil}} 
#1 
\else 
\rlap{\hbox to \dimen1{\hfil$#1$\hfil}} 
/ 
\fi}

\def\longlongrightarrow{
\relbar\joinrel\relbar\joinrel\relbar\joinrel\relbar\joinrel\rightarrow}
\def\longlonglongrightarrow{
\relbar\joinrel\relbar\joinrel\relbar\joinrel\relbar\joinrel\relbar\joinrel\relbar\joinrel\relbar\joinrel\relbar\joinrel\relbar\joinrel\rightarrow}

\begin{document}

\begin{frontmatter}

\vspace{5mm}

\title{
A covariant view on the nucleons' quark core}
\author{G.\ Eichmann, A.\ Krassnigg, M.\ Schwinzerl, and R.\ Alkofer}
\address{ Institut f\"ur Physik, Karl-Franzens-Universit\"at Graz, A-8010 Graz, Austria }

\begin{abstract}
Established results for the quark propagator in Landau gauge QCD, together
with a detailed comparison to lattice data, are used to
formulate a Poincar{\'e}-covariant Faddeev approach to the nucleon.  The
resultant three-quark amplitudes describe the quark core of the nucleon.
The nucleon's mass and its electromagnetic form factors are calculated
as functions of the current quark mass. The corresponding results
together with charge radii and magnetic moments are discussed in connection with
the contributions from various ingredients in a consistent calculation of nucleon
properties, as well as the role of the pion cloud in such an approach.
\end{abstract}

\end{frontmatter}



\section{Introduction}\label{sec:I}

            The structure of the nucleon has been the object of many theoretical
            investigations for decades, see {\it e.\,g.\/} \cite{Thomas:2001kw}.
            Although many details about nucleon properties have emerged, some remain
	    elusive, a fact which can easily be inferred from the surprises caused
            by recent experimental results. One prominent example is
            the discrepancy between the ratio of electromagnetic proton
            form factors extracted via Rosenbluth seperation on the one hand and
            polarisation-transfer experiments on the other hand; a recent summary
            of the experimental situation and possible theoretical explanations is
            {\it e.\,g.\/} given in Ref.~\cite{Arrington:2006zm}. While quark models have
	    pioneered our understanding of the nucleon, they become
            increasingly unlikely to contribute to a more concise picture of
            the nucleon's structure: the complicated
            nature of baryons cannot be precisely portrayed by model constructions like
            constituent quarks or solitonic meson clouds. In this respect it is worth
            noting that efforts to bridge the gap between studies of strong
            (``nonperturbative'') QCD and of hadron structure are on-going and show
            encouraging results. Some examples of such progress can be found in
            the lectures compiled in Ref.~\cite{Schladming06}. Besides and together with
            lattice QCD and chiral perturbation theory functional methods have proven to
            be a method of choice when investigating the infrared properties of QCD,
            for some recent brief overview see {\it e.g.\/} \cite{Alkofer:2006jf}.
            The investigation to be reported here will build on such functional methods in
	    the context of Dyson-Schwinger equations,
            and we will relate our results to corresponding ones obtained by a synthesis of
            lattice and effective field theory calculations. 
	
	    Dyson-Schwinger equations (DSEs) present a nonperturbative continuum approach to QCD.
	    Each Green function in QCD satisfies an integral equation, which involves also
	    higher Green functions; this leads to an infinite tower of coupled integral equations
	    (see e.g. Refs.\,\cite{Roberts:1994dr,Alkofer:2000wg,Fischer:2006ub,Holl:2006ni} for recent reviews of the subject).
	    For numerical hadron studies in QCD one relies on a truncation of the infinite tower of equations to
	    a subset which is solved explicitly. Our study focuses on the DSE of the quark propagator
	    and the Bethe-Salpeter equations (BSEs) for quark-quark and quark-diquark systems, respectively.
	    The amplitudes obtained from the solution of the nucleon BSE are used in subsequent
	    calculations of electromagnetic nucleon properties.

            Originating from the NJL model \cite{Nambu:1961fr,Nambu:1961tp}, the covariant
	    quark-diquark description has often been applied                            
            to the investigation of nucleon and $\Delta$ properties \cite{Buck:1992wz,Weiss:1993kv,Ishii:1995bu}.
            Nucleon form factors were obtained either by employing ans\"atze for the quark-diquark amplitudes \cite{Bloch:1999ke,Bloch:1999rm}
            or by solving their BSE using ans\"atze for the ingredients of its kernel,
            i.\,e., the dressed-quark propagator, diquark propagator and diquark amplitudes
            \cite{Oettel:1998bk,Oettel:1999gc,Hecht:2002ej,Oettel:2000jj,Oettel:2002wf,Alkofer:2004yf,Holl:2005zi}.
            The present work extends these studies insofar as those ingredients are obtained consistently from dynamic
	    equations at a more fundamental level of the approach.
            It is an ab-initio calculation in the sense that after specifying a truncation and fixing the pseudoscalar meson sector to the experiment
            as well as interaction parameters to lattice data of the quark propagator,
	    no further observables need be used as an input of the calculation.

            To make this evident: the main objective of the studies to be reported here is
            to develop a QCD-based understanding of the nucleon structure. Given the
            experimental results, or more precisely, the complicated nature of the
            nucleon, the aim is as ambitious as the potential outcome is rewarding.

	    This article is organized as follows:
            Section \ref{sec:boundstateeq} gives a brief introduction to bound state equations in quantum field theory, in particular
	    for a three-quark state. Section \ref{sec:dsebse} introduces the quark propagator and the quark-gluon interaction, i.\,e., the
	    dynamical setup of the approach. The concept of diquarks is introduced in Sec.\,\ref{sec:dq} and applied
	    to the nucleon as a two-body system of quark and diquark in Sec.\,\ref{sec:qdqbse}.
            The nucleon's electromagnetic form factors in this framework are developed in Sec.\,\ref{sec:em}.
	    Section \ref{sec:results} contains our results for the nucleon mass and electromagnetic properties as well as
	    a discussion; we conclude in Sec.~\ref{sec:conclusions}.
	
	    In order to make this paper self-contained as well as to provide all details necessary for reproducing our
	    results we have added appendices containing solution strategies for the QCD gap equation in App.~\ref{app:dsesupplement},
	    all necessary detail on Bethe-Salpeter amplitudes of mesons and diquarks in App.~\ref{app:mesondiquark}
	    as well as quark-diquark systems in App.~\ref{app:qdqamp}, the construction of the electromagnetic current in App.~\ref{app:em},
	    and limitations on the numerical implementation coming from singularities of the Green functions in our approach in App.~\ref{app:sing}.
	    These also serve to avoid confusion of different
	    notations and conventions in the literature.


\section{Bound-state equations in QCD}\label{sec:boundstateeq}

\begin{figure}[hbt]
            \begin{center}
            \includegraphics[scale=0.48,angle=-90]{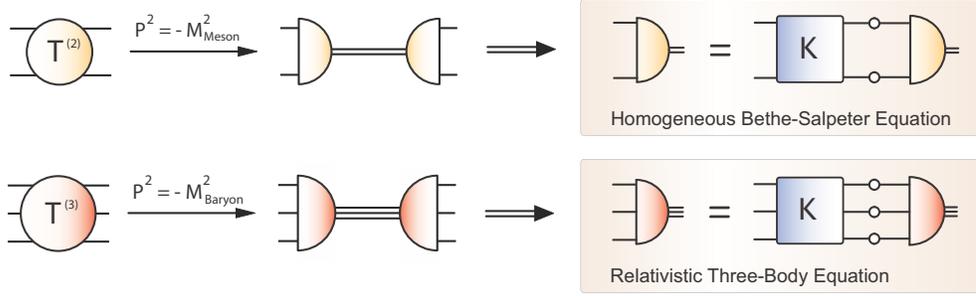}
            \caption[Bound-state equations]{The pole ansatz for the 2- and 3-quark scattering matrices and the resulting bound-state equations.} \label{fig:BSE+Fadeev}
            \end{center}
            \end{figure}

            As bound states of quarks, hadrons correspond to poles in the quark 4- or 6-point
	    functions $G^{(2)}$ and $G^{(3)}$ or their amputated connected parts, the scattering
	    matrices $T^{(2)}$ and $T^{(3)}$ defined via
            \begin{equation}\label{bs:tmatrix}
                G^{(n)} = G_0^{(n)} + G_0^{(n)} T^{(n)} G_0^{(n)}\;.
            \end{equation}
            Up to antisymmetrization, $G_0^{(n)}$ is the product of $n$ dressed quark propagators $S_i$.
            We work in Euclidean momentum space, i.\,e.,
            we use $\left\{\gamma^\mu,\gamma^\nu\right\}=2\delta^{\mu\nu}$ and $\gamma^{\mu\dag}=\gamma^\mu$,
            and dropped all Dirac, color and flavor indices of each quark leg in the above quantities for brevity.
            The product in Eq.\,\eqref{bs:tmatrix} is understood
            as a summation over all occurring indices as well as integration over all internal momenta.
            At the pole corresponding to the bound-state mass $M$, bound-state amplitudes $\Psi$ are
            introduced as the residues of the scattering matrix:
            \begin{equation}\label{bs:poleansatz}
            T \stackrel{P^2\rightarrow -M^2}{\longlongrightarrow} \frac{\Psi\,\bar{\Psi}}{P^2+M^2}\;,
            \end{equation}
            where $P$ is the total momentum of the $n$ quarks.
            The scattering kernel $K^{(n)}$ ($n\geq 2$), constructed out of $l$-quark irreducible components ($l=2\dots n$), is
            related to the scattering matrix $T^{(n)}$ via Dyson's equation:
            \begin{equation}\label{bs:dysonseq}
            \begin{split}
                T = K + K\,G_0\,K + K\,G_0\,K\,G_0\,K + \cdots &= K + K\,G_0\,T \\
                \Longleftrightarrow\quad T^{-1} &= K^{-1} - G_0\;.
            \end{split}
            \end{equation}
            Inserting the pole ansatz and comparing the residues leads to a bound-state equation and a canonical normalization condition
            at the bound-state pole $P^2=-M^2$ which completely determine the amplitudes on the mass shell (see Fig.~\ref{fig:BSE+Fadeev})
            \begin{align}
                & \Psi = K\,G_0\,\Psi\;, \label{bs:boundstateeq}\\
                & \bar{\Psi} \,\frac{d\,(K^{-1}-G_0)}{d P^2}\,\Psi  = 1\;.  \label{bs:normalization}
            \end{align}
            Eq.\,\eqref{bs:boundstateeq} is a homogeneous integral equation and fully relativistic:
            it is the BSE in the two-body case and its quantum-field theoretical analogue for a three-body system.
            It can be solved if the quark propagator $S_i$ and the kernel $K^{(n)}$ are known.
            Both ingredients can in principle be obtained from the infinite coupled set of
	    Dyson-Schwinger equations (see, e.\,g., \cite{Holl:2006ni} and references therein).
            Feasible present-day numerical DSE solutions include 2- and 3-point functions within certain truncations,
            but the complexity of DSEs for 4-point functions has so far prevented a direct numerical approach.
            The construction of appropriate kernels is restricted by the underlying symmetries of the theory.
            In quantum-field theory symmetries are implemented by Ward-Takahashi (WTIs) and Slavnov-Taylor identities (STIs)
            which relate different $n$-point functions to each other.
            The most prominent example is the axial-vector Ward-Takahashi identity (AVWTI) \cite{Maris:1997hd}, which
	    relates the two-quark kernel $K^{(2)}$ to the kernel of the quark DSE. It is imperative to satisfy these
	    identities in the truncation used in a numerical study; in this respect, the rainbow-ladder truncation
	    has emerged as the lowest order in a symmetry-preserving truncation scheme \cite{Bender:2002as,Bender:1996bb}.

\begin{figure}[hbt]
            \begin{center}
            \includegraphics[scale=0.47,angle=-90]{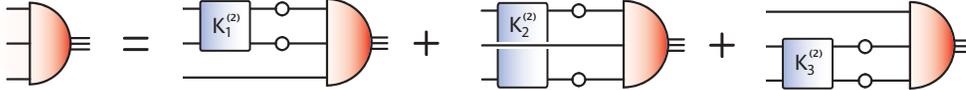}
            \caption[Relativistic Faddeev equations]{The relativistic Faddeev equations, Eq.\,\eqref{bs:faddeevtruncated}.}
            \label{fig:faddeevtruncated}
            \end{center}
            \bigskip
            \end{figure}

            For baryons, the 3-body kernel $K^{(3)}$ in the relativistic 3-body equation
            can be written as the sum of a 3-quark irreducible contribution and the 2-body
	    kernels $K_i^{(3)} = K_i^{(2)} \otimes \,S_i^{-1}$ \cite{Buck:1992wz,Ishii:1995bu,Cahill:1988dx},
            where the subscript $i$ identifies the spectator quark:
            \begin{equation}
            K^{(3)} = K^\text{(3)}_\text{irr} + \sum_{i=1}^3 K_i^{(3)}\;.
            \end{equation}
            Neglecting the 3-body irreducible part yields relativistic 3-body equations of the Faddeev
	    type, commonly referred to as \textsl{relativistic Faddeev equations} (see Fig.~\ref{fig:faddeevtruncated})
            \begin{equation}\label{bs:faddeevtruncated}
            \Psi = \sum_{i=1}^3 \Psi_i \quad \text{with} \quad \Psi_i = K_i^{(3)}\,G_0^{(3)}\,\Psi = K_i^{(2)}\,S_j\,S_k\,\Psi\;,
            \end{equation}
            where $\{i,j,k\}$ denotes a cyclic permutation of $\{1,2,3\}$, i.\,e.\ of all Dirac, color, and flavor indices.
            The resulting relativistic Faddeev amplitude does therefore not contain explicit three-quark correlations.

\section{Quark propagator and quark-antiquark scattering kernel}\label{sec:dsebse}


\subsection{Elementary QCD Green functions}\label{sec:qcdgreenfunctions}

\begin{figure}[hbt]
            \begin{center}
            \includegraphics[scale=0.48,angle=-90]{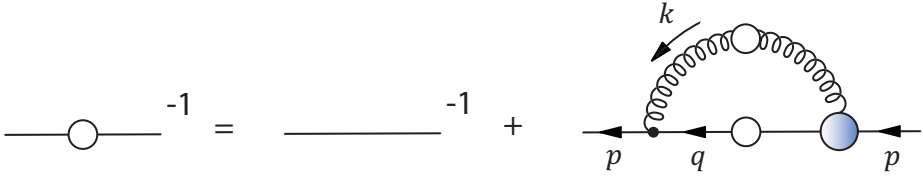}
            \caption[Quark DSE]{The quark DSE in pictorial form. } \label{fig:DSE}
            \end{center}
            \end{figure}

            Equation \eqref{bs:faddeevtruncated} is the starting point of this work. Its solution requires know\-ledge of the
            quark propagator and the two-quark kernel, which are given in this section.
	    The renormalized dressed-quark propagator $S(p,\mu)$ contains two dressing
            functions $\sigma_v$ and $\sigma_s$ which correspond to a general fermion propagator's
	    vector and scalar Lorentz structures.
            Equivalently, one can write $S$ in terms of the quark renormalization function $A$ and the quark mass function $M$,
            which is independent of the renormalization point $\mu$:
            \begin{equation}\label{dse:qprop}
                S(p,\mu) = \left( A(p^2,\mu^2) \left( i \Slash{p} + M(p^2) \right)\right)^{-1}
                              = -i \Slash{p} \,\sigma_v(p^2,\mu^2) + \sigma_s(p^2,\mu^2)\;.
            \end{equation}
            These dressing functions are obtained by solving the quark Dyson-Schwinger equation, also referred
	    to as the \textsl{QCD gap equation} (see Fig.~\ref{fig:DSE})
            \begin{equation}\label{dse:qdse}
                S^{-1}(p,\mu) =  Z_2(\mu^2,\Lambda^2) \left( i \Slash{p} + M(\Lambda^2) \right) + \Sigma(p,\mu,\Lambda) \; ,
            \end{equation}
            where $Z_2(\mu^2,\Lambda^2)=A(p^2=\Lambda^2,\mu^2)$ %
	     is the quark renormalization constant
            and $\Lambda$ is a translationally invariant ultra-violet regularization scale.
            The quark self-energy $\Sigma$ is defined via
            \begin{equation}\label{dse:qselfenergy}
                \Sigma(p,\mu,\Lambda) = - \frac{4 g^2}{3}\,Z_{1F}(\mu^2,\Lambda^2) \int^\Lambda \frac{d^4 q}{(2\pi)^4} \,\,
                                         i\gamma^\mu \, S(q,\mu) \, D^{\mu\nu}(k,\mu)\, \Gamma^\nu(q,p,\mu)\;,
            \end{equation}

            where the prefactor $(N_C^2-1)/(2 N_C)=4/3$ stems from the color trace.
            $\Sigma$ contains one bare ($g Z_{1F} \,i\gamma^\mu$) and one dressed ($g  \Gamma^\mu$) quark-gluon vertex and
            the gluon propagator $D^{\mu\nu}$ with gluon momentum $k=p-q$ which have to be either known a priori
            or determined in the course of a self-consistent solution of the DSEs of the quark- and gluon-propagators
	    together with the quark-gluon vertex.
            Technical details of the solution of Eq.\,\eqref{dse:qdse} are sketched in App.\,\ref{app:dsesupplement}.

            In Landau gauge, the dressed-gluon propagator characterized by a dressing function $Z$
            is transverse with respect to the gluon momentum $k=p-q$:
            \begin{equation}
                D^{\mu\nu}(k,\mu)=\frac{Z(k^2,\mu^2)}{k^2}\,T^{\mu\nu}_k, \quad  T^{\mu\nu}_k := \delta^{\mu\nu}-\frac{k^\mu k^\nu}{k^2} \;.
            \end{equation}
            The general dressed quark-gluon vertex consists of 12 tensor structures,
            \begin{equation}\label{qgv:structure}
                \Gamma^{\nu}(p,q,\mu) = \sum_{i=1}^{12} \Gamma_i(p^2,q^2,k^2,\mu^2)\,\tau_i^\nu(k,l)\;,
            \end{equation}
            with the simplest possible representation of the Dirac basis elements given by
            $\tau_i^\nu(k,l) = \gamma^\nu \{\mathds{1}, \Slash{k}, \, \Slash{l}, \, \Slash{l}\,\Slash{k} \}$,
            $l^\nu\{\mathds{1}, \Slash{k}, \, \Slash{l}, \, \Slash{l}\,\Slash{k} \}$ and $k^\nu\{\mathds{1}, \Slash{k}, \, \Slash{l}, \, \Slash{l}\,\Slash{k} \}$,
            where $l=-(p+q)/2$.
            Both gluon propagator and quark-gluon vertex satisfy their own DSEs.
            Progress on a consistent solution of this coupled system has been made both analytically
            in terms of infrared exponents of the Green functions\,\cite{Zwanziger:2001kw,Lerche:2002ep,Alkofer:2006gz},
            as well as numerically for general momenta in certain truncations \cite{Fischer:2002hna,Fischer:2003rp}.
	    However, the implementation of such a scheme in our problem is clearly beyond the scope of the present
	    study. Therefore we use underlying symmetry properties of QCD to arrive at a consistent truncation
	    whose numerical implementation is feasible in our approach to the nucleon.

\subsection{Effective quark-gluon interaction}\label{sec:effectiveinteraction}

\begin{figure}[hbt]
            \begin{center}
            \includegraphics[scale=0.47,angle=-90]{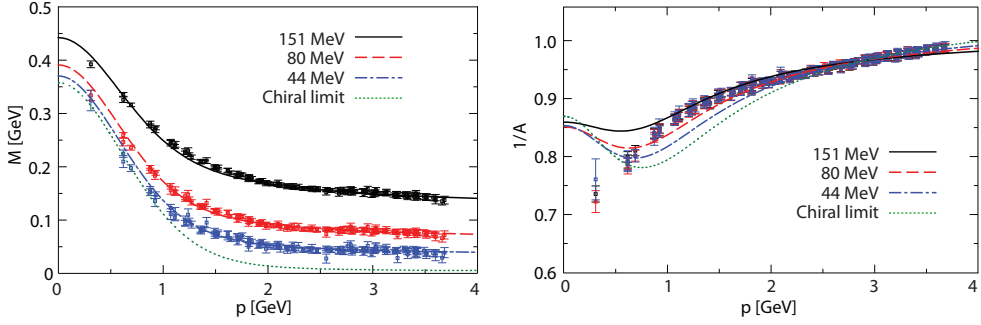}
            \caption[Quark propagator and lattice data]{ Quark dressing functions: Quenched lattice data \cite{Bowman:2002bm} vs.
                                                         the DSE result employing the interaction \eqref{dse:maristandy} with $N_f=0$.
                                                         The quark masses in the legend are given at $p^2=(2.9\,\text{GeV})^2$.
                                                         Similar results using different parameterizations of $\alpha(k^2)$
                                                         have been obtained in \cite{Bhagwat:2003vw,Fischer:2005nf,Fischer:2007ea}.}
            \label{fig:dse-lattice}
            \end{center}
            \bigskip
            \end{figure}

            The correct implementation of chiral symmetry and its dynamical breaking
            is guaranteed if the AVWTI is satisfied,
	    which is achieved, e.\,g., in the rainbow-ladder truncation.
            Using the STIs in Landau gauge: $Z_{1F} = Z_2/\tilde{Z}_3$ and $Z_g \, \tilde{Z}_3\, Z_3^{1/2} = 1$,
            where $\tilde{Z}_3$, $Z_3$ and $Z_g$ are ghost, gluon and charge renormalization constants,
            the quark self-energy is truncated via the replacement
            \begin{equation}
                g^2 \,Z_{1F}\,D^{\mu\nu}(k,\mu)\, \Gamma^\nu(q,p,\mu) \rightarrow Z_2^2\,\frac{4\pi \alpha(k^2)}{k^2}
		\,T^{\mu\nu}_k \, i\gamma^\nu\;,
            \end{equation}
            which leaves the effective interaction $\alpha(k^2)$ to be determined as the single unknown function
	    in the approach. $\alpha$ is a combination of the
            gluon dressing function and a purely $k^2$-dependent dressing of the vector part of the quark-gluon vertex:
            \begin{equation}
                \alpha(k^2) = \frac{g^2}{4\pi} \frac{1}{Z_2 \tilde{Z}_3} \, Z(k^2,\mu^2) \, \Gamma_1(k^2,\mu^2)\;.
            \end{equation}
            From $Z_g \, \tilde{Z}_3\, Z_3^{1/2} = 1$ and the renormalization-scale dependence
            of these quantities, given by $g \sim 1/Z_g$, $Z \sim 1/Z_3$, $\Gamma_1 \sim Z_2/\tilde{Z}_3$, it follows
            that $\alpha(k^2)$ is independent of the renormalization point.
            Asymptotically, it has to approach the perturbative behavior of QCD's running coupling,
            \begin{equation}\label{dse:asympcoupling}
                \alpha(k^2) \stackrel{k^2\rightarrow\infty}{\longlongrightarrow} \frac{\pi\gamma_m}{\ln{k^2/\Lambda_{QCD}^2}}\;,
            \end{equation}
            where $\gamma_m=12/(11 N_C-2 N_f)$ is the anomalous dimension of the quark propagator (in our calculation we use
	    $N_f=4$). For $k^2\rightarrow 0$, an infrared analysis of the DSEs
            leads to the power laws $Z(k^2)\rightarrow (k^2)^{2 \kappa}$ and $\Gamma_1(k^2)\rightarrow (k^2)^{-1/2-\kappa}$
	    \cite{Alkofer:2006gz,Alkofer:2004it} with $0.5<\kappa<0.7$
	    \cite{Zwanziger:2001kw,Lerche:2002ep,Alkofer:2003jj}. Accordingly, the coupling, as infered from Yang-Mills vertex
	    functions becomes constant.
            On the other hand, the interaction should exhibit sufficient strength at small gluon momenta
            to enable dynamical chiral symmetry breaking and the generation of a constituent-mass scale for the quark.
            This translates into notable non-perturbative enhancements of the quark dressing functions
	    $A(p^2)$ and $M(p^2)$ at infrared momenta, see, e.\,g., \cite{Roberts:1994dr}. Several models for $\alpha$ combining the UV limit with
	    an ansatz in the infrared have been employed in the past and applied to detailed studies of meson physics
	    \cite{Jain:1991pk,Munczek:1991jb,Frank:1995uk,Alkofer:1995jx,Maris:1997tm,Maris:1999nt,Alkofer:2002bp}.
            In our present study we start from the interaction of Ref.\,\cite{Maris:1999nt}
            which was fitted to light pseudoscalar meson data and provides an efficient description of
	    pseudoscalar and vector mesons and their properties (see, e.\,g., \cite{Maris:1997tm,Maris:1999nt,Maris:1999bh,%
Maris:2000sk,Ji:2001pj,Maris:2002mz,Jarecke:2002xd,Holl:2004fr,Krassnigg:2004if,Holl:2005vu,Maris:2005tt,%
Bhagwat:2006pu,Maris:2006ea,Bhagwat:2006xi,Bhagwat:2007rj}).
	    In an attempt to combine Dyson-Schwinger and lattice approaches, several parametrizations for the quark-gluon interaction have been tested
            which are capable of reproducing recent lattice data at higher quark masses \cite{Bowman:2002bm,Bowman:2005vx} for
	    the quark dressing functions $M$ and $A$ \cite{Bhagwat:2003vw,Fischer:2005nf,Fischer:2007ea}. In the same way
            we tune the parameters of the effective interaction of Ref.~\cite{Maris:1999nt} such that the resulting mass function agrees with the quenched
	    lattice results from Ref.~\cite{Bowman:2002bm}. The parametrization of the interaction reads
            \begin{equation}\label{dse:maristandy}
                \alpha(k^2) = \frac{c_\xi\pi}{\omega_\xi^7} \,\left(\frac{k^2}{\Lambda_0^2}\right)^2 \,e^{-k^2/(\omega_\xi^2 \,\Lambda_0^2)} +
                              \frac{\pi \gamma_m \left(1-e^{-k^2/\Lambda_0^2}\right)}{\frac{1}{2}\ln \left\{ e^2 - 1 + \left(
			      1+k^2/\Lambda_{QCD}^2\right)^2 \right\} }\,.
            \end{equation}
            The first part of (\ref{dse:maristandy}) provides the characteristic infrared strength while the second term
            accounts for the ultraviolet behavior of \eqref{dse:asympcoupling}.
	    Furthermore, we use $\Lambda_{QCD} = 0.234\,\text{GeV}$ and $\Lambda_0 = 1\,\text{GeV}$. We note that
            in contrast to the expected infrared behavior this ansatz behaves as $\alpha(k^2)/k^2 \rightarrow \emph{const.}$
            for $k^2\rightarrow 0$. This facilitates the numerics as the self-energy integral is of lesser divergence
	    (for details, see App.~\ref{app:dsesupplement}), but does not play an important role,
	    since the quark DSE is dominated by the interaction at intermediate momenta \cite{Fischer:2007ze}.
            For the light-quark mass, all parameters have been chosen according to the original values
            of the one-parameter model of Ref.\,\cite{Maris:1999nt} to reproduce the phenomenological quark condensate and meson observables at the $u/d$
	    quark mass. There and in subsequent meson studies, the parameter $\omega$ is used to vary the \textsl{width} of the IR
	    part of the interaction, while the infrared strength is kept the same, which always leads to the correct amount of
	    dynamical chiral symmetry breaking. In our present study, we fixed also the value of $\omega$ via the comparison to lattice
	    data for the quark propagator.
	    In order to match the lattice data available at larger quark masses (cf.\,Fig.\,\ref{fig:dse-lattice}),
            both coupling strength $c_\xi$ and coupling width $\omega_\xi$ must diminish with increasing quark masses.
            This anticipates to some extent a further quark-mass dependent structure in the quark-gluon vertex which is missing in
            rainbow-ladder truncation. Our fit ot data in Ref.~\cite{Bowman:2002bm} provides the parametrizations
	    $c_\xi = 0.443/(1 + 4.079\,\xi - 1.026\,\xi^2 )$ and $\omega_\xi = 0.377 + 0.295\, c_\xi $ with the quark mass parameter
            $\xi=M(2.9\,\text{GeV})/\Lambda_{QCD}$. For the fitting procedure only we have replaced our usual value of $N_f=4$ by $N_f=0$.
            The renormalization scale $\mu = 2.9\,\text{GeV}$ has been chosen to allow for an optimal fit to lattice data. In
	    these lattice calculations the scale has been fixed by requiring the string tension to be $\sigma = (440\;\mbox{MeV})^2$.

\subsection{$q\bar{q}$ and $qq$ bound states}\label{sec:2bodyboundstates}

	    The solution of the homogeneous BSE is the on-shell Bethe-Salpeter amplitude (BSA) $\Gamma(p,P)$
	    for a quark-antiquark bound state with relative momentum $p$, total momentum $P$, and mass $M$ (at $P^2=-M^2$):
            \begin{equation}\label{bse:bse}
                \Gamma_{\alpha\beta}(p,P) = \int^\Lambda \frac{d^4 q}{(2\pi)^4} \, K_{\alpha\gamma,\delta\beta}(p,q,P)
		\big\{ S(q_+) \Gamma(q,P) S(-q_-) \big\}_{\gamma\delta}\;,
            \end{equation}
            where Greek indices represent Dirac, color and flavor indices, and the quark and
	    antiquark momenta are $q_+= q+\sigma P$ and $q_-= -q+(1-\sigma) P$. The amplitude's dependence on $p,P$ can
	    be formulated in terms of the three Lorentz invariants $p^2$, $P^2$, and $p\cdot P$, where the latter depends on the
	    angular variable $z:=p\cdot P/\sqrt{p^2 P^2}$. Calculations are simplified if one expands the scalar amplitudes in
	    $\Gamma$ in Chebyshev polynomials in $z$ (for details, see App.~\ref{app:mesondiquark}).
            Due to Lorentz invariance, the momentum partitioning parameter $\sigma \in [0,1]$ can be chosen arbitrarily;
	    for equal quark and antiquark masses
	    a value of $1/2$ technically simplifies the BSA.
            The arguments $q_\pm^2$ of the quark propagator dressing functions in (\ref{bse:bse}) are complex for timelike $P^2$;
	    methods to evaluate the quark propagator in the complex plane of Euclidean four-momentum-squared are discussed in App.\,\ref{app:dsesupplement}.
            The kernel $K$ is the amputated quark-antiquark scattering kernel which is irreducible with respect to a pair of $q\bar{q}$ lines.
            The kernel consistent with the rainbow-truncated kernel of the quark self-energy is
	    a ladder kernel and written as
            \begin{equation}\label{bse:rlkernel}
                K_{\alpha\gamma,\beta\delta} (p,q,P) = Z_2^2 \,\frac{4\pi\alpha(k^2)}{k^2}
                                                      \left(\frac{\lambda^i}{2}\right)_{\!AC} \! \left(\frac{\lambda^i}{2}\right)_{\!BD}
                                                      (i\gamma^\mu)_{\alpha\gamma} \, T^{\mu\nu}_k \, (i\gamma^\nu)_{\beta\delta}\;,
            \end{equation}
            where the $SU(3)_C$ Gell-Mann matrices have been explicitly stated (in our case, the flavor structure in the BSE gives no contribution
	    since we consider an equal-mass system and assume isospin symmetry). This truncation describes an iterated dressed-gluon exchange between the quark and the antiquark and
	    respects the AVWTI and chiral symmetry:
            in the chiral limit it ensures that the ground-state pseudoscalar mesons are the massless Goldstone bosons
            linked to dynamical chiral symmetry breaking, while for finite current quark masses
            it leads to a generalization of the Gell-Mann-Oaks-Renner relation\,\cite{Maris:1997hd}.
            The spin structure of the amplitudes for different sets of quantum numbers and the method for
	    solving the BSE are presented in App.\,\ref{app:mesondiquark}.


            \begin{table}
                \begin{center}

                \caption{Mass, strength and width parameters in the coupling Eq.\,\eqref{dse:maristandy} at the physical point.
                         The meson masses were obtained by solving pseudoscalar and vector meson BSEs, the pion decay constant
			 from Eq.\,\eqref{bse:pdc}.}
                \begin{tabular}{rrrrrr} \label{tab:dse1}
                 $\xi$     &  $c_\xi$   &   $\omega_\xi$  &   $m_\pi$              &   $f_\pi$            &   $m_\rho$           \\ \hline
                 $0.036$   &  $0.387$   &   $0.491$       &   $138\,\text{MeV}$    &   $92\,\text{MeV}$   &   $727\,\text{MeV}$
                \end{tabular}

                 \caption{
                          Possible constituent-quark mass scales: $M(0)$, $M_E$ defined by $M_E^2:=M^2(p^2)=p^2$,
			  and $m_q=|\text{Im}\sqrt{p^2_\text{sing}}|$,
                          where $p^2_\text{sing}$ is the location of the singularity (here: a pair of complex conjugate poles)
                          in $\sigma_v$, $\sigma_s$ which is closest to the origin ---
                          see also the discussion in App.\,\ref{app:sing}.
                          The values for the current quark mass and quark condensate at $2\,\text{GeV}$ were obtained
                          by evolving the renormalization-point independent current mass and condensate as determined from the perturbative tails, Eq.\,\eqref{dse:asymptoticmassf},
                          according to the 1-loop formulas employing the quenched scale
                          $\Lambda_{QCD}^{\overline{MS}}=0.225\,\text{GeV}$ \cite{Fischer:2005nf,Capitani:1998mq}.
                         }
                \begin{tabular}{rrrrr} \label{tab:dse2}
    $m_{2\,\text{GeV}}^{\overline{MS}}$   &   $M(0)$ & $M_E$  &  $m_q$     &   $\langle\bar{q}q\rangle_{2\,\text{GeV}}^{\overline{MS}}$    \\ \hline
    $4.8\,\text{MeV}$       &   $0.37\,\text{GeV}$   & $0.32\,\text{GeV}$   &  $0.52\,\text{GeV}$   &   $-(255\,\text{MeV})^3$
                \end{tabular}

                \end{center}
            \end{table}

            \begin{figure}[hbt]
            \begin{center}
            \includegraphics[scale=0.38,angle=-90]{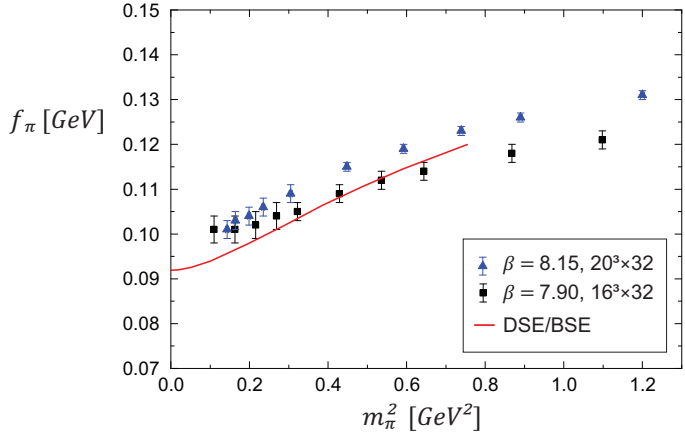}
            \caption[Pion decay constant]{The pion decay constant $f_\pi$ determined from Eq.\,\eqref{bse:pdc} vs. quenched lattice results \cite{Gattringer:2005ij}.}
            \label{fig:fpi}
            \end{center}
            \bigskip
            \end{figure}

            By employing the rainbow-ladder truncation of the DSEs, we link our study of nucleon properties to sophisticated, consistent
	    meson studies. Our results for diquark amplitudes are connected directly to corresponding meson results in this truncation
	    which has provided a good description of pseudoscalar and vector meson observables.
            Table\,\ref{tab:dse1} shows the parameters appearing in the effective coupling at the physical light-quark mass, defined
	    by $m_\pi = 138\,\text{MeV}$, together with basic meson observables.
            Further tensor structure in the quark-gluon vertex (especially a scalar part $\sim l^\nu$ in Eq.~\eqref{qgv:structure})
	    seems to be necessary to successfully describe other meson channels \cite{Alkofer:2002bp,Watson:2004kd,Cloet:2007pi}.
            Efforts towards going beyond rainbow-ladder truncation have been made, e.\,g.,
	    in \cite{Fischer:2007ze,Watson:2004kd,Bhagwat:2004hn,Watson:2004jq,Matevosyan:2006bk}, but are still
	    numerically unaffordable in our study of the nucleon.
	
	    By adjusting the coupling to lattice data, the current-mass (or, equivalently, pion-mass) dependence of meson
	    and nucleon properties are completely determined by
            the mass dependence of $c_\xi$ and $\omega_\xi$ in the effective coupling and therefore directly relatable to quenched lattice
	    results for the quark propagator. In this way, all parameters are fixed and nucleon properties are predictions in the model.
	    As an example we have plotted our results for the pseudoscalar meson leptonic decay constant vs.~corresponding lattice
	    data in Fig.~\ref{fig:fpi}.
	    In Tab.~\ref{tab:dse2} we list numbers characterizing the quark propagator dressing functions.


\section{The diquark ansatz for the 2-quark scattering matrix}\label{sec:dq}
            After having specified the quark propagator, the quark-antiquark and therefore also the quark-quark kernel in
	    terms of a rainbow-ladder truncation, the relativistic Faddeev equation \eqref{bs:faddeevtruncated} can be solved numerically.
            However, in practice the Faddeev amplitude $\Psi_{\alpha\beta\gamma}$ as a 3-particle amplitude
            depends on three four-momenta and in the case of the nucleon consists of 32 Dirac tensors,
            which renders a direct solution of Eq.\,\eqref{bs:faddeevtruncated} numerically expensive.

\subsection{Diquark correlations}
            A possible simplification is to reduce the three-body to a two-body problem.
            The underlying assumption of the Faddeev truncation in Sec.\,\ref{sec:boundstateeq} was to view correlations of two quarks as the
            dominant structure in the nucleon. This is motivated by the observation that colored two-quark states
            can appear in an $SU(3)_C$ anti-triplet or sextet configuration, allowing the formation of
            a color-singlet nucleon out of the anti-triplet state together with a color-triplet quark.
	    In the following, we identify the dominant structures
            of these two-quark correlations in the form of diquarks, leading to a description
            of the nucleon as a bound state of quark and diquark. Further motivation of the significance of the
	    diquark concept has recently come from investigations of diquark confinement in Coulomb-gauge QCD \cite{Alkofer:2005ug}.

            \begin{figure}[hbt]
            \begin{center}
            \medskip
            \includegraphics[scale=0.46,angle=-90]{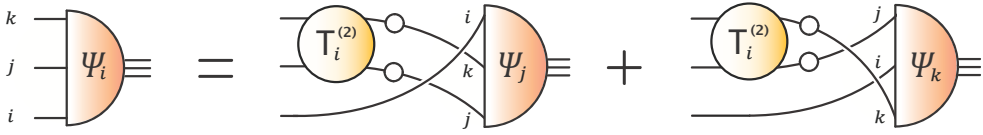}
            \caption[Relativistic Faddeev equation 2]{
                                         Relativistic Faddeev equation \eqref{dq:faddeev} which involves the 2-quark scattering matrix
                                         instead of the 2-quark kernel.} \label{fig:faddeevtruncated2}
            \end{center}
            \medskip
            \end{figure}

\begin{figure}[hbt]
            \begin{center}
            \includegraphics[scale=0.51,angle=-90]{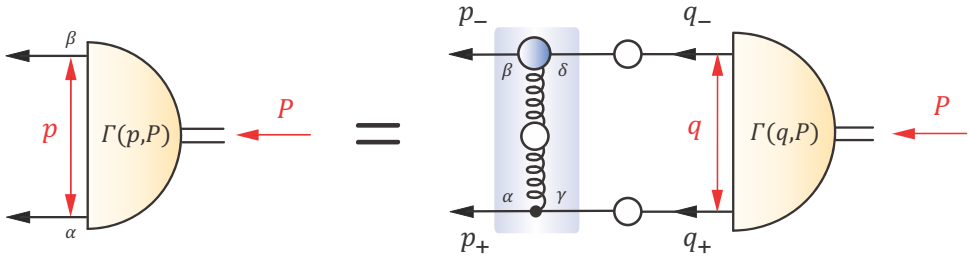}
            \caption[Diquark BSE]{The Diquark BSE \eqref{dq:bse}.} \label{fig:BSE}
            \end{center}
            \end{figure}

            The first step in such a construction is to rewrite the relativistic Faddeev equation by replacing
	    the 2-quark kernels $K_i^{(3)} = K_i^{(2)} \otimes \, S_i^{-1}$
            by 2-quark scattering matrices $T_i^{(3)} = T_i^{(2)} \otimes \, S_i^{-1}$ ($i$ denotes the spectator quark).
            In general, they are connected to each other by the relation $T_i^{-1} = K_i^{-1}-G_0$, i.\,e.~Dyson's equation from Eq.\,\eqref{bs:dysonseq}.
            Then it follows from Eq.\,\eqref{bs:faddeevtruncated} that 
            \begin{equation}
                \Psi = T_i \, (K_i^{-1}-G_0)\,\Psi = T_i\,K_i^{-1}\,(\Psi-\Psi_i) = (1+T_i\,G_0)(\Psi-\Psi_i)\;,
            \end{equation}
            where for the moment we omitted the superscripts in $T_i^{(3)}$, $K_i^{(3)}$ and $G_0^{(3)}$.
            The relativistic Faddeev equation then becomes a set of coupled integral equations for the Faddeev
	    components $\Psi_i$ (c.f. Fig.\,\ref{fig:faddeevtruncated2})
            \begin{equation}\label{dq:faddeev}
                \Psi_i = T_i^{(3)}G_0^{(3)}(\Psi_j+\Psi_k) = T_i^{(2)}S_j\,S_k \,(\Psi_j+\Psi_k)\;,
            \end{equation}
            where the components $\Psi_i$ are still three-body amplitudes.
            For distinguishable quarks, $\{ijk\}$ is an even permutation of $\{123\}$; for identical quarks an antisymmetrization
            in the total Faddeev amplitude $\Psi$ is required.
            Determining the two-body T-matrices $T_i^{(2)}$
            from the corresponding two-body kernels $K_i^{(2)}$ according to Eq.\,\eqref{bs:dysonseq}, e.\,g.~by inserting rainbow-ladder kernels,
            would make Eq.\,\eqref{dq:faddeev} simply an intricate way of reexpressing the original equation \eqref{bs:faddeevtruncated}.
            Instead, we follow a different path and aim at an \textit{ansatz} for the T-matrix which is
            (a) based on the assumption that it contains diquark poles at timelike values of the total momentum $P^2$,
            but (b) still retains some of the characteristic features of Eq.\,\eqref{bs:dysonseq}.
            This is realized as a separable sum over
            diquark correlations. We restrict ourselves to the $0^+$ and $1^+$ channels,
            i.\,e.~to scalar and axial-vector diquarks for reasons explained below:
            \begin{equation}\label{dq:tmatrixansatz}
                T_{\alpha\gamma, \beta\delta}(p,q,P) = \sum_{(\mu\nu)}\Gamma^{(\mu)}_{\alpha\beta}(p,P)\,D^{(\mu\nu)}(P^2) \,\bar{\Gamma}^{(\nu)}_{\delta\gamma}(q,-P)\;.
            \end{equation}
            $\Gamma^{(\mu)}_{\alpha\beta}$ are diquark amplitudes analogous to the corresponding meson amplitudes
            and the $D^{(\mu\nu)}$ denote diquark propagators.
            The assumed poles of the T-matrix in the ansatz \eqref{dq:tmatrixansatz} are embedded in the diquark propagators and define the corresponding diquark masses:
            \begin{equation}
                D(P^2) \stackrel{P^2\rightarrow -M_\text{sc}^2}{\longlongrightarrow} \frac{1}{P^2+M_\text{sc}^2}, \quad
                D^{\mu\nu}(P^2) \stackrel{P^2\rightarrow -M_\text{av}^2}{\longlongrightarrow} \frac{T^{\mu\nu}_P}{P^2+M_\text{av}^2}\;,
            \end{equation}
            where $T^{\mu\nu}_P$ again denotes a transverse projector.
            In the same manner as described in Sec.\,\ref{sec:boundstateeq}
            this leads to a homogeneous diquark BSE, $\Gamma = K^{(2)}G^{(2)}_0\Gamma$,
            for a diquark bound-state on the mass shell $P^2=-M^2$ which resembles the meson BSE, Eq.\,\eqref{bse:bse}, and was
	    used for detailed studies of diquarks, e.\,g., in \cite{Maris:2002yu,Maris:2004bp}. The equation is shown
	    diagrammatically in Fig.~\ref{fig:BSE} and reads
            \begin{equation}\label{dq:bse}
                \Gamma^{(\mu)}_{\alpha\beta}(p,P) = \int^\Lambda \frac{d^4 q}{(2\pi)^4} \, K_{\alpha\gamma,\beta\delta}(p,q,P) \left\{ S(q_+) \Gamma^{(\mu)}(q,P) S^T(q_-) \right\}_{\gamma\delta}\;,
            \end{equation}
            where the momenta have been defined in the discussion of Eq.\,\eqref{bse:bse}.
            The replacements $S(-q_-) \rightarrow S^T(q_-)$ and $K_{\alpha\gamma,\delta\beta} \rightarrow K_{\alpha\gamma,\beta\delta}$
            originate from substituting an antiquark- with a quark leg.
            For the sake of consistency the kernel $K$ has to be identified with the rainbow-ladder kernel \eqref{bse:rlkernel}.
            By working out the color factors (see App.\,\ref{app:mesondiquark-general}) one finds that
            the resulting equation for $\Gamma\,C^\dag$ with the quantum numbers $J^P$
            is identical to the equation for a color-singlet $J^{-P}$ meson except for the diquark's coupling strength
            which is reduced by a factor of $2=N_c-1$. This confirms that the interaction in the color anti-triplet diquark
            channel is strong and attractive. The same analysis also shows that the interaction is strong and repulsive in the color
	    sextet channel \cite{Cahill:1987qr,Hecht:2000jh}.
            Comparison with meson phenomenology therefore suggests that the lowest-mass diquarks are the
	    scalar diquarks (the partners of the pseudoscalar mesons),
            followed by axial-vector, pseudoscalar and vector diquarks.
            This was also observed in Bethe-Salpeter \cite{Maris:2002yu} and lattice \cite{Alexandrou:2006cq} investigations
	    and justifies the restriction to
            scalar and axial-vector diquarks for describing light baryons composed of quark and diquark.

            The ansatz \eqref{dq:tmatrixansatz} is useful only if the full 2-quark T-matrix,
            as obtained from the quark-quark kernel via Dyson's equation \eqref{bs:dysonseq},
            indeed contains timelike scalar and axial-vector diquark poles.
            Since asymptotic colored diquark states correspond to timelike poles in the diquark propagators,
            one could suspect a violation of diquark confinement in this case.
            However, the absence of a Lehmann representation of a certain propagator
            is a sufficient but not necessary criterion for confinement of the corresponding state
            due the associated violation of reflection positivity \cite{Alkofer:2000wg,Osterwalder:1973dx,Haag:1992hx}.
            In this way, two-point correlations of colored fields may contain
            real timelike poles in momentum space without contradicting confinement,
            a statement which is also true for the quark propagator \cite{Alkofer:2003jj,Oehme:1994pv,Roberts:2000aa}.
            Free-particle quark and diquark propagators
            can also yield quantitatively meaningful results for hadronic observables (see, e.\,g., \cite{Oettel:2000jj}).
	    The existence of a solution of \eqref{dq:bse} implies that a rainbow-ladder truncation induces such
	    diquark poles, and for our particular approach and methods this pole structure
	    is in fact a necessary prerequisite.
	    On the other hand, the introduction of interaction terms beyond rainbow-ladder truncation in the skeleton
	    expansion of the quark-quark kernel in the diquark BSE removes diquark states from the physical mass spectrum
            because of large repulsive corrections \cite{Bender:2002as,Bender:1996bb,Bhagwat:2004hn,Hellstern:1997nv}.
            Nevertheless, also kernels that do not produce diquark bound states support a physical interpretation
            of the poles in the diquark propagators either in terms of mass scales or inverse correlation length scales
            of the diquark correlations inside a baryon.

            \begin{table}
                \begin{center}
                \caption{Scalar-axialvector diquark mass splitting in the chiral limit. The BSE value is compared
                         to several lattice-QCD results. The units are GeV.}
                \begin{tabular}{rrrrr} \label{tab:dqmass}
                  BSE    & Ref.\,\cite{Wetzorke:2000ez}  &  Ref.\,\cite{Alexandrou:2006cq}  &   Ref.\,\cite{Babich:2007ah} &   Ref.\,\cite{Orginos:2005vr}                \\ \hline
                  $0.21$ & $0.10(5)$                     &  $0.14(1)$                       &   $0.29(4)$                  &
                  $0.36(7)$
                \end{tabular}
                \end{center}
            \end{table}

            \begin{figure}[hbt]
            \begin{center}
            \includegraphics[scale=0.42,angle=-90]{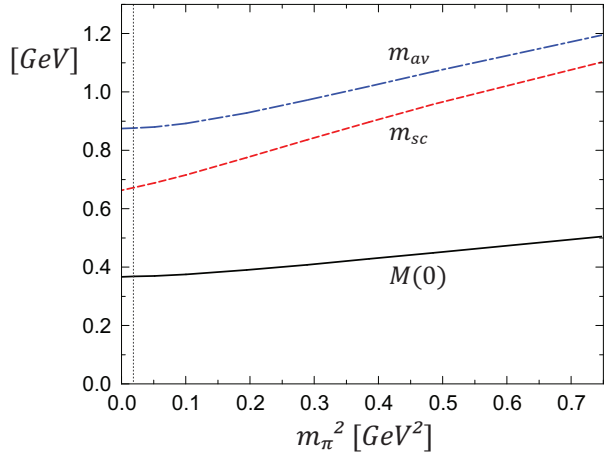}
            \caption[Diquark masses]{Scalar and axial-vector diquark masses together with the quark constituent mass scale $M(0)$
                                     vs.~squared pion mass. The vertical line marks the physical point.} \label{fig:dqmasses}
            \end{center}
            \end{figure}

            Diquark masses are gauge-dependent, but gauge-independent mass differences can be determined from lattice calculations.
            Several such investigations have been performed over the last years
            in different approaches and with various fermion actions
            \cite{Alexandrou:2006cq,Wetzorke:2000ez,Babich:2007ah,Orginos:2005vr,Hess:1998sd}. While they exhibit quite different quantitative
            results, the common qualitative feature is that the mass splitting between scalar and axial-vector diquark
            in the chiral limit is of the size of hundred to several hundred MeV (see Tab.~\ref{tab:dqmass}) and decreases with increasing
            current quark mass; Fig.\,\ref{fig:dqmasses} shows scalar and axial-vector masses together with a measure of the quark
            constituent mass as result of the DSE-BSE approach employed in this work.


\subsection{Offshell and asymptotic behavior of the T-matrix}

            By solving the scalar and axial-vector diquark BSEs, Eq.\,\eqref{dq:bse},
            the quark-quark scattering matrix is known at the diquark mass poles,
            i.\,e., at $P^2= -M_\text{sc}^2$ and $P^2=-M_\text{av}^2$.
            The description of baryons as composites of quark and diquark also requires the knowledge of the T-matrix
            for general diquark momenta. The ansatz \eqref{dq:tmatrixansatz} dictates its off-shell behavior
            via both off-shell diquark amplitudes and diquark propagators.
            Inverting this ansatz and inserting Dyson's equation \eqref{bs:dysonseq} determines
            the inverse scalar and axial-vector diquark propagators for all values of $P^2$
            (the superscripts in $T^{(2)}$, $K^{(2)}$ and $G_0^{(2)}$ are omitted):
            \begin{equation}\label{dq:prop1}
                D^{-1} = \bar{\Gamma}\,T^{-1}\,\Gamma = \bar{\Gamma}\,\left( K^{-1}-G_0 \right)\,\Gamma\;.
            \end{equation}
            Eq.\,\eqref{dq:prop1} necessitates knowledge of the diquark amplitudes' off-shell behavior.
            This cannot be determined without further information; a possible ansatz
	    is constructed in App.\,\ref{app:mesondiquark-offshell}.

            From Eq.\,\eqref{bs:dysonseq} one infers that $T$ is asymptotically dominated
            by the gluon ladder exchange which is constant in $P^2$: the ladder diagram is the lowest term in a skeleton expansion of
            $K$ and independent of $P^2$ (all higher contributions drop off for $P^2 \rightarrow \infty$) and, as a product of two quark propagators,
            $G_0\sim P^{-2}$ for $P^2 \rightarrow \infty$.
            Eq.\,\eqref{dq:tmatrixansatz}, as a sum of two separable terms, is probably no accurate representation of the ladder kernel in the ultraviolet,
            but the correct asymptotic power behavior $\Gamma\,D\,\bar{\Gamma} \rightarrow const.$ is guaranteed via Eq.\,\eqref{dq:prop1}.
            If the diquark amplitudes asymptotically behave as $P^\lambda$, then $D^{-1}\rightarrow P^{2\lambda}$ such that $T$ becomes
            constant in the ultraviolet.
            Eq.\,\eqref{dq:prop1} is explicitly worked out in App.\,\ref{app:mesondiquark-dqprop}.


\section{The quark-diquark BSE}\label{sec:qdqbse}

            All elements of the relativistic Faddeev equation \eqref{dq:faddeev}, i.\,e., the dressed-quark propagator and
	    the quark-quark scattering matrix, have been specified in the previous sections.
            The separability of the diquark ansatz \eqref{dq:tmatrixansatz} for the T-matrix allows for the reduction of the original three-body equation
            for the nucleon to a two-body Bethe-Salpeter equation for a quark-diquark bound state.
            Up to symmetrization, quark-diquark amplitudes are introduced by removing a diquark amplitude and
            propagator from the Faddeev component $\Psi_{\alpha\beta\gamma}$ of Eq.\,\eqref{dq:faddeev}:
            \begin{equation}\label{nuc:defqdqamp}
                \Psi_{\alpha\beta\gamma}(p,p_r,P) = \sum_{a,b} \Gamma_{\beta\gamma}^a(p_r,p_d) \, D^{ab}(p_d) \, \left\{ \Phi^b(p,P)\,u(P)
		\right\}_\alpha\;.
            \end{equation}
            $a,b,c$ are diquark indices: $\Gamma^5$ denotes the scalar and $\Gamma^{a=1\dots 4}$ the axial-vector diquark amplitude;
            the diquark propagator $D^{ab}$ is either scalar ($a,b=5$) or axial-vector ($a,b=1\dots 4$).
            The momenta $p$ and $p_r$ are the relative momenta between quark and diquark and within the diquark, $p_d$ and $P$
            are total diquark and nucleon momenta.
            The spinors $u_\alpha(P)$ are solutions of the Dirac equation in Euclidean space.
            The remaining matrix-valued Bethe-Salpeter amplitudes $\Phi^b_{\alpha\beta}(p,P)$ contain one (scalar or axial-vector) diquark and two fermion legs and
            feature a decomposition constructed from the same Dirac basis elements as used in the meson and diquark case (see App.\,\ref{app:qdqamp}).
            The spin- and isospin-1/2 nucleon is therefore a sum of scalar and axial-vector diquark correlations, and the quark-diquark
            amplitude describes the relative momentum correlation between quark and diquark.
            Inserting the ansatz \eqref{dq:tmatrixansatz} for the T-matrix together with \eqref{nuc:defqdqamp}
            into the relativistic Faddeev equation \eqref{dq:faddeev} yields a quark-diquark Bethe-Salpeter equation
            \begin{equation}\label{nuc:bse}
                \Phi_{\alpha\beta}^a(p,P) = \int \frac{d^4 k}{(2\pi)^4} \, \left\{ K^{ab}(p,k,P) \, S(k_q) \, \Phi^c(k,P) \right\}_{\alpha\beta} \, D^{bc}(k_d)\;,
            \end{equation}
            where the quark-diquark kernel is given by
            \begin{equation}\label{nuc:kernel}
                K^{ab}_{\alpha\beta}(p,k,P) = \left\{ \Gamma^b(k_r,k_d) \, S^T(q) \, \bar{\Gamma}^a(p_r,-p_d) \right\}_{\alpha\beta}\;,
            \end{equation}
            which couples scalar and axial-vector diquark amplitudes (i.e., $a,b=1\dots 5$).
            It describes an iterated exchange of roles between the spectator quark and the quarks which constitute the diquark;
            this quark exchange generates the attractive interaction that binds quarks and diquarks to a nucleon.

            \begin{figure}[hbt]
            \begin{center}
            \medskip
            \includegraphics[scale=0.47,angle=-90]{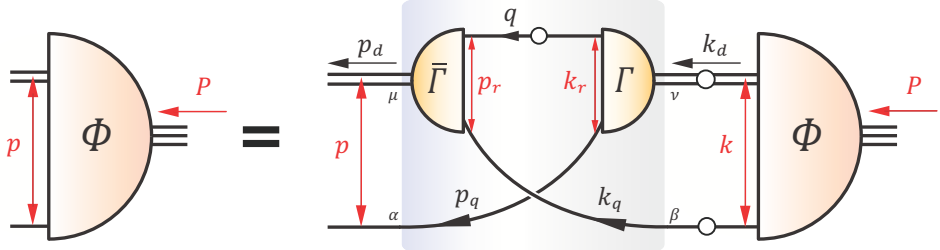}
            \caption[Quark-Diquark BSE]{The quark-diquark BSE \eqref{nuc:bse} in
            pictorial form. } \label{fig:qdqbse}
            \medskip
            \end{center}
            \end{figure}

            The momenta in Eqs.\,(\ref{nuc:bse},\,\ref{nuc:kernel}) are (cf.\,Fig.\,\ref{fig:qdqbse}):
            \begin{equation} \label{nuc:momenta}
                \begin{aligned}
                    p_q &= p+\,P\;,  \\
                    k_q &= k+\eta\, P\;,
                \end{aligned} \qquad
                \begin{aligned}
                    p_d &= -p+(1-\eta)\,P\;,   \\
                    k_d &= -k+(1-\eta)\,P\;,
                \end{aligned} \qquad
                \begin{aligned}
                    q &= p_d-k_q,  \\
                    k_r &= (1-\sigma)\,p_q-\sigma\,q\;, \\
                    p_r &= (1-\sigma)\,k_q-\sigma\,q\;.
                \end{aligned}
            \end{equation}
            Again, the momentum partitioning parameters $\sigma,\,\eta \in [0,1]$ for diquark and quark-diquark amplitudes
            are arbitrary since in a covariant description there is no unique definition of a relative momentum.
            Translation invariance implies that, e.\,g., for each BSE solution $\Phi(p,P,\eta)$ a family of solutions
            of the form $\Phi(p+(\eta'-\eta)P,P,\eta')$ exists \cite{Oettel:2002wf}.
            We set $\sigma=1/2$
            but keep $\eta$ as a variable since it can be used to ease the constraints caused by singularity structures in the complex
            plane (cf. App.\,\ref{app:sing}).
            By working out the color and flavor factors of the quark-diquark amplitudes, given in Eq.\,\eqref{nuc:amplitudes},
            and of the diquark amplitudes (\ref{dq:sc},\,\ref{dq:av}), the BSE kernel picks up a color-flavor factor
            \begin{equation}
            -\frac{1}{2}
            \left(%
            \begin{array}{cc}
              1         & -\sqrt{3} \\
              -\sqrt{3} & -1        \\
            \end{array}%
            \right)\;,
            \end{equation}
            where the first row (column) represents the scalar part of the
            kernel and the second row (column) the axial-vector part.

            Eq.\,\eqref{nuc:bse} is solved using similar methods
            to the meson- and diquark-BSE cases, in particular: (a) reduction to a system of homogeneous one-dimensional integral equations by
            a Chebyshev expansion of the coefficient functions in their angular variable, and (b) introducing an artificial eigenvalue
            which becomes $1$ at the bound-state mass. These procedures are explained in detail e.\,g.~in \cite{Oettel:2001kd}.
            Solving the quark-diquark BSE yields the nucleon mass and quark-diquark amplitudes on the nucleon's mass shell.
            Since the calculation can in principle be performed at arbitrary current masses and the corresponding pion mass is easily obtained,
            one can determine the nucleon mass as a function of the pion mass and directly compare it with lattice
	    investigations and chiral extrapolations.
            Such results are presented in Sec.\,\ref{sec:results}.


\section{Electromagnetic current}\label{sec:em}

            The construction of an electromagnetic current operator in the framework of Bethe-Salpeter equations was first treated
            by Mandelstam \cite{Mandelstam:1955sd}. In our approach, incoming and outgoing nucleon states are described by quark-diquark
            amplitudes. In terms of the interaction with an external current, the baryon is resolved into its constituents, quark and diquark
            and the interaction between them, to each of which the current can couple\cite{Oettel:1999gc}. A systematic procedure for the construction
            of the nucleon-photon vertex based on electromagnetic current conservation is the "gauging of equations" prescription
            \cite{Haberzettl:1997jg,Kvinikhidze:1998xn,Kvinikhidze:1999xp}.
            In this context, "gauging", formally denoted by $T \rightarrow T^\mu$, is a derivative: it is linear and satisfies Leibniz' rule.
            For a general $n$-body bound-state amplitude $\Psi$
            the current matrix is defined as the residue of the gauged scattering matrix at the bound-state pole with mass $M$:
            \begin{equation}
                T \stackrel{P^2\rightarrow -M^2}{\longlongrightarrow} \frac{\Psi \bar{\Psi}}{P^2+M^2}, \quad
                T^\mu \stackrel{\tilde{P}^2\rightarrow -(M^2+Q^2/4)}{\longlonglongrightarrow} \frac{\Psi J^\mu \bar{\Psi}}{(\tilde{P}^2+M^2+Q^2/4)^2}\;,
            \end{equation}
            where $K$ is the interaction kernel, $T$ is the scattering matrix
            and $G_0$ the product of $n$ propagators; $P$ is the total momentum,
	    \begin{equation}
	    \tilde{P}=(P_i+P_f)/2
	    \end{equation}
	     is the Breit momentum
            (i.\,e., the average of incoming and outgoing total momenta), and $Q=P_f-P_i$ the photon momentum.
            From $T^\mu = - T \left( T^{-1} \right)^\mu T$, $T^{-1}=K^{-1}-G_0$ and $\left( K^{-1}\right)^\mu = -K^{-1} K^\mu K^{-1}$
            one finds
            \begin{equation}
                T^\mu \stackrel{\tilde{P}^2\rightarrow -(M^2+Q^2/4)}{\longlonglongrightarrow} T \left( G_0^\mu + G_0\, K^\mu G_0 \right) T
            \end{equation}
            and therefore
            \begin{equation}\label{ff:current}
                J^\mu = \bar{\Psi} \left( G_0^\mu + G_0 K^\mu G_0 \right) \Psi\;.
            \end{equation}

            In our present context, $\Psi$ has to be identified with the quark-diquark amplitude $\Phi$, $T$ is the \textit{quark-diquark} scattering matrix,
            $G_0 = S\, D$ the product of a dressed quark and diquark propagator, and
            $K = \Gamma \,S\, \bar{\Gamma}$ the quark-diquark kernel describing the quark exchange.
            The quark-photon and diquark-photon vertices are defined as the gauged inverse propagators:
            $\Gamma^\mu_\text{q} := -\left(S^{-1}\right)^\mu$ and $\Gamma^\mu_\text{dq} := -\left(D^{-1}\right)^\mu$.
            The gauged diquark amplitudes (or \textit{seagulls}) $\Gamma^\mu =: M^\mu$ describe the photon coupling to the diquark amplitudes. The ingredients of the current matrix are therefore
            written as
            \begin{align}
                G_0^\mu &= \left( S \,D \right)^\mu = S\,\Gamma^\mu_\text{q}\, S \, D + S\, D \,\Gamma^\mu_\text{dq} \,D\;, \\
                K^\mu &= \left( \Gamma \,S \,\bar{\Gamma} \right)^\mu = M^\mu \,S \,\bar{\Gamma} + \Gamma \,S \,\Gamma^\mu_\text{q} \,S \,\bar{\Gamma} + \Gamma \,S \,\bar{M}^\mu\;.  \label{ff:gaugedkernel}
            \end{align}
            These diagrams are worked out in detail in App.\,\ref{app:em} and depicted in Fig.\,\ref{fig:current}.
            The current matrix elements related to the gauged quark-diquark propagator $G_0^\mu$ constitute the impulse-approximation diagrams
            where the photon couples to either quark or diquark line. The premise of current conservation also entails the necessity of including the diagrams
            containing the gauged kernel \cite{Oettel:1999gc}. The electromagnetic current of a baryon in a quark-diquark framework is therefore completely specified by identifying the quark-photon vertex,
            the scalar and axial-vector diquark-photon vertices and an ansatz for the seagull terms. These quantities are constrained
            by Ward-Takahashi identities and thereby related to quark and diquark propagators and diquark amplitudes which have already
            been determined previously. This is demonstrated in Apps.\,\ref{app:vertices} and \ref{app:seagulls}.

            In order to establish a link between the electromagnetic current \eqref{ff:current} and electromagnetic form factors,
            one recalls that the most general form of the nucleon-photon current is obtained by sandwiching the fermion-photon vertex
            between spinors $\bar{u}(P_f,s_f)$, $u(P_i,s_i)$ which are solutions of the Dirac equation: $\Lambda_+(P_i) u(P_i,s_i)=u(P_i,s_i)$,
            $\bar{u}(P_f,s_f) \Lambda_+(P_f) = \bar{u}(P_f,s_f)$, where $P_i$ and $P_f$ with $P_i^2=P_f^2=-M^2$ are initial and final nucleon momenta,
            $s_i,\,s_f=\pm$ are the spin labels, and e.g. $\Lambda_+(P_i)=\sum_{s_i} u(P_i,s_i)\bar{u}(P_i,s_i)=\{\mathds{1} + \Slash{P_i}/(i M)\}/2$ is
            the positive-energy projector of the incoming nucleon. The matrix-valued current of Eq.\,\eqref{ff:current} is obtained by
            removing these spinors via taking the spin sums, i.e. contracting the vertex with the projectors themselves:
            \begin{equation}\label{ff:curr}
            \begin{split}
                J^\mu(Q^2) &= \Lambda_+(P_f) \big( F_1 \,i\gamma^\mu - F_2 \,i\sigma^{\mu\nu} Q^\nu \big)
                             \Lambda_+(P_i) \\
                           &= \Lambda_+(P_f) \big( (F_1+2M F_2)\, i\gamma^\mu - 2F_2 \,\tilde{P}^\mu \big)
                             \Lambda_+(P_i)\;,
            \end{split}
            \end{equation}
            with $\sigma^{\mu\nu} = -i [\gamma^\mu, \gamma^\nu ]/2$.
            The second row was obtained by using the Gordon identity
            \begin{equation}
                \Lambda_+(P_f) \left( \frac{i \sigma^{\mu\nu}}{2}Q^\nu - \tilde{P}^\mu + i M \gamma^\mu \right) \Lambda_+(P_i)=0\;.
            \end{equation}

            The two dressing functions $F_1$ and $F_2$ are the Dirac and Pauli form factors and depend only on $Q^2$
            since on the nucleon's mass shell $\tilde{P}^2=-(M^2+Q^2/4)$ and $\tilde{P}\cdot Q=0$. $F_1$ is dimensionless and $F_2$ has the dimension of
            an inverse mass; for $Q^2=0$ they reduce to the nucleon's charge and anomalous magnetic moment.
            Note that we did not introduce the Pauli form factor in terms of nuclear magnetons, $F_2 = \hat{F}_2/(2M)$ because
            the nucleon mass $M$ employed here is the result of the quark-diquark BSE and usually not identical to the physical
            mass $M^\text{exp}=0.94\,\text{GeV}$. Furthermore, $M$ will increase when going to larger current masses and
            it is advantageous to use magnetons of a fixed mass scale for comparison with lattice
            calculations.

            The requirement $F_1(Q^2=0)=1$ for the proton is automatically satisfied if the quark-diquark amplitudes are
            canonically normalized via Eq.\,\eqref{nuc:normalization} \cite{Oettel:1999gc}.
            The Sachs electric and magnetic form factors, defined as
            \begin{equation}\label{ff:sachs}
                   G_E = F_1 - 2M \tau F_2\;, \qquad  2M G_M = F_1 + 2M F_2\;,
            \end{equation}
            with $\tau=Q^2/(4M^2)$, can be extracted from the current \eqref{ff:curr} via
            \begin{equation}
                G_E = \frac{M}{2\tilde{P}^2} \,\text{Tr} \left\{ J^\mu \tilde{P}^\mu \right\}\;, \qquad
                2 M G_M = \frac{i M^2}{Q^2} \, T^{\mu\nu}_{\tilde{P}} \,\text{Tr} \left\{ J^\mu \gamma^\nu \right\}\;.
            \end{equation}
            Given a form factor $F(Q^2)$, the corresponding electromagnetic radius $r_F$ is defined as the slope
            at zero momentum transfer via the Taylor expansion
            \begin{equation}\label{ff:radii}
                F(Q^2) = F(0) \left\{ 1-\frac{r_F^2}{6} Q^2 + \dots \right\} \quad \Leftrightarrow \quad r_F^2 = -\frac{6}{F(0)} \left.\frac{dF}{dQ^2} \right|_{Q^2=0}\;,
            \end{equation}
            $F(0)$ is the corresponding electric charge or magnetic moment which
            in the case of the neutron's electric form factor, $G_E^n(0)=0$,
            is omitted in Eq.\,\eqref{ff:radii}.


\section{Numerical results and discussion}\label{sec:results}

            To facilitate the discussion, we shortly recapitulate the steps employed in this work.
            Starting from the general relativistic three-body equation for the baryon's amplitude \eqref{bs:boundstateeq},
            we neglected the three-body irreducible kernel, assuming that correlations of two quarks are dominant in the nucleon,
            and arrived at a relativistic Faddeev equation \eqref{bs:faddeevtruncated}, which we converted to \eqref{dq:faddeev} such
	    that it involves the two-quark scattering matrix $T$.
	    Instead of determining $T$ from Dyson's equation \eqref{bs:dysonseq}, we employed a separable ansatz \eqref{dq:tmatrixansatz}
	    involving diquark amplitudes and a diquark propagator which exhibits timelike poles at scalar and axial-vector diquark masses.
	    With this simplification Eq.\,\eqref{dq:faddeev}
            could be reformulated in terms of a quark-diquark Bethe-Salpeter equation \eqref{nuc:bse} for the baryon amplitude.
            The ingredients of this equation are:
            \begin{itemize}
            \item The dressed-quark propagator determined from its Dyson-Schwinger equation \eqref{dse:qdse} via a rainbow truncation.
                  We employed a parametrization for the quark-gluon interaction \eqref{dse:maristandy} which was adjusted to basic meson observables
                  and to quenched lattice results for larger current-quark masses.
            \item A diquark propagator that is obtained from Eq.\,\eqref{dq:prop1}, which immediately follows from the ansatz for the T-matrix
                  and needs as an input the
            \item diquark amplitudes obtained by solving a diquark BSE \eqref{bse:bse} at the poles in the T-matrix.
	          For offshell diquark momenta, an ansatz
                  is employed for these amplitudes (\ref{dq:scoffshell}-\ref{dq:avbasis}).
            \end{itemize}

            The quark-diquark BSE yields quark-diquark amplitudes which are needed to calculate the electromagnetic current
	    diagrams (\ref{ff:current}-\ref{ff:gaugedkernel}).
            They depend on the quark-photon vertex, the diquark-photon vertices and the seagull vertices.
            For the quark-photon interaction we employed the Ball-Chiu vertex (i.\,e., Eq.\,\eqref{ff:qpv} without the transverse part).
            The diquark-photon vertices are given by \eqref{ff:quarkloop} and the seagulls by Eqs.\,(\ref{ff:seagullssc}-\ref{ff:seagullsav}).
            All of the subsequent results were obtained by retaining only the leading (zeroth) Chebyshev moments of the diquark amplitudes.
            The necessity of this simplification is argued in App.\,\ref{app:sing}. Concerning the off-shell dependence of the diquark
            amplitudes, we use $n=1$ in Eq.\,\eqref{dq:offshellansatz2}, i.\,e., a moderate $\sqrt{P^2}$-like suppression of the non-leading amplitudes
            at perturbative momenta. Occasionally we will also compare with the choice $n=2$.

            \begin{table}
                \begin{center}

                \caption{Results for pion decay constant and meson, diquark and nucleon masses (in GeV) at the physical point, defined by $m_\pi=138$ MeV.
                         "Full" denotes the complete calculation and "dom." the case where only the dominant diquark amplitudes $\sim\gamma^5 C$, $\gamma^\mu C$ are retained.
                         The deviations compared to Table \ref{tab:dse1} are due to the omission of the angular dependence in meson and diquark amplitudes.}
                \begin{tabular}{r|rrrrr} \label{tab:res1}
                                   &   $f_\pi$     &   $m_\rho$     &   $m_{sc}$     &   $m_{av}$     &   $M_N$           \\ \hline
                 \textbf{full}     &   $0.101$      &   $0.736$      &   $0.676$      &   $0.889$      &   $0.931$         \\[-0.2cm]
                 dom.              &   $0.072$      &   $0.773$      &   $0.644$      &   $0.894$      &   $0.968$         \\ \hline
                 exp.              &   $0.092$      &   $0.770$      &                &                &   $0.94$         \\ \hline
                \end{tabular}
                \bigskip

                \end{center}
            \end{table}

\subsection{Nucleon mass and pion-cloud effects}\label{sec:results:nuclenmass}

\begin{figure}[hbt]
            \begin{center}
            \includegraphics[scale=0.41,angle=-90]{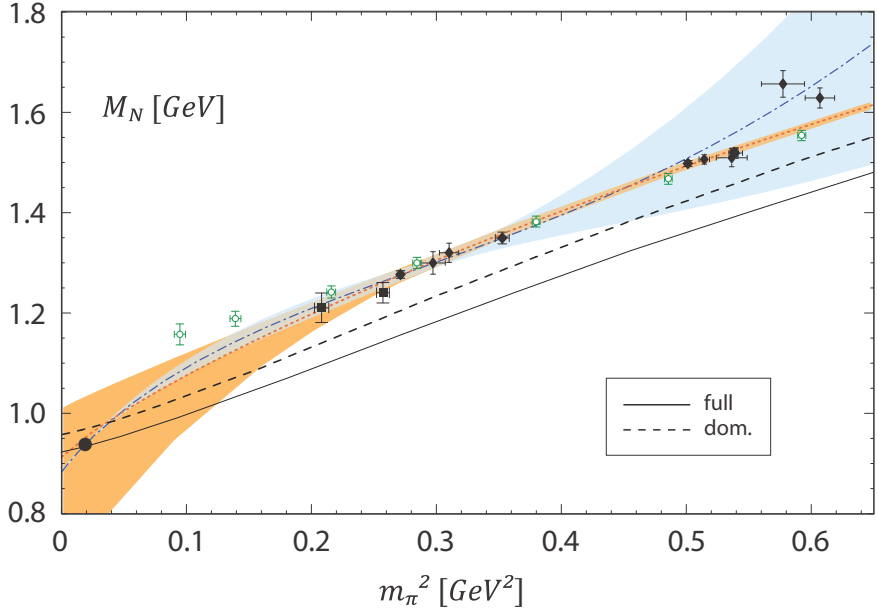}
            \caption[Nucleon mass]{The DSE/BSE result for the nucleon mass (dashed line: dominant diquark amplitudes only,
                                   solid line: full calculation) versus a compilation of contemporary lattice data, both quenched
                                   (open circles \cite{Boinepalli:2006xd}) and unquenched
                                   (filled squares \cite{Frigori:2007wa}, filled diamonds \cite{AliKhan:2003cu}).
                                   Dotted \cite{Leinweber:2003dg} and dash-dotted lines \cite{Procura:2006bj} with respective error band represent
                                   a chiral extrapolation and an interpolation between the physical point and selected lattice data, respectively.
                                   } \label{fig:nucleonmass}
            \end{center}
            \end{figure}

            Table \ref{tab:res1} shows the pion decay constant and meson, diquark and nucleon masses at the physical point
            as obtained from Eqs.\,\eqref{bse:pdc}, \eqref{bse:bse}, \eqref{dq:bse} and \eqref{nuc:bse}.
            The results for $m_\rho$ and the diquark masses are comparable to the values obtained in Refs.\,\cite{Maris:2002yu} and \cite{Maris:1999nt} for $\omega=0.5$ GeV
            (in the present work, $\omega_\xi=0.49$ at the physical point, cf. Table\,\ref{tab:dse1}).
            At first sight, the result $M_N=0.93$ GeV for the nucleon mass seems quite remarkable since
            after fixing the coupling strength \eqref{dse:maristandy} no further parameters (in particular, no observables related to the nucleon)
            have been used as an input of the calculation.
            Matching the experimental nucleon mass therefore implies that the corrections to the truncations made in our calculation
	    cancel. In particular, one would expect corrections
            from dropping the quark-diquark assumption (i.e., reverting Eq.\,\eqref{nuc:bse} to a relativistic Faddeev
	    equation with a ladder truncation in the
            quark-quark channel), corrections from going beyond rainbow-ladder truncation (towards a full 2-quark kernel and quark-gluon vertex),
            and also corrections from irreducible 3-quark contributions.
            An investigation of the large quark-mass behavior (cf. Fig.\,\ref{fig:nucleonmass}) indicates that the cancellation of these
	    effects at the physical point is accidental.
            Still, it shows that the quark-diquark picture accounts for more than $\sim 90\%$ of the values of the nucleon mass
            as obtained from lattice investigations at different current quark masses.

\begin{figure}[hbt]
            \begin{center}
            \includegraphics[scale=0.48,angle=-90]{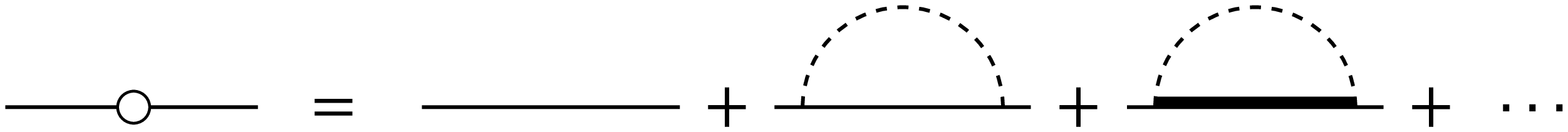}
            \caption[Chiral expansion]{Expansion of the nucleon propagator in chiral perturbation theory.
                                       Solid, dashed and thick solid lines correspond to nucleon, pseudoscalar meson and $\Delta$ degrees of freedom.} \label{fig:chiralexpansion}
            \end{center}
            \end{figure}

            From a phenomenological point of view, an interesting contribution comes from the so-called pion cloud of the nucleon.
            In terms of quarks and gluons, it corresponds to the long-range part of $q\bar{q}$ correlations interacting with the nucleon.
            Established in the cloudy bag model \cite{Theberge:1980ye,Thomas:1982kv,Lu:1997sd}, where the pion field is coupled to a
	    constituent-quark bag \cite{Chodos:1974je},
            it is considered to be an important component of the nucleon's structure at low energies and small quark masses.
            Pion effects can be treated in a systematic way by chiral effective field theory,
            which is an approach to describe low-energy QCD with effective pion, nucleon, and $\Delta$ degrees of freedom \cite{Gasser:1983yg,Bernard:1995dp}.
            In combination with lattice simulations, it has proven to be an efficient tool for describing masses and electromagnetic properties of hadrons \cite{Young:2002cj,Hemmert:2002uh,Gockeler2005}.
            In this framework the nucleon mass near the chiral limit is obtained from the chiral expansion \cite{Young:2002ib}
            \begin{equation}\label{xpt:nucmass1}
                M_N(m_\pi^2) = a_0(\Lambda) + a_2(\Lambda) m_\pi^2 + a_4(\Lambda) m_\pi^4 + \dots + \Sigma(m_\pi^2,\Lambda)\;,
            \end{equation}
            where $a_i$ are the (bare and a priori unknown) parameters appearing in the effective Lagrangian that correspond to the "nucleon core",
            and $\Sigma$ denotes the sum of all meson-loop contributions, i.\,e., to 1-loop order the sum of $N\pi$, $N\Delta\pi$ etc. self-interactions (cf. Fig.\,\ref{fig:chiralexpansion}).
            $\Lambda$ is a generic regularization parameter that appears because effective field theory treats baryons and pions as point particles.
            This is appropriate for long-distance physics but leads to divergences in the loop integrals since the composite substructure is
            not accurately implemented \cite{Donoghue:1998bs}.
            A non-pointlike nucleon-pion interaction is physically equivalent to
            using a momentum cutoff corresponding to the baryon size ($\Lambda \gtrsim 0.2$ GeV $\sim 1$ fm$^{-1}$)
            which picks out the long-distance or low-energy part of the self-energy integrals.
            Then, by expanding in $m_\pi^2$, $\Sigma$ is split into
            cutoff-independent non-analytic terms, $\Sigma_{NN\pi}\sim m_\pi^3$, $\Sigma_{N\Delta\pi}\sim m_\pi^4 \ln{m_\pi}$,
            and cutoff-dependent terms that are even in $m_\pi^2$. Since
            the $\Lambda$-dependence of the latter is of the same type as in the bare coefficients $a_i$, both can be combined
            to renormalized coefficients such that the nucleon mass finally reads \cite{Young:2002ib}
            \begin{equation}\label{xpt:nucmass2}
            \begin{split}
                M_N(m_\pi^2) &= M_N(0) + c_2 m_\pi^2 + c_4 m_\pi^4 + \dots \\
                             &- \frac{3g_A^2}{32\pi f_\pi^2} m_\pi^3 + \frac{3g_A^2}{25\pi f_\pi^2}\frac{3}{4\pi\Delta} m_\pi^4 \ln{m_\pi} + \dots\;,
            \end{split}
            \end{equation}
            where usually the experimental numbers for the pion decay constant ($f_\pi=92.4$ MeV), axial coupling ($g_A=1.26$)
	    and nucleon-delta mass splitting ($\Delta=0.292$ GeV) are inserted.
            The final physical result is expressed in terms of renormalized low-energy constants $M_N(0)$, $c_2$ and $c_4$ which
	    are determined by a fit to lattice data. Through this renormalization procedure the regulator dependence is removed,
	    at least if one would work to all orders of the chiral expansion. To finite order a small residual cutoff dependence
	    is left, therefore a convenient choice of the regulator can be utilized to improve the convergence of the chiral
	    series \cite{Donoghue:1998bs}.

            In order to study the (cutoff-dependent) separation
            of the nucleon mass into a "core" and a "meson cloud" contribution one has to examine Eq.\,\eqref{xpt:nucmass1}, since
            in \eqref{xpt:nucmass2} the meson cloud effects in the chiral limit, $\Sigma(0,\Lambda)$, have already been absorbed into $M_N(0)$.
            For a dipole regulator with $\Lambda\sim 0.8$ GeV, the typical value is of the size $\Sigma(0,\Lambda)\sim -250$ MeV
            \cite{Leinweber:2003dg,Young:2002cj,Young:2002ib}, therefore leading to an expected core mass of $a_0(\Lambda)\sim 1.2$ GeV.
            Different shapes of the regulator can vary this result considerably;
            a dipole form mimics the physical shape of the meson-baryon vertex, e.\,g., in terms of
            the observed axial form factor of the nucleon \cite{Thomas:2001kw}.
            Similar numbers have been reported from a Dyson-Schwinger solution of the nucleon propagator dressed by pions ($\sim -200$ MeV) \cite{Hecht:2002ej,Oettel:2002cw},
            in studies of the cloudy bag model ($-300$ to $-400$ MeV) \cite{Pearce:1986du},
            and in a perturbative quark-diquark study with pointlike pion exchange ($-150$ to $-300$ MeV) \cite{Pearce:1986du}.
            This confirms that nucleon-pion loops are attractive and the binding energy has the effect of lowering the nucleon's mass.
            In agreement with the presumption that pion-loop effects should be suppressed in the heavy-quark regime,
            $\Sigma(m_\pi^2,\Lambda)$ is seen to fall off with increasing pion mass \cite{Young:2002cj}.

            \begin{figure}[hbt]
            \begin{center}
            \includegraphics[scale=0.41,angle=-90]{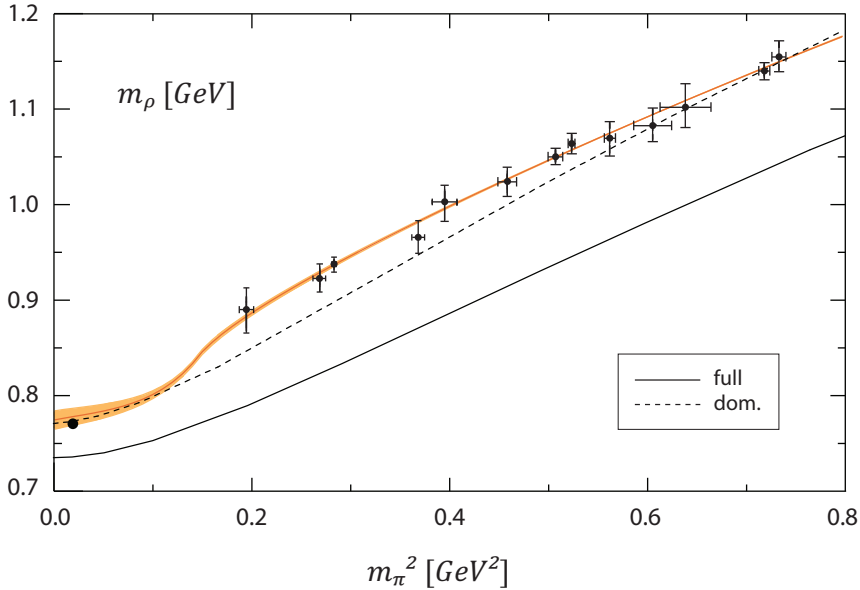}
            \caption[Vector meson mass]{The DSE/BSE result for the $\rho$ mass versus CP-PACS data \cite{AliKhan:2001tx},
                                        together with the chiral extrapolation of Ref.\,\cite{Allton:2005fb}.} \label{fig:rhomass}
            \end{center}
            \end{figure}

            From Fig.\,\ref{fig:nucleonmass} it is clear that pion-cloud effects cannot be the only source of disagreement between the DSE/BSE curves
            and the lattice results, since we underestimate the data, which for large $m_\pi^2$ should eventually reveal the nucleon's quark core, by $\sim 10 \%$.
            On the other hand, a remarkable feature is that the slopes $dM_N/dm_\pi^2$ of both approaches are identical above a certain value of $m_\pi^2$. This can be
            traced back to the coupling strength of Eq.\,\eqref{dse:maristandy} which was fixed to (quenched) lattice results of the quark mass function.
            To evolve from the "quark-diquark core" to the full quark core of the nucleon,
            we consequently have to search for repulsive contributions that are constant in the quark mass.
            This is even more imperative when considering irreducible 3-quark correlations which were neglected through the Faddeev truncation \eqref{bs:faddeevtruncated}:
            the leading diagram in this case is the 3-gluon vertex coupling to all three quark lines.
            It is expected to provide a further attractive contribution to the nucleon structure which,
            at least at the perturbative level, is independent of the current-quark mass.

            It is enlightning that the same qualitative mismatch between BSE/Faddeev and lattice approach
            is already visible in the vector meson case (Fig.\,\ref{fig:rhomass}) where the rainbow-ladder truncation
            is the only effective restriction. The conclusion is that
            dressed gluon exchange alone binds too strongly, and that
            a quark-gluon vertex and quark-(anti)quark kernel beyond rainbow-ladder truncation
            should provide the necessary amount of repulsion in the meson and diquark channels.
            As mentioned in Sec.\,\ref{sec:dq}, such a generalization will probably remove the
            timelike diquark poles in the quark-quark T-matrix but could still maintain a measure
            for the diquark's mass scales which enter the nucleon mass.
            In this respect we note that in Ref.\,\cite{Oettel:2002cw} the full Dyson-Schwinger solution for the nucleon propagator, dressed with pions,
            again \textit{reduced} the nucleon mass shift $\Sigma(0,\Lambda)$ by increasing the width
            of the nucleon-pion vertex above $\sim 1$ GeV (loosely speaking: by shifting more and more short-range "pionic" contributions from the nucleon core into the pion cloud).
            This might be a sign for the short-range $q\bar{q}$ correlations
            producing the repulsive effect that is missing in rainbow-ladder truncation.

            Interestingly, while there seems to be virtually no difference
            between quenched and dynamical lattice data for the nucleon mass at large $m_\pi^2$, unquenched lattice calculations
	    exhibit sizeable effects at the propagator level, e.\,g.~for the quark mass function \cite{Bowman:2005vx}.
            A fit to these dynamical data probably would have diminished the slope $dM_N/dm_\pi^2$ considerably
            but exceeded the parameter bounds within which the ansatz \eqref{dse:maristandy} could be safely applied.
            However, the issue of quenching/unquenching at the \textit{rainbow-ladder} level is elusive, since
            unquenching effects in the gluon propagator alone (i.\,e., via solving the full gluon DSE)
            only affect the interaction $\alpha(k^2)$ in the quark DSE
            in terms of a slight change of its intermediate-momentum shape \cite{Fischer:2005en}.
            More drastic consequences for hadronic observables arise from "unquenching" the $qq$ kernel and quark-gluon vertex
            and have been studied by implementation of an additional effective pion exchange \cite{Fischer:2007ze}.
            Finding an appropriate symmetry-preserving truncation beyond rainbow-ladder
            might not only be the way to provide sufficient repulsive strength in the two-body channel, but
            possibly also the key direction for implementing the pion cloud in the Faddeev approach.

            \begin{table}
                \begin{center}

                \caption{Proton, neutron and isoscalar magnetic moments in units of physical magnetons, i.e. $G_M^{p,n}(Q^2=0)=\mu_{p,n}/(2 M_N^\text{exp})$,
                         and according to Eq.\,\eqref{ff:sachs}: $\mu_{p,n} = M_N^\text{exp}/M_N^\text{calc} + \kappa_{p,n}$, with $\kappa_s=\kappa_p+\kappa_n$.
                         The electric and magnetic radii of proton and neutron are given in units of fm; $r_E^n$ denotes $\sqrt{-(r_E^n)^2}$.
                         "Param." denotes the use of analytic parametrizations for the diquark-photon vertices.}
                \begin{tabular}{r|rrrrrrr} \label{tab:ff1}
                                  &$\mu_p$    &   $\mu_n$      &   $\kappa_s$      &   $r_E^p$      &   $r_E^n$      &   $r_M^p$    &   $r_M^n$       \\ \hline
                 \textbf{full }   &$2.52$     &   $-1.55$      &   $-0.04$         &   $0.67$       &   $0.13$       &   $0.58$     &   $0.57$      \\[-0.2cm]
                 dom.             &$2.70$     &   $-1.68$      &   $ 0.05$         &   $0.69$       &   $0.08$       &   $0.60$     &   $0.60$      \\[-0.2cm]
                 param.           &$3.91$     &   $-2.01$      &   $ 0.89$         &   $0.63$       &   $0.26$       &   $0.45$     &   $0.49$      \\ \hline
                 exp.             &$2.79$     &   $-1.91$      &   $-0.12$         &   $0.87$       &   $0.34$       &   $0.86$     &   $0.88$
                \end{tabular}

                \end{center}
            \end{table}

\subsection{Magnetic moments and electromagnetic radii}\label{sec:results:emstatic}

\begin{figure}[hbt]
            \begin{center}
            \includegraphics[scale=0.46,angle=-90]{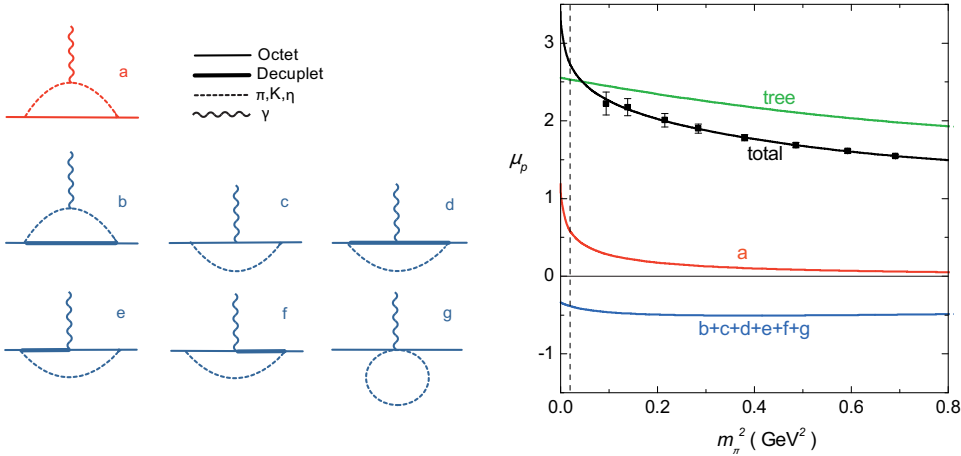}
            \caption[Magnetic moments]{Meson-loop contributions to the proton's magnetic moment in Heavy Baryon Chiral Perturbation Theory \cite{Wang:2007iw}
                                       with lattice data from \cite{Boinepalli:2006xd}. Picture taken from Ref.\,\cite{Wang:2007iw}.} \label{fig:hbxpt}
            \end{center}
            \end{figure}

\begin{figure}[hbt]
            \begin{center}
            \includegraphics[scale=0.49,angle=-90]{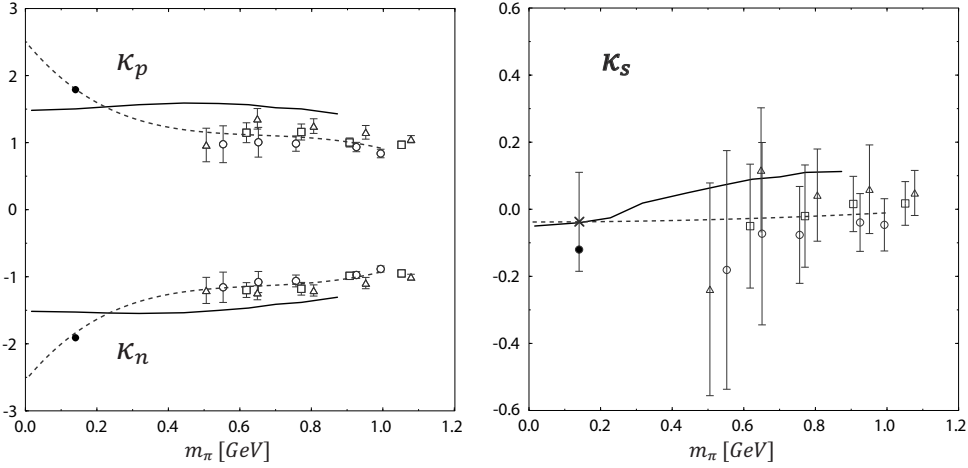}
            \caption[Magnetic moments]{Anomalous magnetic moments of proton and neutron (left) and isoscalar (right).
                                       The DSE/BSE results (solid curves) are compared to quenched QCDSF lattice data and
                                       their chiral extrapolations (dashed curves) \cite{Gockeler2005}.
                                       The dots mark the physical values.} \label{fig:magneticmoments}
            \end{center}
            \end{figure}

            The first and last rows of Table\,\ref{tab:ff1} show results for the magnetic moments and charge and magnetic radii together with the experimental values.
            The calculated magnetic moments of proton and neutron are somewhat smaller than seen in experiment.
            All radii are sizeably underestimated, with the magnetic radii being yet smaller than the electric ones.
            On physical grounds this is exactly what is to be expected from a calculation missing contributions from the pion cloud.
            The chiral expansion for the magnetic moments in chiral perturbation theory when taking into account only the $N\pi$-loop reads \cite{Young:2004tb}
            \begin{equation}
                \mu^{p,n}(m_\pi^2) = \mu^{p,n}(0) \mp \frac{g_A^2 M_N}{8\pi f_\pi^2}  m_\pi + \dots\;, \label{ff:xpt-mom}\\
            \end{equation}
            In the same manner as Eq.\,\eqref{xpt:nucmass2} for the nucleon mass,
            Eq.\,\eqref{ff:xpt-mom} does not reflect the cut-off dependent separation into quark core and pion contributions:
            the pion cloud gives a non-vanishing contribution in the chiral limit which has been absorbed into $\mu^{p,n}(0)$.
            Such a decomposition has been explicitly performed in Ref.\,\cite{Wang:2007iw} using a dipole regulator and a cutoff $\Lambda=0.8$ GeV.
            The result is displayed in Fig.\,\ref{fig:hbxpt} and exhibits a slowly decreasing magnetic moment of the proton's quark core,
            a finite and positive contribution from the pion loop in the chiral limit (and also at the physical point),
            and a finite pion contribution at large quark masses. The decomposition for the neutron looks similar.
            The DSE/BSE results for the anomalous magnetic moments of proton and neutron are shown in the left panel of Fig.\,\ref{fig:magneticmoments}.
            The qualitatively similar behavior of both findings strongly suggests the long-range pion cloud to be the main missing contribution to the present approach.

            The right panel of Fig.\,\ref{fig:magneticmoments} displays the isoscalar combination $\kappa_s=\kappa_p+\kappa_n$.
            From \eqref{ff:xpt-mom} it is clear that the leading non-analytic parts $\sim m_\pi$
            cancel in $\kappa_s$, i.e., up to leading order in the chiral expansion the isoscalar magnetic moment is not quark-mass dependent.
            A sizeable mass dependence of $\kappa_s$ for small $m_\pi^2$ would therefore definitely indicate the absence of other effects apart from pionic corrections.
            The plot shows that this is not the case for small masses but appears above $m_\pi \sim 0.2$ GeV.

\begin{figure}[hbt]
            \begin{center}
            \includegraphics[scale=0.46,angle=-90]{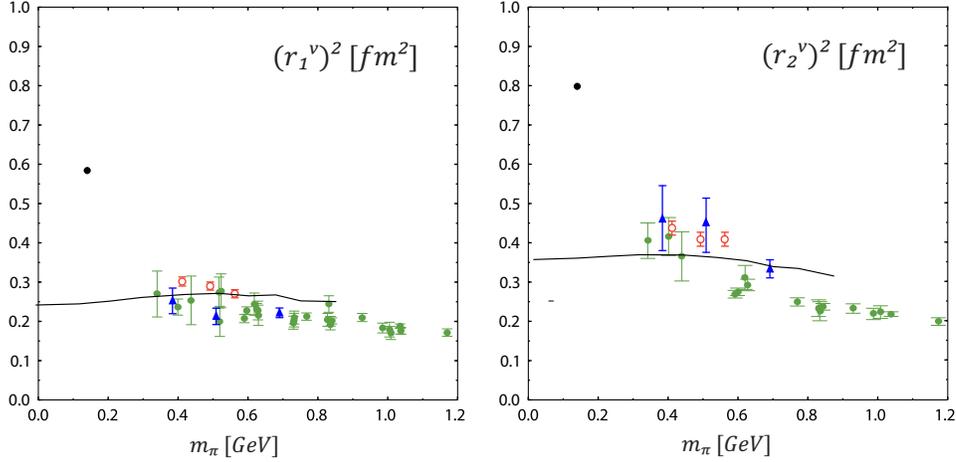}
            \caption[Magnetic moments]{Isovector radii corresponding to the Dirac and Pauli form factors $F_{1,2}^v=F_{1,2}^p-F_{1,2}^n$
                                       compared to quenched (open circles \cite{Alexandrou:2006ru})
                                       and unquenched (filled circles \cite{Gockeler:2007ir}, filled triangles \cite{Alexandrou:2006ru})
                                       lattice data.} \label{fig:radii}
            \end{center}
            \end{figure}

            A chiral expansion similar to \eqref{ff:xpt-mom} exists for the electromagnetic radii of proton and neutron \cite{Bernard:1995dp}.
            In contrast to the magnetic moments the charge radii diverge in the chiral limit since a massless pion can propagate over infinite distances.
            The results shown in Fig.\,\ref{fig:radii} provide further support for a interpretation in terms of a missing pion cloud
            as the DSE/BSE results are found to constitute a flat plateau at the level of the lattice data which are available at large pion masses.

            In Table \ref{tab:ff1} and Figs.\,\ref{fig:nucleonmass}, \ref{fig:rhomass} and \ref{fig:ff} we also compare with the results
            obtained by taking only the dominant scalar and axial-vector diquark amplitudes into account,
            i.\,e., retaining only an $s$ wave contribution in the diquark rest frame.
            To maintain consistency throughout the whole calculation this simplification must be made already at the level of the diquark BSE.
            In this case the general diquark amplitudes (\ref{dq:sc}-\ref{dq:av}) and correspondingly \eqref{mesamp} for the meson case reduce to
            \begin{align}
                \Gamma(q,P) &= f_1^\text{sc}(q^2,P^2)\,i\gamma^5 C\;,   \\
                \Gamma^\mu(q,P) &= f_1^\text{av}(q^2,P^2) \,i\gamma^\mu C\;.
            \end{align}
            All of the previous quark-diquark studies (e.\,g.,
	    \cite{Bloch:1999ke,Bloch:1999rm,Oettel:1998bk,Oettel:1999gc,Hecht:2002ej,Oettel:2000jj,Oettel:2002wf,Alkofer:2004yf,Holl:2005zi}) were carried out under this truncation.
            Since it changes most of the considered observables by $\lesssim 10\%$ compared to the full result
	    (see also Ref.\,\cite{Maris:2002yu}), it can safely be viewed as a reasonable approximation to the overall problem,
            although it may obscure the interpretation of $M_N$ and $m_\rho$ when comparing to lattice data.
            Exceptions are the pion decay constant which is reduced by $\sim 30\%$ and the structure of the electric form factors in Fig.\,\ref{fig:ff}
	    around $Q^2\sim 1\,\mbox{GeV}^2$ .

            Another observation concerns the use of ans\"atze for the diquark-photon vertices.
            For instance, the general axial-vector diquark-photon vertex can be written in the form of Eq.\,\eqref{ff:avdqpv}, of which
            $f_1^\text{av-dq}$ and $f_3^\text{av-dq}$ are the dominant transverse components and correspond to the axial-vector diquark magnetic moment
            \begin{equation}
               \mu_{dq}=f_1^\text{av-dq}(-m_{av}^2,-m_{av}^2,0)=-f_3^\text{av-dq}(-m_{av}^2,-m_{av}^2,0)\;.
            \end{equation}
            Therefore, a possibility to circumvent the intricate vertex \eqref{ff:quarkloop} is
            to use version \eqref{ff:avdqpv} and neglect all the transverse terms except the one involving the magnetic moment,
            $\mu_{dq} \left( \delta^{\mu\beta} Q^\alpha - \delta^{\mu\alpha} Q^\beta \right)$.
            $\mu_{dq}$ may be generalized to include an appropriate $Q^2$ dependence,
            and if only the on-shell form of the vertex is of interest,
            one can additionally contract the vertex with transverse projectors from the external diquark lines \cite{Alkofer:2004yf}.
            Reasonable values for $\mu_{dq}$ are $\sim 2-3$; the full quark-loop vertex \eqref{ff:quarkloop} yields $\mu_{dq} = 2.7$.
            We apply the same procedure to the scalar diquark-photon vertex where we use \eqref{ff:scdqpv} and set $f_1^\text{sc-dq}$ to zero, and
            the scalar-axialvector transition vertex which may be expressed by
            \begin{equation}
                \Gamma^{\mu,5\beta}(p,q) = i \varepsilon^{\mu\beta\rho\lambda} p^\rho q^\lambda \frac{\kappa_{sa}}{M_N}\;,
            \end{equation}
            with $\kappa_{sa}= 2.3$ extracted from the general form \eqref{ff:quarkloop}.
            Table \ref{tab:ff1} shows that these parametrizations lead to inflated und unbalanced magnetic moments of proton and neutron
            since $\kappa_s$ is now large and positive.
            The reason is that the on-shell values of the transversal dressing functions are not enough to describe the full vertex:
            at off-shell momenta (especially large spacelike momenta) there is a considerable amount of attenuation which is not captured
            by the above forms. The main part of the discrepancy is produced by the axial-vector diquark-photon vertex whose contribution to the magnetic moments
            is blown up by an order of magnitude due to the missing off-shell structure.
            If parametrizations of the above types are employed and desired to reproduce the magnetic moments, it is therefore mandatory
            to include a functional dependence on both the photon momentum $Q^2$ \textit{and} the average in- and outgoing momentum $P^2$.

\subsection{Electromagnetic form factors}\label{sec:results:emff}

            \begin{figure}[hbt]
            \begin{center}
            \includegraphics[scale=0.67]{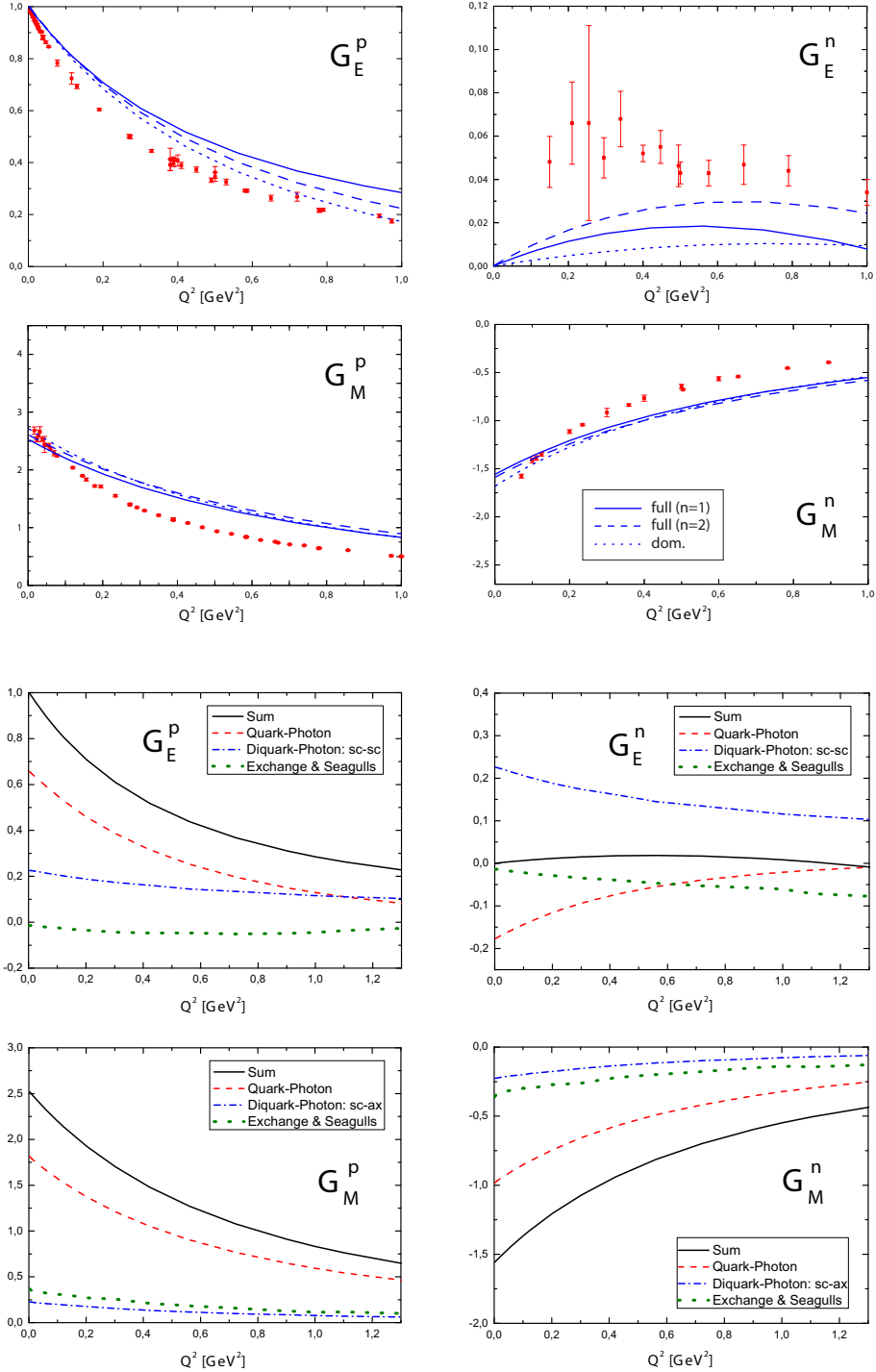}
            \caption[Form factors]{For the experimental points the data sets as selected in Ref.\,\cite{Friedrich:2003iz} are used [Data compiled
	    by P.~Grabmayr, Univ.~T\"ubingen]. } \label{fig:ff}
            \end{center}
            \end{figure}

            Fig.\,\ref{fig:ff} shows the proton and neutron's electromagnetic Sachs form factors at the physical point compared to experimental results (upper part),
            and split into the most dominant contributions from which the relative strengths can be read off (lower part).
            In the case of the neutron form factor one can see that the contributing diagrams are not small by themselves:
            it is current conservation that ensures the cancelation at $Q^2=0$ such that the neutron charge is exactly zero.
            The unit charge of the proton is the result of the quark-diquark normalization condition \eqref{nuc:normalization}.
            The strongest contribution is usually the direct coupling of the photon to the quark line.
            However, the shift of spectator quark into diquark degrees of freedom necessitates the remaining diagrams of Fig.\,\ref{fig:current}.
            For the electric form factors it is the photon coupling to the scalar diquark which is important;
            for the magnetic form factors the scalar-axialvector transition, which is a purely transverse term and does not contribute to current conservation,
            provides a small contribution. The exchange and seagull contributions tend to be small but are necessary to guarantee a conserved current.

            At larger momentum transfers $Q^2\gtrsim 2$ GeV$^2$, pion cloud effects should vanish
            since the structure of the nucleon at small distances ($\lesssim 0.2$ fm) is probed.
            The large-$Q^2$ behavior of the proton's form factor ratio $\mu_p G_E^p/G_M^p$ is therefore a genuine rendition of the nucleon's quark core.
            Since quark and diquark propagators obtained from the quark DSE or via the T-matrix ansatz necessarily contain singularities,
            our applicable $Q^2$ range, depending on model details, is limited to below $1-2$ GeV$^2$,
	    i.\,e., the region where the pion cloud is still effective (for a detailed discussion, see App.\,\ref{app:sing}).
            Access to large $Q^2$ and comparison to the true "quark core" as seen in experiment can therefore only be established by finding appropriate methods to
            evaluate these quantities beyond their dominant singularities and to include the respective residue contributions in the form factor integrals.

\begin{figure}[hbt]
            \begin{center}
            \includegraphics[scale=0.43,angle=-90]{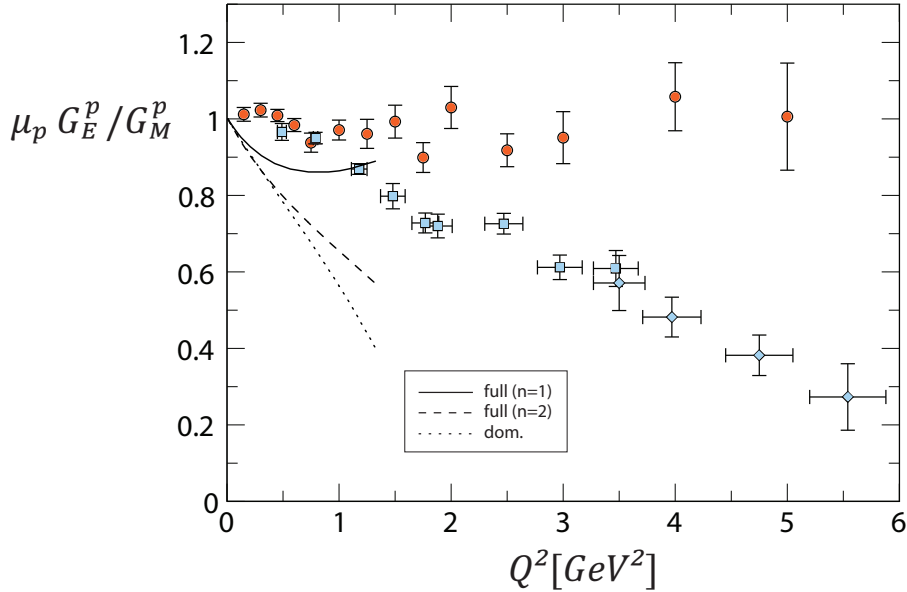}
            \caption[Form factor ratio]{The proton's form factor ratio for different relative behavior of the diquark amplitudes.
                                        Rosenbluth experimental data (circles) are taken from \cite{Walker:1993vj},
                                        polarization transfer data from \cite{Jones:1999rz} (squares) and \cite{Gayou:2001qd} (diamonds).} \label{fig:ratio}
            \end{center}
            \end{figure}

            In Fig.\,\ref{fig:ratio} for the proton's form factor ratio $\mu_p G_E^p/G_M^p$,
            the calculation involving the dominant diquark amplitudes points towards an early zero crossing at $Q^2 \sim 2$ GeV$^2$.
            The curve rises again when including the full diquark substructure, and the common non-vanishing slope at $Q^2=0$
            is caused by the deviation between electric and magnetic radii (Table\,\ref{tab:ff1}),
            since the Taylor expansion at small $Q^2$ entails \cite{Alkofer:2004yf}
            \begin{equation}
                \mu_p \frac{G_E^p}{G_M^p} = 1 - \frac{Q^2}{6}\left( (r_E^p)^2-(r_M^p)^2 \right) + \dots
            \end{equation}
            One however encounters a notable sensitivity to the
            perturbative behavior of the $qq$ T-matrix, precisely: to the off-shell dependence of the diquark amplitudes.
            The cases $n=1$ and $n=2$ correspond to a suppression of the subleading  amplitudes at large spacelike diquark momenta by a power of
            $\sqrt{P^2}$ and $P^2$, respectively. Larger $n$ implies being closer to the dominant-amplitude result.
            This sensitivity is already visible in the plain electric form factors $G_E^p$, $G_E^n$,
            and it is again strong in the axial-vector diquark-photon contribution which depends on the diquark amplitudes via Eq.\,\eqref{ff:quarkloop}.
            In the form factor ratio, the systematic error band at increasing $Q^2$ (basically the region between full and dashed curve) induced by this uncertainty is quite large.
            Additionally, at the present level of sophistication the off-shell ansatz for the diquark amplitudes is just a surrogate
            for the correct perturbative behavior of the T-matrix (in terms of the bare ladder kernel which has been truncated by the diquark ansatz).
            This perturbative behavior is clearly important at large $Q^2$, and even if one could eventually access that region the predictive power of the findings might be questioned,
            at least within use of the diquark ansatz for the T-matrix. Such uncertainties could be removed by calculating the scattering matrix directly from the kernel in Eq.\,\eqref{bs:dysonseq}.

\section{Conclusions and outlook}\label{sec:conclusions}

             The main goal of this work was to extend previous nucleon studies in a quark-diquark approach
             by implementing results obtained from Dyson-Schwinger and Bethe-Salpeter equations at a more fundamental level of QCD.
             The quark-diquark model can be derived from the full relativistic three-body equation by a series of well-defined truncations
             that preserve Poincar\'e covariance.
             We have studied the nucleon mass as well as charge radii and magnetic moments as functions of the quark or pion mass to
	     a point well beyond the strange-quark mass.
	     Furthermore, we calculated electromagnetic nucleon form factors up to a photon momentum-squared of
	     $Q^2\approx 1-2$ GeV${}^2$. The former allows for a comparison with
             lattice results and their chiral extrapolations; the latter is of large interest in the current experimental situation
	     regarding the ratio of $\mu G_E/G_M$.

             In order to link our investigation of nucleon properties to existing advanced meson studies
             we employed a rainbow-ladder truncation in the quark DSE and the $q\bar{q}$ and $qq$ scattering kernels.
	     These correctly implement characteristic non-perturbative features of QCD
             such as spontaneous chiral symmetry breaking and yields effective confinement of the quark propagator.
             By adjusting the shape and quark-mass dependence of the effective coupling $\alpha(k^2)$ to pseudoscalar meson observables and
             lattice results for the quark propagator, meson properties are readily obtained and all parameters are fixed, yielding
	     all nucleon observables as predictions of our setup.
             An analysis of the vector meson mass confirms that our results from rainbow-ladder truncated DSEs are within $10\%$ of the
	     corresponding lattice data.

             A similar situation exists for the nucleon mass, where in addition to the rainbow-ladder truncation a diquark ansatz
	     is used. In principle, the diquark ansatz is a reasonable
	     approximation in the infrared region where contributions from diquark correlations are dominant.
             In the UV region the truncation causes systematic uncertainties in the electric form factors at finite $Q^2$ (and
	     therefore also in the form factor ratio $\mu G_E/G_M$),
             whereas it does not affect the nucleon's charge radii and magnetic moments.
	     Clearly, this problem could be remedied by omitting the diquark ansatz in favor of a direct solution of the relativistic
	     Faddeev equation in a gluon ladder truncation,
             thereby advancing the current treatment of baryons to that of mesons in the DSE framework.

             Other issues are connected to the use of the rainbow-ladder truncation of the DSEs:
	     the similar characteristics in the quark-mass dependence of both vector meson and nucleon masses is remarakble. It suggests
	     that effects beyond rainbow-ladder truncation account for the main discrepancy
             between the results presented in this work and those obtained from a combination of lattice and chiral effective field theory methods.
             Part of such effects come from the pion cloud, i.\,e., the long-range $\bar{q}q$ interactions with the nucleon.
             The absence of such contributions is clearly visible in our results for the nucleon's charge radii and magnetic moments.
	
	     In this respect, the interplay of pion cloud- and other contributions beyond rainbow-ladder truncation plays an important role.
             From chiral effective field theory and studies within the cloudy bag model it is known that pion loops are attractive and reduce
             the nucleon mass near the chiral limit, whereas its effect becomes small at larger quark masses.
             The present constant underestimation of the lattice data encountered in the vector-meson and nucleon masses makes clear that a
	     truncation of the DSEs beyond rainbow-ladder must also provide further repulsive effects.
	
             It is conceivable that both contributions could be taken into account by directly implementing a more sophisticated quark-quark
	     interaction kernel in the relativistic Faddeev equation. Despite the need for such future improvement,
             the results presented in this work show that the quark-diquark approach
             is able to capture the major part of the nucleon's quark core, if its Green function content is consistently
	     obtained from the dynamic equations of QCD.

\section*{Acknowledgements}

             We acknowledge helpful discussions with C.\,S.~Fischer, T.\,R.~Hemmert,
	     R.~Krenn, D.~Nicmorus, and C.\,D.~Roberts. We would also like to thank
             C.\,D.~Roberts for a critical reading of the manuscript.

             This work has been supported in part by the Deutsche Forschungsgemeinschaft
             under Grant No.\ Al279/5-1 and 5-2  as well as the Austrian Science Fund FWF under
             Grant No.\ W1203 (Doctoral Program ``Hadrons in vacuum, nuclei and stars'') and
             Schr\"odinger-R\"uckkehr-Stipendium Nr.~R50-N08.

\newpage
\begin{appendix}


\section{Supplements to the quark DSE}\label{app:dsesupplement}

            The quark DSE \eqref{dse:qdse} can be reexpressed in terms of two coupled integral equations for $A(p^2,\mu^2)$ and $M(p^2)$,
            \begin{align}
                \qquad &&         A(p^2,\mu^2) &=& \!\!\!\!\!               Z_2(\mu^2,\Lambda^2) &+ \Sigma_A(p^2,\mu^2,\Lambda^2), && \qquad \label{dse:Adse}\\
                \qquad && M(p^2)\,A(p^2,\mu^2) &=& \!\!\!\!\! M(\Lambda^2)\,Z_2(\mu^2,\Lambda^2) &+ \Sigma_M(p^2,\mu^2,\Lambda^2), && \qquad \label{dse:Mdse}
            \end{align}
            where $\Sigma_A$ and $\Sigma_M$ are obtained from the quark self-energy via
            \begin{equation}
                \Sigma(p,\mu,\Lambda) = i \Slash{p} \, \Sigma_A(p^2,\mu^2,\Lambda^2) + \Sigma_M(p^2,\mu^2,\Lambda^2).
            \end{equation}
            In rainbow truncation they read:
            \begin{align}
                \Sigma_A(p^2,\mu^2,\Lambda^2) &= \frac{16 \pi Z_2^2}{3 p^2} \int^\Lambda \frac{d^4 q}{(2\pi)^4}\, \sigma_v(q^2) \,\frac{\alpha(k^2)}{k^2}
                                                 \left( p \!\cdot\! q + \frac{2 \,p \!\cdot\! k \, q \!\cdot\! k }{k^2} \right),   \label{dse:A}\\
                \Sigma_M(p^2,\mu^2,\Lambda^2) &= 16 \pi Z_2^2\int^\Lambda \frac{d^4 q}{(2\pi)^4}\, \sigma_s(q^2) \,\frac{\alpha(k^2)}{k^2}.     \label{dse:M}
            \end{align}
            Eqs. (\ref{dse:Adse},\,\ref{dse:Mdse}) can be solved iteratively for chosen values of $Z_2$ and $M(\Lambda^2)$.
            Alternatively one can employ a renormalization condition, e.g. $A(\mu^2,\mu^2)=1$,
            and specify the current mass $M(\mu^2)$ at the renormalization point as an input value.
            Then both $Z_2$ and $M(\Lambda^2)$ are determined together with $A(p^2)$ and $M(p^2)$
            in the course of the iteration via
            \begin{align}
                \qquad &&               Z_2(\mu^2,\Lambda^2) &=& \!\!\!\!\!\!\!\!        1 &- \Sigma_A(\mu^2,\mu^2,\Lambda^2),  && \qquad\\
                \qquad && M(\Lambda^2)\,Z_2(\mu^2,\Lambda^2) &=& \!\!\!\!\!\!\!\! M(\mu^2) &- \Sigma_M(\mu^2,\mu^2,\Lambda^2). && \qquad
            \end{align}

            Asymptotically, the DSE solution for the quark mass function reproduces the behavior predicted from perturbation theory:
            \begin{equation}\label{dse:asymptoticmassf}
                M(p^2) \stackrel{p^2\rightarrow\infty}{\longlongrightarrow} \, \frac{\hat{m}}{\mathcal{F}(p^2)^{\gamma_m}} +
                       \frac{2 \pi^2 \gamma_m}{N_C}  \;\frac{-\langle \bar{q} q\rangle}{\mathcal{F}(p^2)^{1-\gamma_m}\,p^2}
            \end{equation}

            where $\mathcal{F}(p^2)= \frac{1}{2}\ln \left(p^2/\Lambda_{QCD}^2\right)$. The coefficients $\hat{m}$ and $-\langle \bar{q} q \rangle$
            define the renormalization-point independent current mass and chiral condensate. For finite current masses, the second term is suppressed
            by a factor of $p^2$ while in the chiral limit, defined by $\hat{m}=0$, it determines the behavior of the asymptotic mass function.
            The renormalization-point-dependent chiral quark condensate is obtained from
            \begin{equation}\label{dse:qq}
                -\langle \bar{q} q \rangle_\mu = Z_2(\mu^2,\Lambda^2) \,Z_m(\mu^2,\Lambda^2)\,N_C \int^\Lambda \frac{d^4 q}{(2\pi)^4} \,\text{Tr}_D \{ S_\text{chiral}(q,\mu) \},
            \end{equation}
            with $Z_m(\mu^2,\Lambda^2) = M(\Lambda^2)/M(\mu^2) $, evaluated at large current masses.
            For large renormalization points, the condensates are related via
            \begin{equation}
                -\langle \bar{q} q \rangle_\mu = -\langle \bar{q} q \rangle\, \mathcal{F}(\mu^2) ^{\gamma_m}.
            \end{equation}

            For $p^2 \in \mathds{R}_+$, the propagator functions $\sigma_v(q^2)$, $\sigma_s(q^2)$ and the coupling $\alpha(k^2)$ in the integrals $\Sigma_A$ and $\Sigma_M$
            are only needed on the positive real $q^2$ and $k^2$ axes. In this case the coupled system (\ref{dse:Adse},\,\ref{dse:Mdse}) can be solved without complications.
            The straightforward way to evaluate the quark propagator for a complex argument $p^2\in\mathds{C}$ is to
            implement the result for $q^2\in\mathds{R}_+$ and insert the coupling $\alpha(k^2)$ at the complex values $k^2 = p^2+q^2-2 \,p \cdot q$
            which constitute the region within the parabola $(t\pm i\,|\text{Im} \sqrt{p^2}|)^2$, $t\in \mathds{R}_+$.
            Strictly speaking, this is practicable only if the integrand inside this region is free of singularities in $k^2$, i.e.,
            according to \eqref{dse:A} and \eqref{dse:M}, if the coupling is singularity-free and $\alpha(k^2\rightarrow 0) \rightarrow k^4$.

            \begin{figure}[hbt]
            \begin{center}
            \includegraphics[scale=0.27,angle=-90]{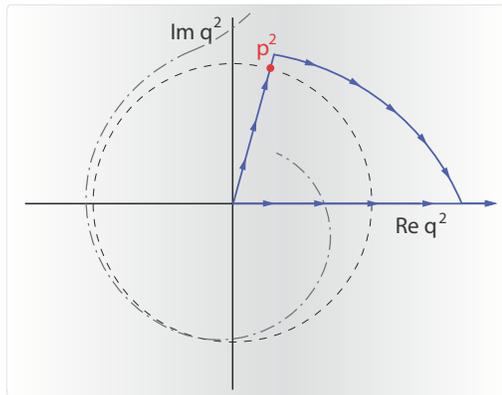}
            \caption[DSE in the complex plane]{Branch cuts in the complex $q^2$ domain of the quark propagator corresponding to a fixed external point $p^2$
					       and a possible integration path (see text).} \label{fig:dsecomplexplane}
            \end{center}
            \end{figure}

            After performing all the integrations except the $q^2$ integral, singular points in $k^2$ lead to branch cuts in the complex $q^2$ plane,
            illustrated in Fig.\,\ref{fig:dsecomplexplane}: for instance, a pole at $k^2=0$ generates a circular branch cut in the complex $q^2$ plane
	    with an opening at $q^2=p^2$ (dashed line).
            The logarithmic one-loop behavior \eqref{dse:asympcoupling} entails that the coupling necessarily exhibits singularities also at $k^2\neq 0$:
            those will generally lead to more complicated branch cut structures (dash-dotted line) which, however, still leave the arc $q^2 = r\,e^{i \arg{p^2}}$, $r\in\mathds{R}_+$, unharmed.
            A possible way to avoid all occurring branch cuts (an alternative method is described in \cite{Fischer:2005en}) is to deform the integration contour $q^2 \in (0,\Lambda^2)$ to a complex arc that passes the point
            $p^2$ and eventually returns to $\Lambda^2 \in \mathds{R}$ in the far spacelike region. Since $\sigma_v(q^2)$ and $\sigma_s(q^2)$
            must already be known on these complex paths, the complex DSE solution is therefore obtained via iteration of
            (\ref{dse:Adse},\,\ref{dse:Mdse}) on a family of deformed complex paths in $p^2$.
            The disadvantage of this method is that successively increasing the angle of the respective arc
            is viable only up to the first pole (pair) of the propagator occurring in the timelike complex plane.

            A virtue of the parameterization \eqref{dse:maristandy} is that these issues are largely avoided by its infrared behavior
            $\alpha(k^2\rightarrow 0)\rightarrow k^2$, and by the large oscillations caused by the exponential parts which effectively shield the complex conjugated
            poles stemming from the logarithmic tail in the coupling. The remaining $1/k^2$ pole in the $\Sigma_A$ integral results in small numerical
            artifacts which are visible in the complex functions $\sigma_v(p^2)$ and $\sigma_s(p^2)$ if they are directly calculated without employing refined methods.
            However, if one employs a different coupling that behaves as $\alpha(0)=const.$ or involves more malicious singularities,
            these artifacts become dominant and one inevitably has to resort to more advanced methods.
            Similar procedures can be used for evaluating meson and diquark amplitudes in the complex plane of the relative momentum between the contributing quarks.


\section{Meson and diquark amplitudes}\label{app:mesondiquark}

\subsection{General decomposition}\label{app:mesondiquark-general}

            The meson BSA has been introduced in \eqref{bse:bse} together with
	    the variables it depends on.
            The Dirac structure of the amplitudes is determined by the Clifford algebra of the Dirac $\gamma$ matrices.
	    As for any fermion-fermion-scalar
            or fermion-fermion-vector vertex, it allows for 4 basis matrices for the pseudoscalar amplitude,
            \begin{equation}\label{bse:scbasisgeneral}
                \tau_{1\dots 4}(q,P) = \{ \mathds{1}, \Slash{P}, \, \Slash{q}, \, \Slash{q}\,\Slash{P} \}\;,
            \end{equation}
            and 12 basis elements for the vector amplitude:
            \begin{equation}\label{bse:vcbasisgeneral}
                \tau_{1\dots 12}^\mu(q,P) = \gamma^\mu  \{\mathds{1}, \Slash{P}, \, \Slash{q}, \, \Slash{q}\,\Slash{P} \}\;, \;
                                                 q^\mu  \{\mathds{1}, \Slash{P}, \, \Slash{q}, \, \Slash{q}\,\Slash{P} \}\;, \;
                                                 P^\mu  \{\mathds{1}, \Slash{P}, \, \Slash{q}, \, \Slash{q}\,\Slash{P} \}\;.
            \end{equation}
            Implementing the negative parity requirement
            \begin{equation}
            \begin{split}
                \Gamma(q,P) &= -\gamma^4 \, \Gamma(\Lambda q,\Lambda P)\,\gamma^4\;, \\
                \Gamma^\mu(q,P) &=  \gamma^4\, \Lambda^{\mu\nu} \Gamma^\nu (\Lambda q,\Lambda P)\,\gamma^4\;,
            \end{split}
            \end{equation}
            where $\Lambda = \text{diag}(-1,-1,-1,1)$, gives rise to the following general structure of pseudoscalar and vector meson amplitudes,
            written with full Dirac, color and flavor dependence:
            \begin{equation}\label{mesamp}
            \begin{split}
                \Gamma(q,P)     &=  \sum_{k=1}^4    f_k^\text{ps}(q^2,z,P^2) \big\{ i\gamma^5 \tau_k(q,P) \big\}_{\alpha\beta}   \,\otimes\, \frac{\delta_{AB}}{\sqrt{3}}  \,\otimes\,  \mathsf{r^e_{ab}}, \\
                \Gamma^\mu(q,P) &=  \sum_{k=1}^{12} f_k^\text{vc}(q^2,z,P^2) \big\{ i\tau^\mu_k(q,P)  \big\}_{\alpha\beta}     \,\otimes\, \frac{\delta_{AB}}{\sqrt{3}}  \,\otimes\,  \mathsf{r^e_{ab}}.
            \end{split}
            \end{equation}
            The respective dressing functions $f_k(q^2,z,P^2)$ only depend on the Lorentz scalars
            $q^2$, $P^2$ and the angular variable $z=\hat{q}\!\cdot\!\hat{P}$ ($\hat{q}$ denotes a normalized 4-vector $q/\sqrt{q^2}$).
            By solving the meson Bethe-Salpeter equation, these dressing functions are obtained on the domains $q^2 \in \mathds{R}_+$
            (by use of the same methods as discussed in App.\,\ref{app:dsesupplement} also for $q^2 \in \mathds{C}$), $P^2=-M^2$ (i.\,e., on the mass shell), and $z \in (-1,1)$.
            Greek indices refer to the Dirac structure.
            The color structure of the meson amplitudes is diagonal, with $A,B=1,2,3$.
            We are working with two flavors and assume isospin symmetry.
            The flavor matrices $\mathsf{r^e_{ab}}$ ($a,b=1,2$)
            then refer to isospin singlet ($e=0$) and triplet states ($e=1,2,3$); in the pseudoscalar case we only consider
	    the triplet states (i.\,e., the pion). We choose the $\mathsf{r^e}$ to be normalized to unity
            via $\text{Tr}\{ \mathsf{{r^e}}^\dagger \mathsf{r^{e'}} \}=\delta_{ee'}$, but mostly omit them from formulae, since they
            give no contribution to the Bethe-Salpeter equation \eqref{bse:bse}:
            for degenerate flavors the propagators in the kernel are not flavor-dependent.
            The charge-conjugated amplitudes are defined by
            \begin{equation}\label{bse:conjugation}
            \begin{split}
                \bar{\Gamma}(q,-P) &= C \,\Gamma^T(-q,-P) \,C^{-1}, \\
                \bar{\Gamma}^\mu(q,-P) &= -C \,{\Gamma^\mu}^T(-q,-P) \,C^{-1}\;.
            \end{split}
            \end{equation}
	    where the superscripted $T$ denotes matrix transposition.

            \begin{figure}[hbt]
            \begin{center}
            \includegraphics[scale=0.41,angle=-90]{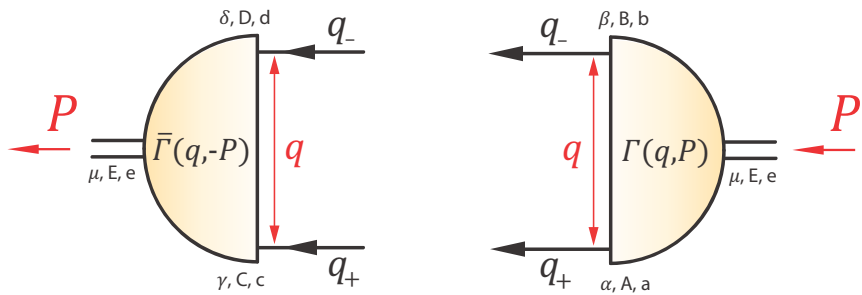}
            \caption[Diquark amplitudes]{Notational conventions for diquark amplitudes. The quark momenta are $q_\pm = \pm q +P/2$
            and the indices are Dirac/Lorentz, color and flavor indices. } \label{fig:dqamp}
            \end{center}
            \end{figure}

            Apart from the opposite parity requirement and different color and flavor tensors,
            scalar and axial-vector diquark amplitudes exhibit the same structure as their pseudoscalar and vector meson counterparts.
            The incoming antiquark momentum $-q_-$ is replaced by an outgoing quark momentum $q_-$ which is reflected by the
            charge conjugation matrix $C=\gamma^4 \gamma^2$.
            We denote diquark amplitudes by the same generic symbol $\Gamma$:
            \begin{align}
                \Gamma(q,P)     &=  \sum_{k=1}^4    f_k^\text{sc}(q^2,z,P^2) \big\{ i\gamma^5\tau_k(q,P)\,C \big\}_{\alpha\beta} \,\otimes\,  \frac{\varepsilon_{ABE}}{\sqrt{2}}  \,\otimes\,  \mathsf{s^0_{ab}}\;, \label{dq:sc}\\
                \Gamma^\mu(q,P) &=  \sum_{k=1}^{12} f_k^\text{av}(q^2,z,P^2) \big\{ i\,\tau^\mu_k(q,P)\,C \big\}_{\alpha\beta}   \,\otimes\,  \frac{\varepsilon_{ABE}}{\sqrt{2}}  \,\otimes\,  \mathsf{s^{1\dots 3}_{ab}}\;.\label{dq:av}
            \end{align}
            The definition of conjugation is the same as for mesons with corresponding $J$ quantum numbers, see Eq.\,\eqref{bse:conjugation}.
            Due to the Pauli principle, diquark amplitudes must be antisymmetric under quark exchange $q_+ \leftrightarrow q_-$,
            \begin{equation}\label{dq:antisymmetry}
                \Gamma(q,P) = -\Gamma^T( \left.-q \right|_{\sigma\leftrightarrow(1-\sigma)} , P)\;,
            \end{equation}
            where the transposition involves all Dirac, color and flavor indices. Because of the antisymmetry of the color anti-triplet diquark,
            the combination
            of flavor and spin structure must be symmetric. Therefore spin and isospin states coincide for the two-flavor case:
            scalar diquarks correspond to an antisymmetric isospin singlet and axial-vector diquarks to a symmetric isospin
            triplet.
            The isospin singlet and triplet matrices $\mathsf{s^0_{ab}}$ and $\mathsf{s^{1\dots 3}_{ab}}$ are explicitly given by
            \begin{align*}
                \mathsf{s^0} &= (\mathsf{ud^\dag-du^\dag})/\sqrt{2} = i\sigma_2/\sqrt{2}\;, \\
                \mathsf{s^1} &= \mathsf{uu^\dag} = (\mathds{1}+\sigma_3)/2\;, \\
                \mathsf{s^2} &= (\mathsf{ud^\dag+du^\dag})/\sqrt{2} = \sigma_1/\sqrt{2}\;,\\
                \mathsf{s^3} &= \mathsf{dd^\dag} = (\mathds{1}-\sigma_3)/2\;,
            \end{align*}
            where $\sigma_i$ are the Pauli matrices, and $\mathsf{u}=(1,0)$, $\mathsf{d}=(0,1)$.
            The $\mathsf{s^e}$ are normalized to unity: $\text{Tr}\{ \mathsf{{s^e}}^\dagger \mathsf{s^{e'}} \}=\delta_{ee'}$
            and do not give a contribution to the BSE integral \eqref{dq:bse} itself. However,
            the color factor $\sim\varepsilon_{ABE}$ representing the diquark anti-triplet configuration leads to a prefactor $1/2$
            in front of the integral if the rainbow-ladder kernel is inserted: diquarks are less bound than mesons.
            The Dirac amplitudes $\sim \mathds{1},\,\gamma^\mu$ in \eqref{bse:scbasisgeneral} and \eqref{bse:vcbasisgeneral}
            are the dominant ones in a solution of the rainbow-ladder BSE
            for the lowest-mass mesons  and diquarks
            and reproduce masses of the full solution within an error of $\lesssim 20 \%$ \cite{Maris:1997tm,Maris:1999nt,Maris:2002yu}.


\subsection{Mesons and diquarks on the mass shell}

            For the actual solution of the meson or diquark BSE,
            eqs.\,\eqref{bse:bse} and \eqref{dq:bse}, it is advantageous to construct orthogonalized versions
            of the general Dirac basis elements (\ref{bse:scbasisgeneral},\,\ref{bse:vcbasisgeneral})
            at the respective mass pole $P^2 = -M^2$.
            A suitable choice for the (pseudo--)scalar case is
            \begin{equation}\label{bse:psbasis}
                \tau_1 = \mathds{1}\;, \quad
                \tau_2 = \Slash{\hat{P}}\;, \quad
                \tau_3 = \hat{q}\!\cdot\!\hat{P}\, \Slash{\hat{q}}_T\;, \quad
                \tau_4 = -i\,[ \Slash{\hat{P}},\Slash{\hat{q}} ]\;,
            \end{equation}
            where $\hat{q}^{\mu}_T=T^{\mu\nu}_P \hat{q}^{\nu}$ and $T^{\mu\nu}_P=\delta^{\mu\nu}-\hat{P}^\mu \hat{P}^\nu$ is the transverse projector with respect to $P$, and for the (axial--)vector case:
            \begin{align}\label{bse:vcbasis}
               \nonumber  \tau_1^{\mu} &= \gamma^{\mu}                                            &   \tau_5^{\mu} &= \hat{q}\!\cdot\!\hat{P}\,( \gamma^{\mu} \Slash{\hat{q}}_T - \hat{q}^{\mu} )                                           \\
                          \tau_2^{\mu} &= \gamma^{\mu}\Slash{\hat{P}}                             &   \tau_6^{\mu} &= \frac{i \gamma^{\mu}}{2} [ \Slash{\hat{P}},\Slash{\hat{q}} ] + i\,\hat{q}^{\mu}\Slash{\hat{P}}         \\
               \nonumber  \tau_3^{\mu} &= i\,\hat{q}^{\mu}                                        &   \tau_7^{\mu} &= \hat{q}^{\mu}\Slash{\hat{q}}_T - \frac{\hat{q}^2_T}{3}\gamma^{\mu}                                \\
               \nonumber  \tau_4^{\mu} &= \hat{q}\!\cdot\!\hat{P}\,\hat{q}^{\mu} \Slash{\hat{P}}  &   \tau_8^{\mu} &= \frac{\hat{q}^{\mu}}{2} [ \Slash{\hat{P}},\Slash{\hat{q}} ] + \frac{\hat{q}^2_T}{3}\gamma^{\mu} \Slash{\hat{P}} .
            \end{align}
            Using normalized $P^\mu$ simplifies the discussion of the diquark amplitudes' off-shell behavior (cf.\,App.\,\ref{app:mesondiquark-offshell});
            using normalized $q^\mu$ is convenient but not further relevant since the diquark BSE is solved for complex $q^2$ in the first place.
            These basis elements are designed such that all corresponding dressing functions $f_k(q^2,z,-M^2)$ are real and even in $z$ for $\sigma=1/2$
            owing to charge conjugation symmetry.
            On the mass shell it is sufficient to consider 8 of the general 12 components since on-shell vector mesons
            (and axial-vector diquarks) are transverse with respect to their total momentum $P$: the basis elements $9\dots 12$
            in this representation correspond to the basis elements $P^\mu\{\mathds{1}, \Slash{P}, \, \Slash{q}, \, \Slash{q}\,\Slash{P} \}$
            which are purely longitudinal. Therefore, for vector mesons and analogously for axial-vector diquarks, one has
            \begin{equation}
                \Gamma^\mu_{\alpha\beta}(q,P) \big|_{P^2=-M_\text{vc}^2} =
                T^{\mu\nu}_P\,\sum_{k=1}^{8} f_k^\text{vc}(q^2,z,-M_\text{vc}^2) \Big\{ i\tau^\nu_k(q,P) \big|_{P^2=-M_\text{vc}^2} \Big\}_{\alpha\beta}\;.
            \end{equation}
            The orthogonality relations for the basis elements are
            \begin{equation}
                \text{Tr}\{\tau_i\, \tau_j\} = 4\,\delta_{ij} a_i(z)\;,\quad
                T^{\mu\nu}_P\,\text{Tr}\{\tau_i^{\mu}\, \tau_j^{\nu}\} = 4\,\delta_{ij} b_i(z)\;,
            \end{equation}
            where
            \begin{equation}
            \begin{split}
                a_1 &= a_2 = \frac{b_1}{3} = -\frac{b_2}{3} = 1\;,  \\
                \frac{a_3}{z^2}  &= \frac{a_4}{4} = -b_3 = \frac{b_4}{z^2} = -\frac{b_5}{2 z^2} = \frac{b_6}{2} = 1-z^2\;, \\
                b_7 &= -b_8 = \frac{2}{3}(1-z^2)^2\;.
            \end{split}
            \end{equation}

            Contraction of \eqref{bse:bse} and \eqref{dq:bse} with the basis elements and using the orthogonality relations leads to coupled homogeneous
            integral equations for the coefficients $f_k(q^2,z,-M^2)$.
            Since they depend only on Lorentz-invariant scalar products, these equations can be solved in any arbitrary frame.
            Exploiting the $O(4)$ symmetry of the problem, one can employ a Chebyshev decomposition for numerical convenience:
            \begin{equation}\label{dq:chebyshev}
                f_k(q^2,z,-M^2) \approx \sum_{n=0}^{n_\text{max}} f_k^n(q^2)\,U_n(z), \quad \int_1^1 dz \sqrt{1-z^2} \,U_m(z)\,U_n(z) = \frac{\pi}{2}\, \delta_{mn}\;,
            \end{equation}
            where the $U_n(z)$ are Chebyshev polynomials of the second kind. For a suitable choice of basis elements,
	    only a few Chebyshev moments $f_k^n(q^2)$ have to be taken into account
            to match the full solution \cite{Maris:1997tm,Maris:1999nt}.
            We will use this observation below and reduce the angular dependence to a constant, see App.\,\ref{app:sing} and Sec.\,\ref{sec:results}.
            By introducing an artificial parameter $\lambda(P^2)$ at the right-hand side of the homogeneous BSE \eqref{dq:bse},
            the equation becomes an eigenvalue problem for $\lambda(P^2)$ where a bound-state solution is obtained for $\lambda(-M^2)=1$.

            As mentioned in Sec.\,\ref{sec:boundstateeq}, bound-state amplitudes can be normalized via a ``canonical'' normalization condition,
	    Eq.\,\eqref{bs:normalization}, derived from the pole condition \eqref{bs:poleansatz}. In the case of a two-body Bethe-Salpeter amplitude obtained via
            rainbow-ladder truncation, the ladder kernel is independent of the total momentum $P$
            such that only the derivatives of the propagators contribute. For pseudoscalar and vector mesons this reads:
            \begin{equation}\label{bse:normalization}
                \left. \frac{d}{d P^2} \right|_{P^2=-M_\text{ps}^2} Q(P^2) = 1\;, \quad
                \left. \frac{d}{d P^2} \right|_{P^2=-M_\text{vc}^2} Q_T(P^2) = 1\;,            
            \end{equation}
            with
            \begin{equation}\label{bse:norm}
                Q^{(\mu\nu)}(P) := \left. \int \frac{d^4 q}{(2\pi)^4} \, \text{Tr}_D \left\{ \bar{\Gamma}^{(\mu)}(q,-K) S(q_+) \Gamma^{(\nu)}(q,K) S(-q_-) \right\}   \right|_{K^2=-M^2}\;.
            \end{equation}
            The notation $Q^{(\mu\nu)}$ covers either the pseudoscalar quantity $Q$ or its vector counterpart $Q^{\mu\nu}$,
            where $Q_T=T^{\mu\nu}_PQ^{\mu\nu}/3$ is the transverse component of the latter.
            $\text{Tr}_D$ denotes a Dirac trace; the color-flavor trace is $1$ since we use normalized color and flavor matrices.
            The Bethe-Salpeter equation \eqref{bse:bse}, together with the normalization (\ref{bse:normalization},\,\ref{bse:norm}), completely determines
            the meson amplitudes on the mass shell.
            The pion decay constant is calculated via
            \begin{equation}\label{bse:pdc}
                f_\text{ps} = \frac{i Z_2}{M_\text{ps}^2}\,\sqrt{\frac{N_C}{N_F}}\,
                \left. \int^\Lambda \frac{d^4 q}{(2\pi)^4}  \, \text{Tr}_D \left\{ i \gamma^5 \,\Slash{P} \,S(q_+) \,\Gamma_\text{ps}(q,P) \, S(-q_-) \right\} \right|_{P^2=-M_\text{ps}^2}\;,
            \end{equation}
            where the prefactor $\sqrt{N_C/N_F}$ is again a consequence of the color-flavor normalization. We note that
	    $N_F=2$ here, since we are dealing with bound states of two light (anti)quarks here.

            The normalization condition for the diquark case can be obtained by differentiation of Eq.\,\eqref{dq:prop1}
            with respect to $P^2$ at the mass pole, insertion of the diquark BSE $( K^{-1}-G_0 )\,\Gamma = 0$ and
            using the fact that the ladder kernel $K$ is independent of $P^2$
            ($G_0 = -\frac{1}{2} \,S\, S$ contains a symmetrization factor $1/2$ for the diquark case \cite{Ishii:1995bu}):
            \begin{equation}\label{dq:normalization}
                \left. \frac{d}{d P^2} \right|_{P^2=-M_\text{sc}^2} Q(P^2) = 1\;, \quad
                \left. \frac{d}{d P^2} \right|_{P^2=-M_\text{av}^2} Q_T(P^2) = 1\;,
            \end{equation}
            with
            \begin{equation}\label{dq:norm}
                Q^{(\mu\nu)}(P) := \left. \frac{1}{2}\int \frac{d^4 q}{(2\pi)^4} \, \text{Tr}_D \left\{ \bar{\Gamma}^{(\mu)}(q,-K) S(q_+) \Gamma^{(\nu)}(q,K) S^T(q_-) \right\}   \right|_{K^2=-M^2}\;.
            \end{equation}


\subsection{Offshell ansatz for the diquark amplitudes}\label{app:mesondiquark-offshell}

            With the methods outlined in this work we cannot gather any information on diquark amplitudes
            for off-shell total diquark momenta $P^2\neq -M^2$.
            On their mass shells, the amplitudes are obtained from the Bethe-Salpeter equation \eqref{dq:bse}
            and equivalent to the residues of a scalar or axial-vector quark-diquark vertex
            at the lowest-lying diquark poles.
            An \textit{inhomogeneous} diquark BSE, to be solved in an analogous way as an inhomogeneous meson BSE \cite{Maris:1997hd},
            also provides only very limited information on the off-shell behavior of these residues.
            Therefore we choose a reasonable ansatz for the diquark's off-shell behavior based on general assumptions.

            We start from Eqs. (\ref{dq:sc},\,\ref{dq:av}) and for convenience discuss the offshell dependence
            of the diquark amplitudes in terms of the basis elements $\tau_k(q,P)$ and $\tau^\mu_k(q,P)$
            while the dressing functions are left unchanged at their mass-shell values:
            $f_k^\text{sc(av)}(q^2,z,P^2)=f_k^\text{sc(av)}(q^2,z,-M^2_\text{sc(av)})$.
            Usage of the orthogonal on-shell bases
            \eqref{bse:psbasis} and \eqref{bse:vcbasis}, where each appearance of $P^\mu$ has been normalized,
            entails that the correct behavior at the mass shell and for $P^2=0$ can be guaranteed by attaching
            a factor $\sqrt{P^2}/(i M)$ to each occurrence of $\hat{P}$ therein,
            or generally by an arbitrary function $h(P^2/M^2)$ satisfying $h(-1)=1$ and $h(0)=0$.
            This then also applies to each transversal projector: $ T^{\mu\nu}_P \longrightarrow \delta^{\mu\nu} - h^2(P^2/M^2)\,L^{\mu\nu}_P. $
            Our ansatz for the full Dirac part of the diquark amplitudes therefore reads
            \begin{align}
                \Gamma_\text{sc}(q,P)     &\,=\, \sum_{k=1}^4 f_k^\text{sc}(q^2,z,-M_\text{sc}^2) \, i\gamma^5 \tau_k(q,P)\,C  \label{dq:scoffshell}\\
                \Gamma^\mu_\text{av}(q,P) &\,=\, \sum_{k=1}^8 f_k^\text{av}(q^2,z,-M_\text{av}^2) \, i\tau^\mu_k (q,P)\,C      \label{dq:avoffshell}
            \end{align}
            where the scalar basis is given by
            \begin{align}\label{dq:scbasis}
                          \tau_1 &= g_\text{sc}\,\mathds{1}         &
                          \tau_3 &= z\,h_\text{sc}\, (\Slash{\hat{q}}-z\,h_\text{sc}^2\, \Slash{\hat{P}})    \\
            \nonumber     \tau_2 &= h_\text{sc}\,\Slash{\hat{P}}     &
                          \tau_4 &= -i\,h_\text{sc}\,[ \Slash{\hat{P}},\Slash{\hat{q}} ]\;,
            \end{align}
            and the axial-vector basis by:
            \begin{align}\label{dq:avbasis}
                          \tau_1^{\mu} &= g_\text{av}\,\gamma^{\mu}                                                                              &
                          \tau_5^{\mu} &= z\,h_\text{av}\,\left\{ \gamma^{\mu} (\Slash{\hat{q}} -z\,h_\text{av}^2\, \Slash{\hat{P}}) - \hat{q}^{\mu} \right\}                                           \\
               \nonumber  \tau_2^{\mu} &= h_\text{av} \, \gamma^{\mu}\Slash{\hat{P}}                                                                    &
                          \tau_6^{\mu} &= h_\text{av}\,\left\{ \frac{i \gamma^{\mu}}{2} [ \Slash{\hat{P}},\Slash{\hat{q}} ] + i\,\hat{q}^{\mu}\Slash{\hat{P}}  \right\}        \\
               \nonumber  \tau_3^{\mu} &= i\,\hat{q}^{\mu}                                                                                       &
                          \tau_7^{\mu} &= \hat{q}^{\mu}(\Slash{\hat{q}} - z\,h_\text{av}^2\, \Slash{\hat{P}}) - \frac{1-z^2\,h_\text{av}^2}{3}\gamma^{\mu}                                \\
               \nonumber  \tau_4^{\mu} &= z\,h_\text{av}^2\,\hat{q}^{\mu} \Slash{\hat{P}}                                         &
                          \tau_8^{\mu} &= h_\text{av}\,\left\{\frac{\hat{q}^{\mu}}{2} [ \Slash{\hat{P}},\Slash{\hat{q}} ] + \frac{1-z^2\,h_\text{av}^2 }{3}\gamma^{\mu} \Slash{\hat{P}} \right\}\;.
            \end{align}

            In addition to the functions $h_\text{sc(av)} := h(x_\text{sc(av)}) = h(P^2/M_\text{sc(av)}^2)$,
            which so far are arbitrary except for the two above requirements,
            we also attached a function $g_\text{sc(av)} := g(x_\text{sc(av)})$ with $g(-1)=1$ to each of the two leading amplitudes
            in order to be able to modulate them separately from the others.
            The transversality condition on the mass shell for the axial-vector amplitudes need not be stated explicitly
            since it is already ensured by the transversal pole in the axial-vector diquark propagator: each diquark amplitude in the
            subsequent calculations appears in
            conjunction with the respective propagator. Likewise, the purely longitudinal components
            related to $\tau^\mu_{9\dots 12}(q,P)$ off the mass shell are generated by the longitudinal contribution to the diquark propagator
            which is suppressed by a factor of $P^2+M_\text{av}^2$ on the mass shell.

            The asymptotic behavior for $P^2\rightarrow\infty$ of the diquark amplitudes is provided by the limits
            $g(x\rightarrow\infty)$ and $h(x\rightarrow\infty)$. We construct these functions
            based on two assumptions: firstly, all subleading amplitudes
            related to $\tau_k(q,P)$ and $\tau^\mu_k (q,P)$, $k>1$, should be suppressed
            by a power of at least $\sqrt{P^2}$ compared to the leading ones $\tau_1(q,P)$ and $\tau^\mu_1 (q,P)$ such that the
            diquark amplitude's perturbative limits are indeed $i\gamma^5\,C$ and $i\gamma^\mu\,C$.
            Secondly, the dominant amplitudes should behave like $\sqrt{P^2}$ in the ultraviolet to ensure via Eq.\,\eqref{dq:prop1}
            that the diquark propagators asymptotically behave as $D(P^2\rightarrow\infty)\rightarrow 1/P^2$.
            Simple parameterizations that fulfill all of these requirements are
            \begin{equation}\label{dq:offshellansatz2}
            g(x) = \sqrt{x+2}\;, \quad h(x) = \frac{1}{i} \sqrt{\frac{x}{(x+2)^n}}\;, \quad n\geq 1\;.
            \end{equation}
            For large $n$ the subleading amplitudes are suppressed at spacelike momenta
            and provide support only in the neighborhood of the mass shell. This resembles
            the case where only the dominant diquark amplitudes have been taken into account.
            Furthermore, a global function depending on $P^2$ which is attached to \textit{all} scalar or axial-vector amplitudes
            does not change the product $\Gamma\,D\,\bar{\Gamma}$ since it would also appear in the diquark propagator and
            leave the T-matrix itself (and therefore baryonic observables) invariant. For instance, one could divide all diquark amplitudes by $g(x)$ such
            that the leading amplitudes become constant in the ultraviolet while all the
            others are asymptotically suppressed: then also the diquark propagator would become constant for $P^2\rightarrow\infty$.
            This has been applied, e.\,g., in Ref.\,\cite{Oettel:2002wf}.


\subsection{Diquark propagator} \label{app:mesondiquark-dqprop}

            The diquark propagator is obtained by Eq.\,\eqref{dq:prop1} for all values of the squared diquark momentum $P^2$.
            This equation can be written as
                  \begin{equation}\label{dq:dqprop3}
                      (D^{-1})^{(\mu\nu)}(P) = K^{(\mu\nu)}(P) + Q^{(\mu\nu)}(P)\;,
                  \end{equation}
            where $K^{(\mu\nu)}$ represents the first term involving the inverse kernel and $Q^{(\mu\nu)}$ the
            quark loop term (in contrast to Eqs.\,\eqref{bse:norm} and \eqref{dq:norm}, now without the $P^2$ dependence of the diquark amplitudes held fixed),
                  \begin{equation}\label{dq:quarkloop}
                      Q^{(\mu\nu)}(P) := \frac{1}{2}\int \frac{d^4 q}{(2\pi)^4} \, \text{Tr}_D \left\{ \bar{\Gamma}^{(\mu)}(q,-P) S(q_+) \Gamma^{(\nu)}(q,P) S^T(q_-) \right\}\;.
                  \end{equation}

            \begin{figure}[hbt]
            \begin{center}
            \includegraphics[scale=0.46,angle=-90]{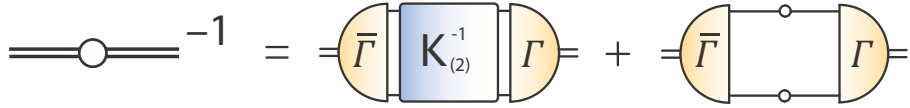}
            \caption[Diquark propagator]{Graphical representation of the diquark propagator of Eq.\,\eqref{dq:dqprop3}.} \label{fig:dqprop}
            \end{center}
            \end{figure}

            $K^{(\mu\nu)}(P)$ involves a 2-loop integral, to be evaluated for complex momenta. We circumvent its explicit calculation
            by observing that the pole and unit residue conditions (in the axial-vector case only for the transverse part)
                  \begin{equation}
                      D^{-1}_{(T)}(-M^2) = 0\;,   \quad   \left(D^{-1}_{(T)}\right)'(-M^2) = 1\;,
                  \end{equation}

            where the derivative $'$ denotes $d/dP^2$ and each quantity $F^{\mu\nu} = (D^{-1})^{\mu\nu}$,
            $K^{\mu\nu}$ and $Q^{\mu\nu}$ is decomposed as
                  \begin{equation}
                      F^{\mu\nu}(P) = F_T(P^2) \,T^{\mu\nu}_P + F_L(P^2)\, L^{\mu\nu}_P\;,
                  \end{equation}
            with $F_T=T^{\mu\nu}_P F^{\mu\nu}/3$, $F_L=L^{\mu\nu}_P F^{\mu\nu}$, $L^{\mu\nu}_P=\hat{P}^\mu \hat{P}^\nu$ and $T^{\mu\nu}_P=\delta^{\mu\nu}-L^{\mu\nu}_P$,
            impose the following constraints on $K(P^2)$ and $K_T(P^2)$:
                  \begin{align}
                      K_{(T)}(-M^2) &= -Q_{(T)}(-M^2)\;,   \\
                      K'_{(T)}(-M^2) &= -Q'_{(T)}(-M^2) +1\;.    
                  \end{align}
            Asymptotically (i.\,e., for $P^2\rightarrow\infty$) $K^{(\mu\nu)}$ becomes dominant in \eqref{dq:dqprop3} since
            the ladder kernel is independent of the diquark momentum whereas $Q^{(\mu\nu)}$ contains the product of two quark
            propagators which behaves as $1/P^2$.
            The ansatz \eqref{dq:offshellansatz2} entails $K^{(\mu\nu)}\rightarrow P^2$ and $Q^{(\mu\nu)}\rightarrow const.$
            Analyticity at $P^2=0$ requires $K_T(0)=K_L(0)$;
            the corresponding relation $Q_T(0)=Q_L(0)$ is guaranteed by the definition \eqref{dq:quarkloop}.
            We employ a parameterization based on the above restrictions:
                  \begin{equation}\label{dq:kansatz}
                  \begin{split}
                      K(P^2) &= \left( 1-Q'(-M_\text{sc}^2) \right) \left( P^2+M_\text{sc}^2 \right) - Q(-M_\text{sc}^2)\;, \\
                      K^{\mu\nu}(P) &= \left\{ \left(1-Q_T'(-M_\text{av}^2)\right) \left(P^2+M_\text{av}^2\right) - Q_T(-M_\text{av}^2)  \right\} \delta^{\mu\nu}\;.
                  \end{split}
                  \end{equation}
            With Eqs.\,\eqref{dq:scoffshell}--\eqref{dq:quarkloop} and Eq.\,\eqref{dq:kansatz}, the off-shell behavior of the
            T-matrix is completely determined.


\section{Quark-diquark amplitudes and quark-diquark BSE} \label{app:qdqamp}

            The matrix-valued Bethe-Salpeter amplitudes $\Phi^{(\nu)}_{\alpha\beta}(p,P)$ which were introduced in Sec.\,\ref{sec:qdqbse}
            feature a decomposition constructed from the same Dirac basis elements as used in the meson and diquark case:
            \begin{equation}\label{nuc:amplitudes}
                \begin{split}
                \Phi_{\alpha\beta}^5(p,P) &=   \sum_{k=1}^2 f_k^\text{sc}(p^2,z) \big\{ \tau_k(p,P) \, \Lambda_+(P) \big\}_{\alpha\beta}
                                               \,\otimes\, \frac{\delta_{AB}}{\sqrt{3}} \,\otimes\, \mathsf{t^0_{ab}}  \\
                \Phi_{\alpha\beta}^\mu(p,P) &= \sum_{k=1}^6 f_k^\text{av}(p^2,z) \big\{ \tau_k^\mu(p,P) \, \gamma^5 \, \Lambda_+(P) \big\}_{\alpha\beta}
                                               \,\otimes\, \frac{\delta_{AB}}{\sqrt{3}} \,\otimes\, \mathsf{t^e_{ab}}\;  .
                \end{split}
            \end{equation}
            Here $p$ is the relative momentum between quark and diquark and $P$ is the total nucleon momentum on the mass shell: $P^2=-M_N^2$.
            The positive parity condition for the full Faddeev amplitude translates into positive parity of the quark-diquark amplitudes.
            The constraint of positive energy for the nucleon, expressed by the positive-energy projector
            \begin{equation}
                \Lambda_+(P) = \frac{1}{2} \left( \mathds{1} + \frac{\Slash{P}}{i M_N} \right)\;,
            \end{equation}
            halves the number of possible Dirac basis elements via $\Slash{P}\Lambda_+ = i M_N \Lambda_+$ from 4+12
            to 2 Dirac matrices $\{\mathds{1},\Slash{p}\}$ for the
            scalar quark-diquark amplitude and 6 matrices $\{\gamma^\mu,\gamma^\mu \Slash{p}, p^\mu, p^\mu \Slash{p}, P^\mu, P^\mu \Slash{p}\}$
            for the axial-vector one.
            A possible orthogonal basis set is $\tau_1=\mathds{1}$, $\tau_2=-i \Slash{\hat{p}}_T$ for the scalar and
            \begin{align}\label{nuc:avbasis}
                \nonumber  \tau_1^{\mu} &= \gamma^\mu_T       &   \tau_4^{\mu} &= i \left( \gamma^\mu_T\, \Slash{\hat{p}}_T - \hat{p}^\mu_T \right)                 \\
                           \tau_2^{\mu} &= i \hat{p}^\mu_T    &   \tau_5^{\mu} &= \hat{p}^\mu_T \, \Slash{\hat{p}}_T - \frac{\hat{p}_T^2}{3}  \gamma^\mu_T          \\
                \nonumber  \tau_3^{\mu} &= \hat{P}^\mu        &   \tau_6^{\mu} &= i \hat{P}^\mu \Slash{\hat{p}}_T
            \end{align}
            for the axial-vector basis. Similarly as before, $p^\mu_T=T^{\mu\nu}_P p^\nu$, where $T^{\mu\nu}_P$ is now the transverse projector
            with respect to the total nucleon momentum. The orthogonality relations are
            \begin{equation}
                \frac{1}{4} \text{Tr}\{\tau_i\, \tau_j\, \Lambda_+\} = \delta_{ij} a_i^\text{sc}(z),\quad
                \frac{1}{4} \text{Tr}\{\tau_i^{\mu}\, \gamma^5\, \tau_j^{\nu}\,\gamma^5 \, \Lambda_+\} = \delta_{ij} a_i^\text{av}(z)
            \end{equation}
            and
            \begin{equation}
            \begin{split}
            a_1^\text{sc}   &= -\frac{a_1^\text{av}}{3} = a_3^\text{av} = 2\;,  \\
            -a_2^\text{sc}  &= -a_2^\text{av} = \frac{a_4^\text{av}}{2} = a_6^\text{av} = 2(1-z^2)\;, \\
            a_5^\text{av} &= -\frac{4}{3}(1-z^2)^2\;.
            \end{split}
            \end{equation}
            Particular linear combinations of the eight Dirac structures $\tau_k(p,P) \, \Lambda_+(P)$ and $\tau_k^\mu(p,P) \, \gamma^5 \, \Lambda_+(P)$
            lead to a partial wave decomposition in terms of quark-diquark total spin and orbital angular momentum eigenstates in the nucleon's
            rest frame \cite{Oettel:1998bk,Alkofer:2005jh}. These combinations correspond to relative $s$, $p$ and $d$ waves which implicates that the nucleon is not spherically symmetric.

            The flavor matrices $\mathsf{t^e}$ in the quark-diquark amplitudes, when applied to proton and neutron isospinors $(1,0)$ and $(0,1)$,
            yield the "quark remainders" of the full proton and neutron flavor wave functions
            constructed from the Clebsch-Gordan prescription after removing the diquark contributions.
            They are given by $\mathsf{t^0}=\mathds{1}$, $\mathsf{t^1}= (\sigma_1 - i\sigma_2)/\sqrt{6}$, $\mathsf{t^2}=-\sigma_3/\sqrt{3}$ and $\mathsf{t^3}=- (\sigma_1 + i\sigma_2)/\sqrt{6}$.
            The flavor traces of the quark-diquark Bethe-Salpeter equation for proton and neutron are obtained by projection on the right and the left with $(1,0)$ and $(0,1)$, respectively.
            Similarly, the antisymmetric color factor of the full Faddeev amplitude is obtained as product of diquark and quark-diquark color factors
            and has its origin in the diquark amplitudes (\ref{dq:sc},\,\ref{dq:av}).

            The quark-diquark amplitudes are normalized by an analogous normalization integral as given in Eq.\,\eqref{bs:normalization}:
            \begin{equation}\label{nuc:normalization}
                \bar{\Phi} \,\frac{d\,(K^{-1}-G_0)}{d P^2}\,\Phi  = 1\;.
            \end{equation}
            In this context, $\Phi$ is the quark-diquark amplitude as solution of the Bethe-Salpeter equation \eqref{nuc:bse},
            $K$ is the quark-diquark kernel \eqref{nuc:kernel} and $G_0$ the product of dressed quark and diquark propagators.
            The normalization condition is equivalent to the normalization of the electric charge of the proton, $G_E^p(0)=1$ \cite{Oettel:1999gc}.
            In contrast to the meson and diquark case, Eqs.\,\eqref{bse:normalization} and \eqref{dq:normalization},
            the quark-diquark kernel depends on the total nucleon momentum $P$ such that $dK^{-1}/dP^2$ cannot be omitted.


\section{Construction of the electromagnetic current diagrams}\label{app:em}

            The electromagnetic current of Eq.\,\eqref{ff:current} has the explicit form
            \begin{equation}\label{ff:current2}
                J^\mu_{\alpha\beta}(Q^2) = \int\! \frac{d^4 p_f}{(2\pi)^4} \!\int\frac{d^4 p_i}{(2\pi)^4} \, \big\{   \bar{\Phi}^a(p_f,-P_f)\, X^{\mu,ab}(p_f,p_i,P_f,P_i)\, \Phi^b(p_i,P_i)  \big\}_{\alpha\beta}\;,
            \end{equation}
            where $P_i$ and $P_f=P_i+Q$ are incoming and outgoing on-shell nucleon momenta. The loop momenta $p_i$ and $p_f$ are arbitrary.
            $\alpha,\beta=1\dots 4$ are quark and $a,b=1\dots 5$ are diquark indices. The quark-diquark amplitudes $\Phi^a$ are the solutions
            of the quark-diquark Bethe-Salpeter equation \eqref{nuc:bse}. The quantity $X^{\mu,ab}$ is given by
            \begin{equation}
            \begin{split}
                X^{\mu,ab} = \,   & X^{\mu,ab}_\text{q} \, (2\pi)^4 \delta^4 \left( p_f-p_i-(1-\eta)\,Q \right)   + \\
                             + \, & X^{\mu,ab}_\text{dq} \,(2\pi)^4 \delta^4 \left( p_f-p_i+\eta \,Q \right)   +  X^{\mu,ab}_\text{K}\;,
            \end{split}
            \end{equation}
            with
            \begin{align}
                \left( X_\text{q} \right)^{\mu,ab}_{\alpha\beta}  &= \left\{ S(p_+)\,\Gamma^\mu_\text{q}(p_+,p_-)\, S(p_-) \right\}_{\alpha\beta} D^{ab}(p_{d-}) \;, \label{ff:J-q}\\
                \left( X_\text{dq} \right)^{\mu,ab}_{\alpha\beta} &=  S_{\alpha\beta}(p_-) \left\{ D^{aa'}(p_{d+}) \, \Gamma^{\mu,a'b'}_\text{dq}(p_{d+},p_{d-}) \, D^{b'b}(p_{d-}) \right\} \;, \label{ff:J-dq}\\
                \left( X_\text{K} \right)^{\mu,ab}_{\alpha\beta}  &= D^{aa'}(p_{d+})\left\{ S(p_+)\,K^{\mu,a'b'}(p_f,p_i,P_f,P_i)\, S(p_-) \right\}_{\alpha\beta}\,D^{b'b}(p_{d-})\;.
            \end{align}

            \begin{figure}[hbt]
            \begin{center}
            \includegraphics[scale=0.46,angle=-90]{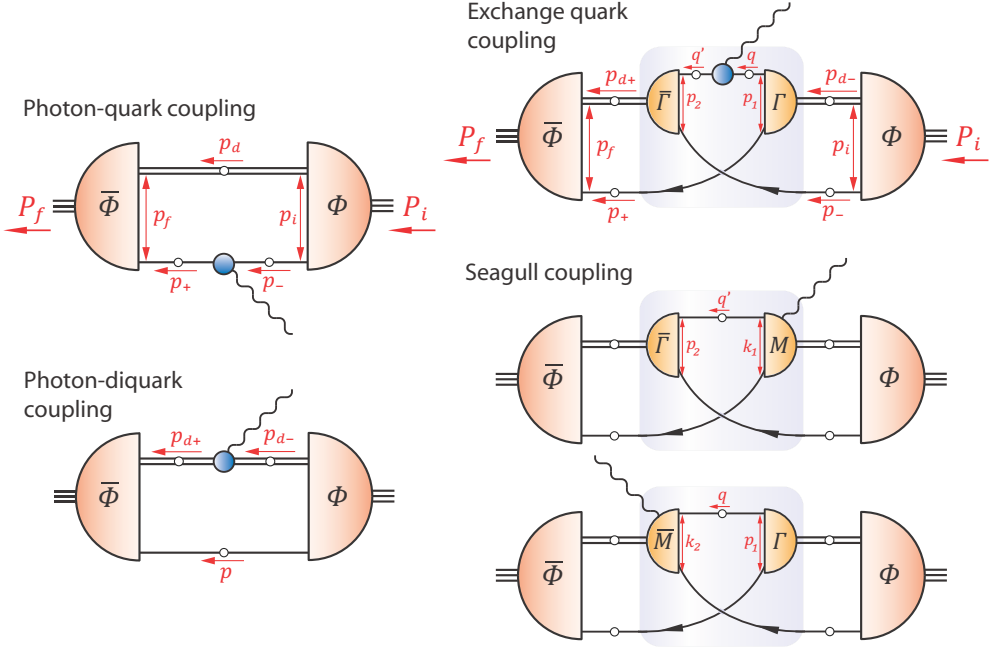}
            \caption[Electromagnetic current diagrams]{
                     The five diagrams that contribute to the nucleon's electromagnetic current, corresponding to Eqs.~\eqref{ff:J-q}--\eqref{ff:J-sgbar}.} \label{fig:current}
            \end{center}
            \end{figure}

            The first two contributions correspond to the impulse approximation and contain quark-photon and diquark-photon vertices.
            The third term is a two-loop diagram and reflects the gauged quark-diquark kernel.
            The diquark propagator and therefore also the quark-photon coupling $X_\text{q}^{\mu,ab}$ is either scalar ($a,b=5$)
            or axial-vector ($a,b=1\dots 4$) while $X_\text{dq}^{\mu,ab}$ and $X_\text{K}^{\mu,ab}$ mix scalar and axial-vector quantities.
            The quark and diquark momenta are:
            \begin{align*}
                p_- &= p_i+\eta\,P_i    &   p_{d-} &= -p_i + (1-\eta)\,P_i\;,  \\
                p_+ &= p_f+\eta\,P_f    &   p_{d+} &= -p_f + (1-\eta)\,P_f\;.
            \end{align*}
            The gauged kernel contains the exchange quark diagram and the seagull contributions:
            \begin{equation}
                K^{\mu,ab} = K^{\mu,ab}_\text{EX} + K^{\mu,ab}_\text{SG} + K^{\mu,ab}_{\overline{\text{SG}}}\;,
            \end{equation}
            with
            \begin{align}
                K_\text{EX}^{\mu,ab}              &= \Gamma_{dq}^b(p_1,p_{d-})\left\{ S(q') \,\Gamma^\mu_\text{q}(q',q)\, S(q) \right\}^T \bar{\Gamma}^a(p_2,-p_{d+})  \label{ff:J-ex}\\
                K_\text{SG}^{\mu,ab}              &= M^{\mu,b}(k_1,p_{d-},Q)\, S^T(q')\,\bar{\Gamma}^a(p_2,-p_{d+}) \label{ff:J-sg}\\
                K_{\overline{\text{SG}}}^{\mu,ab} &= \Gamma_{dq}^b(p_1,p_{d-})\,S^T(q)\,\bar{M}^{\mu,a}(k_2,-p_{d+},Q) \label{ff:J-sgbar}
            \end{align}
            and momenta:
            \begin{align*}
                q  &= p_{d-}-p_+\;, &
                p_1 &= \frac{p_+-q}{2}\;, &
                p_2 &= \frac{p_--q'}{2}\;, \\
                q' &= p_{d+}-p_-\;, &
                k_1 &= \frac{p_+-q'}{2}\;, &
                k_2 &= \frac{p_--q}{2}\;.
            \end{align*}
            For explicit calculations we work in the Breit frame where
            \begin{equation}
                \tilde{P} = \frac{P_f+P_i}{2} = \left\{ \, 0, \,0, \,0, \,i M \sqrt{1+\tau} \,\right\}\;, \qquad Q = \left\{ \,0, \,0, \,|Q|, \,0 \,\right\}\;,
            \end{equation}
            with $\tau=Q^2/(4M^2)$. The electromagnetic current is completely determined by specifying the quark-photon vertex $\Gamma^\mu_\text{q}$,
            the diquark-photon vertex $\Gamma^\mu_\text{dq}$, and the diquark-quark-photon vertex or seagull $M^\mu$.


\subsection{Quark-photon and diquark-photon vertex}\label{app:vertices}

            With the notation $P=(p+q)/2$ and $Q=p-q$, where $p$ and $q$ are outgoing and incoming quark or diquark momenta,
            the general form of the quark-photon vertex and the scalar and axial-vector diquark-photon vertices is
                  \begin{align}
                      \Gamma^\mu_\text{q}(p,q) &=  \sum_{k=1}^{12} f^\text{q}_k(p^2,q^2,Q^2) \left(\tau_\text{q}\right)_k^\mu(P,Q)\;, \\
                      \Gamma^\mu_\text{dq}(p,q) &= \sum_{k=1}^{2} f^\text{sc-dq}_k(p^2,q^2,Q^2) \left(\tau_\text{dq}\right)_k^\mu(P,Q)\;, \\
                      \Gamma^{\mu,\alpha\beta}_\text{dq}(p,q) &= \sum_{k=1}^{14} f^\text{av-dq}_k(p^2,q^2,Q^2) \left(\tau_\text{dq}\right)_k^{\mu,\alpha\beta}(P,Q)\;,
                  \end{align}
            with the basis elements:
                  \begin{align}\label{ff:qphvertex}
                       & \qquad\qquad\left(\tau_\text{q}\right)_{k=1\dots 12}^\mu(P,Q) & = \quad & \gamma^\mu \{\mathds{1}, \Slash{Q}, \, \Slash{P}, \, \Slash{P}\,\Slash{Q} \}\;, &&\\
                                                    &&   & P^\mu \{\mathds{1}, \Slash{Q}, \, \Slash{P}, \, \Slash{P}\,\Slash{Q} \}\;, &&\nonumber\\
                                                    &&   & Q^\mu \{\mathds{1}, \Slash{Q}, \, \Slash{P}, \, \Slash{P}\,\Slash{Q} \}  &&\nonumber\\
                       & \qquad\qquad\left(\tau_\text{dq}\right)_{k=1,2}^\mu(P,Q) & = \quad & P^\mu, Q^\mu &\\
                       & \qquad\qquad\left(\tau_\text{dq}\right)_{k=1\dots 14}^{\mu,\alpha\beta}(P,Q) & = \quad & \delta^{\mu\alpha} \{Q^\beta, P^\beta \}, \, \delta^{\mu\beta} \{Q^\alpha, P^\alpha \}\;, &&\\
                                                                &&     & P^\mu \{\delta^{\alpha\beta}, P^\alpha P^\beta, \, Q^\alpha Q^\beta, \, P^\alpha Q^\beta, \,Q^\alpha P^\beta \}\;, &&\nonumber\\
                                                                &&     & Q^\mu \{\delta^{\alpha\beta}, P^\alpha P^\beta, \, Q^\alpha Q^\beta, \, P^\alpha Q^\beta, \,Q^\alpha P^\beta \}\;. &\qquad&\nonumber
                  \end{align}
            All vertices satisfy vector Ward-Takahashi identities which reflect electromagnetic current conservation.
            They constrain the longitudinal contributions $\sim Q^\mu$
            by relating them to the corresponding quark or diquark propagators:
                  \begin{align}
                      Q^\mu \,\Gamma^\mu_\text{q}(p,q) &= S^{-1}(p)-S^{-1}(q)  \\
                      Q^\mu \, \Gamma^\mu_\text{dq}(p,q) &= D^{-1}(p^2)-D^{-1}(q^2)  \label{wti:dqsc}\\
                      Q^\mu \, \Gamma^{\mu,\alpha\beta}_\text{dq}(p,q) &= D^{-1}_{\alpha\beta}(p) - D^{-1}_{\alpha\beta}(q) \label{wti:dqav}
                  \end{align}
            Implementation of the WTI for the quark-photon vertex with the general quark propagator \eqref{dse:qprop} leads to the expression
                  \begin{equation}\label{ff:qpv}
                  \begin{split}
                      \Gamma^\mu_\text{q}(p,q) = \,\, & \frac{A(p^2)+A(q^2)}{2} i\gamma^\mu   +  \Delta A(p^2,q^2) \frac{i (\Slash{p}+\Slash{q})}{2} (p+q)^\mu   \, +\\
                                                +\,\, & \Delta B(p^2,q^2)\,(p+q)^\mu + T^{\mu\nu}_Q \sum_{k=1}^8 f^\text{q}_k(p^2,q^2,Q^2) \left(\tau_\text{q}\right)_k^\nu(P,Q)\;,
                  \end{split}
                  \end{equation}
            where $B(p^2)=M(p^2)A(p^2)$ and
                  \begin{equation*}
                      \Delta F(p^2,q^2) := \frac{F(p^2)-F(q^2)}{p^2-q^2}\;,
                  \end{equation*}
            i.\,e., a sum of the Ball-Chiu vertex \cite{Ball:1980ay} and 8 unknown transverse components.
            The scalar diquark-photon vertex then reads
                  \begin{equation}\label{ff:scdqpv}
                      \Gamma^\mu_\text{dq}(p,q) =  (p+q)^\mu \,\Delta D^{-1}(p^2,q^2) + f^\text{sc-dq}_1(p^2,q^2,Q^2)\, T^{\mu\nu}_Q P^\nu
                  \end{equation}
            and the axial-vector diquark-photon vertex:
                  \begin{equation}\label{ff:avdqpv}
                  \begin{split}
                      \Gamma^{\mu,\alpha\beta}_\text{dq}(p,q) = \,\, & (p+q)^\mu \left\{ \Delta D^{-1}_T(p^2,q^2) \, \delta^{\alpha\beta} - \Delta \sigma_D(p^2,q^2) \, p^\alpha \,q^\beta \right\}  \\
                                                    - \,\, & \left\{ \sigma_D(p^2) \, \delta^{\mu\beta} \,p^\alpha  + \sigma_D(q^2) \, \delta^{\mu\alpha} \,q^\beta  \right\} \\
                                                    - \,\, & T^{\mu\nu}_Q \sum_{k=1}^9 f^\text{av-dq}_k(p^2,q^2,Q^2) \,\left(\tau_\text{dq}\right)_k^{\nu,\alpha\beta}(P,Q)\;,
                  \end{split}
                  \end{equation}
            with $\sigma_D(p^2):=\left(D^{-1}_T(p^2)-D^{-1}_L(p^2)\right)/p^2$
            such that the inverse axial-vector diquark propagator is written as
                  \begin{equation}
                      D^{-1}_{\alpha\beta}(p) = D^{-1}_T(p^2)\,\delta^{\alpha\beta} - \sigma_D(p^2)\,p^\alpha p^\beta\;.
                  \end{equation}

            The transverse parts of the vertices (\ref{ff:qpv}--\ref{ff:avdqpv}) must vanish for $Q^2\rightarrow 0$,
            either by a $Q^\mu$ dependence of the basis elements or
            by a vanishing amplitude at $Q^2=0$. This guarantees the differential Ward identities
                  \begin{align}
                      \Gamma^\mu_\text{q}(p,p) &= \frac{d S^{-1}(p)}{d p^\mu}= i\gamma^\mu A(p^2) +2 p^\mu \left( i\Slash{p} \,\frac{dA}{dp^2}+\frac{dB}{dp^2} \right)\;, \\
                      \Gamma^\mu_\text{dq}(p,p) &= \frac{dD^{-1}(p^2)}{dp^\mu} \, \stackrel{p^2=-m_\text{sc}^2}{\longlongrightarrow} \, 2 p^\mu\;,  \\
                      \Gamma^{\mu,\alpha\beta}_\text{dq}(p,p) &= \frac{d D^{-1}_{\alpha\beta}(p)}{dp^\mu} \,
                                                                 \stackrel{p^2=-m_\text{av}^2}{\longlongrightarrow} \, 2 p^\mu \delta^{\alpha\beta} -
                                                                 \frac{d}{dp^\mu} \left(  \sigma_D(p^2)\,p^\alpha p^\beta  \right)\;.
                  \end{align}
            In the axial-vector case the symmetry requirement $\Gamma^{\mu,\alpha\beta}_\text{dq}(p,q) = \Gamma^{\mu,\beta\alpha}_\text{dq}(q,p)$ entails
            $f_3=-\tilde{f}_1$, $f_4=\tilde{f}_2$, $f_9=-\tilde{f}_8$, and $f_{1,2,5,6,7,8} = \tilde{f}_{1,2,5,6,7,8}$
            where $\tilde{f}_i(p^2,q^2,Q^2) = f_i(q^2,p^2,Q^2)$, therefore 6 unknown dressing functions are left \cite{Salam:1964zk}.

            \begin{figure}[hbt]
            \begin{center}
            \includegraphics[scale=0.46,angle=-90]{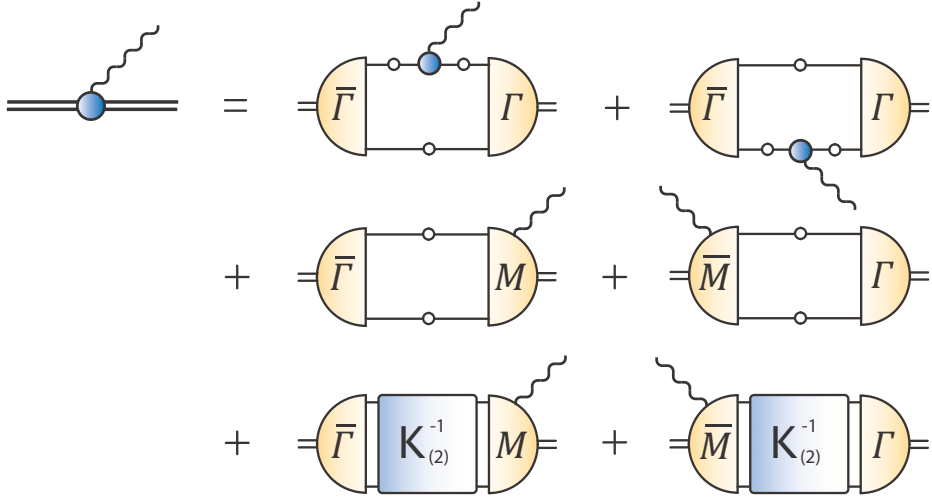}
            \caption[Diquark-photon vertex]{The full diquark-photon vertex. The first two rows correspond to Eq.\,\eqref{ff:quarkloop}.
                                         The last row can be dropped for the scalar-axialvector transition vertex;
                                         for scalar and axial-vector diquark-photon vertices an ansatz is employed,
                                         see Eq.\,\eqref{ff:dqphotonvertexfull}.} \label{fig:dqphotonvertex}
            \end{center}
            \end{figure}

            In the actual calculation we follow Ref.\,\cite{Oettel:2002wf} and express the diquark-photon vertices
            by the "gauged" diquark propagators, cf. Eqs.\,(\ref{dq:dqprop3}-\ref{dq:quarkloop}) and the discussion in Sec.\,\ref{sec:em}:
                  \begin{align} \label{ff:quarkloop}
                      \tilde{\Gamma}^{\mu,ab}_\text{dq}(p,q) = -\frac{1}{2} \int \frac{d^4 k}{(2\pi)^4}\,\text{Tr} & \,  \Big\{  \Lambda_\uparrow^{\mu,ab}(p,q,k) +  \\
                                                &+ \bar{\Gamma}^a(k,-p)\,S(k_+')\,M^{\mu,b}(k,q,Q)\,S^T(k_-') + \nonumber\\
                                                &+ \bar{M}^{\mu,a}(k,-p,Q)\,S(k_+)\,\Gamma^b(k,q)\,S^T(k_-) \Big\}     \nonumber
                  \end{align}
            with the momenta $k_\pm=\pm k+q/2$ and $k'_\pm=\pm k+p/2$.
            The quantity
                  \begin{equation}\label{dq:quarkloop2}
                  \begin{split}
                      \Lambda_\uparrow^{\mu,ab}(p,q,k)  &= \bar{\Gamma}^a(k-Q/2,-p)\,S(k_+)\,\Gamma^b(k,q) \times \\
                                                        & \times \left[ S(k_-\!+\!Q)\,\Gamma^\mu_\text{q}(k_-\!+\!Q,k_-)\, S(k_-) \right]^T
                  \end{split}
                  \end{equation}
            denotes the quark loop where the photon couples to the upper quark line (see Fig.\,\ref{fig:dqphotonvertex}).
            The seagull contributions $M^{\mu,b}$ and $\bar{M}^{\mu,a}$
            that appear in Eq.\,\eqref{ff:quarkloop} are defined in App.\,\ref{app:seagulls}.
            They are necessary to guarantee the Ward-Takahashi identities for the
            scalar ($a,b=5$) and axial-vector ($a,b=1\dots 4$) vertices and transversality for the
            scalar-axialvector transition vertices: $Q^\mu \,\Gamma^{\mu,5b}_\text{dq} = Q^\mu\,\Gamma^{\mu,a5}_\text{dq} = 0$.
            To show this one uses the relations $\Gamma^T(k,P)=-\Gamma(-k,P)$, $\Gamma^{\mu\,T}(k,P)=\Gamma^\mu(-k,P)$
            for the Dirac parts of the diquark amplitudes which follow from the general antisymmetry relation \eqref{dq:antisymmetry}.
            Eq.\,\eqref{ff:quarkloop} furthermore ensures the correct symmetry behavior $\Gamma^{\mu,5b}_\text{dq}(p,q)=-\Gamma^{\mu,b5}_\text{dq}(-q,-p)$.
            The color and flavor factors in Eq.\,\eqref{ff:quarkloop} have already been worked out. They entail
            $e_-=e_+=e_{dq}/2=1/2$ in the scalar and axial-vector case
            and $e_-=-e_+=1/2$, $e_{dq}=0$ for the transition vertex
            ($e_\pm$ and $e_{dq}$ appear in the definition of the seagulls, cf. App\,\ref{app:seagulls}).
            The color trace for all contributions is $1$.
            For instance, using the diquark flavor matrices $\mathsf{s_i}$ ($i=0\dots 3$)
            of App.\,\ref{app:mesondiquark-general} entails that the quark-loop contributions
            of Eq.\,\eqref{ff:quarkloop} are given by
                  \begin{equation}
                      \text{Tr}\left\{ \mathsf{s_i^\dag\,s_j\,Q}\right\} \Lambda_\uparrow^{\mu,ab} +
                      \text{Tr}\left\{ \mathsf{s_i^\dag\,Q\,s_j}\right\} \Lambda_\downarrow^{\mu,ab} =
                      2\,\text{Tr}\left\{ \mathsf{s_i^\dag\,s_j\,Q}\right\} \Lambda_\uparrow^{\mu,ab}\;,
                  \end{equation}
            where $\mathsf{Q}=\text{diag}(q_u,q_d)$ is the quark electric charge matrix, and
            $\Lambda_\downarrow$ denotes the counterpart to Eq.\,\eqref{dq:quarkloop2} where the photon couples to the
            lower quark line.
            In App.\,\ref{app:color-flavor-charge} the diquark charge factors
            \begin{equation}
                2\,\text{Tr}\{\mathsf{s_i^\dag \,s_j \,Q}\} = \left(
                \begin{array}{c|ccc}
                    q_u+q_d    &    0    &    q_d-q_u    &    0    \\ \hline
                    0          &   2q_u  &    0          &    0    \\
                    q_d-q_u    &    0    &    q_u+q_d    &    0    \\
                    0          &    0    &    0          &   2q_d  \\
                \end{array}
                \right)
            \end{equation}
            are explicitly attached to the current matrix diagrams at each occurrence of the diquark-photon vertex.

            In order to satisfy the Ward-Takahashi identities (\ref{wti:dqsc}-\ref{wti:dqav}) for the full diquark propagators \eqref{dq:dqprop3},
            also the diagram involving the gauged inverse ladder kernel
            has to be taken into account when constructing the scalar and axial-vector diquark-photon vertices.
            It depends on the diquark momentum only via the diquark amplitudes, therefore
            its generic form is $\bar{\Gamma}^a\, K^{-1} M^{\mu,b} + \bar{M}^{\mu,a}\,K^{-1} \Gamma^b$.
            Since we employed a parameterization for $K^{(\mu\nu)}$ we have to choose an ansatz for this contribution.
            We use the generic vertices of Eqs.\,(\ref{ff:scdqpv}-\ref{ff:avdqpv}), drop the transverse parts
            and replace $D^{-1}\rightarrow K$, $\sigma_D \rightarrow \sigma_{K^{-1}}$. With the ansatz \eqref{dq:kansatz} for $K$
            the full diquark-photon vertices read
            \begin{equation}\label{ff:dqphotonvertexfull}
            \begin{split}
                \Gamma^\mu_\text{dq}(p,q) &=  \tilde{\Gamma}^\mu_\text{dq}(p,q) + (p+q)^\mu \,\Delta K(p^2,q^2)\;, \\
                \Gamma^{\mu,\alpha\beta}_\text{dq}(p,q) &= \tilde{\Gamma}^{\mu,\alpha\beta}_\text{dq}(p,q) + (p+q)^\mu  \, \Delta K_T(p^2,q^2) \,
		\delta^{\alpha\beta}\;.
            \end{split}
            \end{equation}


\subsection{Seagulls} \label{app:seagulls}

\begin{figure}[hbt]
            \begin{center}
            \includegraphics[scale=0.42,angle=-90]{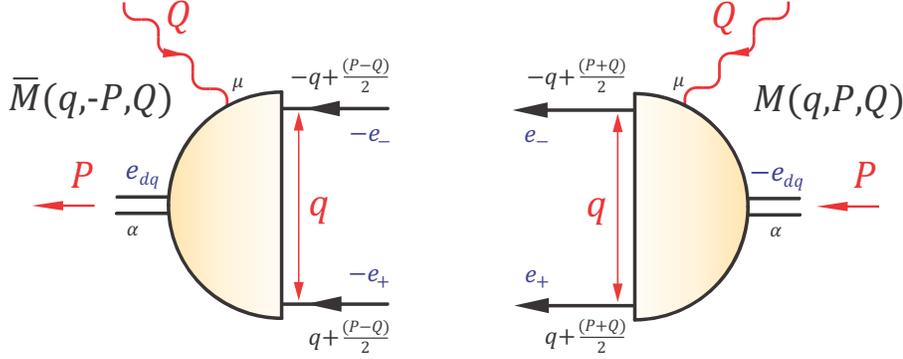}
            \caption[Seagulls]{Conventions for the seagulls. In accordance with the Ward-Takahashi identity \eqref{ff:seagullwti},
                                         outgoing charges are taken to be positive, incoming charges as negative. } \label{fig:seagulls}
            \end{center}
            \end{figure}

            Seagull contributions represent the photon coupling to the diquark amplitudes and reflect the diquark's
            internal substructure. The Ward-Takahashi identities for the seagulls are given by \cite{Oettel:1999gc,Wang:1996zu}:
            \begin{equation}\label{ff:seagullwti}
            \begin{split}
            Q^\mu M^\mu(q,P,Q) \,=\, &+  e_- \, \big\{ \Gamma(q_+,P)-\Gamma(q,P) \big\}  \\
                              &+  e_+ \, \big\{ \Gamma(q_-,P)-\Gamma(q,P) \big\} \\
                              &-  e_{dq}\,  \big\{ \Gamma(q,P_+)-\Gamma(q,P) \big\}\;, \\
            Q^\mu \bar{M}^\mu(q,-P,Q) \,=\, &-  e_+ \, \big\{ \bar{\Gamma}(q_+,-P)-\bar{\Gamma}(q,-P) \big\} \\
                                    &-  e_- \, \big\{ \bar{\Gamma}(q_-,-P)-\bar{\Gamma}(q,-P) \big\}   \\
                                    &+  e_{dq}\,  \big\{ \bar{\Gamma}(q,-P_-)-\bar{\Gamma}(q,-P) \big\}\;,
            \end{split}
            \end{equation}
            where $q$ is the relative momentum, $P$ is the diquark's total momentum, $Q$ is the photon momentum,
            $q_\pm = q \pm Q/2$, and $P_\pm = P \pm Q$. The charges $e_+$, $e_-$ and $e_{dq}$ correspond to
            quark and diquark legs (cf. Fig.\,\ref{fig:seagulls}).
            For $Q\rightarrow 0$ the WTIs reduce to the differential Ward identities:
            \begin{equation}
            \begin{split}
                  M^\mu(q,P,0) &\,=\,  \left( \frac{e_--e_+}{2} \frac{d}{dq^\mu}-e_{dq}\frac{d}{dP^\mu}\right) \Gamma(q,P)\;, \\
            \bar{M}^\mu(q,-P,0) &\,=\,  \left( \frac{e_--e_+}{2} \frac{d}{dq^\mu}-e_{dq}\frac{d}{dP^\mu}\right) \bar{\Gamma}(q,-P)\;.
            \end{split}
            \end{equation}

            To simplify the discussion, we rewrite
            the diquark amplitudes (\ref{dq:scoffshell}--\ref{dq:avbasis}) in terms of the general basis elements
            $\tau_k^{(\mu)}(q,P)$ of Eqs.\,(\ref{bse:scbasisgeneral}--\ref{bse:vcbasisgeneral}), i.\,e.
            \begin{align*}
                \Gamma_\text{sc}(q,P)     &\,=\, \sum_{k=1}^4 f_k^\text{sc}(q^2,z,P^2) \, i\gamma^5 \tau_k(q,P)\,C\;,  \\
                \Gamma^\mu_\text{av}(q,P) &\,=\, \sum_{k=1}^8 f_k^\text{av}(q^2,z,P^2) \, i\tau^\mu_k (q,P)\,C\;,
            \end{align*}
            where $z=\hat{q}\cdot \hat{P}$. The scalar and axial-vector diquark dressing functions $f_k^\text{sc}$ and $f_k^\text{av}$ are the respective
            linear combinations of the coefficients in Eqs.\,(\ref{dq:scoffshell}--\ref{dq:avoffshell}) and therefore carry now
            a dependence on the total diquark momentum $P^2$. That being said, we also drop the superscripts 'sc' and 'av' for the purpose of brevity.
            A possible ansatz for the seagulls corresponding to the scalar diquark amplitudes
            that satisfies the WTI and is free of singularities is then given by
            \begin{equation}\label{ff:seagullssc}
            \begin{split}
            -i \gamma^5 & \,M^\mu(q,P,Q) \,C^\dag = \\
                     =  & \sum_{k=1}^4 \Big\{  e_- (V_+)_k^\mu \,\tau_k(q_+,P)  + e_+ (V_-)_k^\mu \,\tau_k(q_-,P) - e_{dq} \tilde{V}_k^\mu \,\tau_k(q,P_+) \Big\}\\
                        & + \frac{e_--e_+}{2} \gamma^\mu (f_3 + f_4 \Slash{P}) - e_{dq} (f_2 + f_4 \Slash{q}) \gamma^\mu\;,
            \end{split}
            \end{equation}
            and the corresponding ansatz for the axial-vector seagull contribution is
            \begin{equation}\label{ff:seagullsav}
            \begin{split}
            -i & \, M^{\mu,\alpha}(q,P,Q) \, C^\dag = \\
             = &   \sum_{k=1}^8 \Big\{ e_- (V_+)_k^\mu \,\tau_k(q_+,P)
                 + e_+ (V_-)_k^\mu \,\tau_k^\alpha(q_-,P)
                 - e_{dq} \tilde{V}_k^\mu \,\tau_k^\alpha(q,P_+) \Big\}\\
               & + \frac{e_--e_+}{2} \Big\{  \gamma^\alpha \gamma^\mu (f_3 + f_4 \Slash{P})
                                           + \delta^{\mu\alpha} (f_5+f_6 \Slash{P})
                                           + (\delta^{\mu\alpha} \Slash{q} + q^\alpha \gamma^\mu)(f_7+f_8\Slash{P}) \Big\} \\
               & + \frac{e_-+e_+}{4} \,Q^\alpha \gamma^\mu (f_7+f_8\Slash{P})
                 - e_{dq} \Big\{ \gamma^\alpha(f_2 + f_4 \Slash{q}) +q^\alpha(f_6+f_8\Slash{q}) \Big\} \gamma^\mu\;.
            \end{split}
            \end{equation}
            The conjugated seagulls are obtained from
            \begin{align*}
                \bar{M}^\mu(q,-P,Q) &= -C(M^\mu)^T(-q,-P,Q)\,C^{-1}\;, \\
                \bar{M}^{\mu,\alpha}(q,-P,Q) &= C (M^{\mu,\alpha})^T(-q,-P,Q)\,C^{-1}\;.
            \end{align*}
            The quantities $(V_\pm)_k^\mu$ and $\tilde{V}_k^\mu$ are defined by
            \begin{align*}
             (V_\pm)_k^\mu  = \,& X_\pm^\mu \, \left\{ f_k(q_\pm^2,z,P^2) - f_k(q^2,z,P^2) \right\} \\
                            + \,& \frac{P^\mu - 2 \sigma_\pm q\!\cdot\! P \,X_\pm^\mu}{P\!\cdot\! Q - 2 \sigma_\pm q\!\cdot\! P} \,
                                \left\{ f_k(q_\pm^2,z_\pm,P^2) - f_k(q_\pm^2,z,P^2) \right\}\;, \\[0.2cm]
             \tilde{V}_k^\mu  = \,& \tilde{X}^\mu \, \left\{ f_k(q^2,z,P_+^2) - f_k(q^2,z,P^2) \right\} \\
                                    + \,& \frac{q^\mu - 2 \tilde{\sigma} \,q\!\cdot\! P \,\tilde{X}^\mu}{q\!\cdot\! Q - 2 \tilde{\sigma} \,q \!\cdot\! P} \,\left\{ f_k(q^2,\tilde{z},P_+^2) - f_k(q^2,z,P_+^2) \right\}\;, \\[0.2cm]
            X_\pm^\mu  =\,& \frac{(q\pm Q/4)^\mu}{(q\pm Q/4)\!\cdot\! Q}\;, \;\; \tilde{X}^\mu = \frac{(P+ Q/2)^\mu}{(P+ Q/2)\!\cdot\! Q}\;,  \\
            z_\pm  = \,& \hat{q}_\pm \!\cdot\! \hat{P}\;,   \;\;  \tilde{z} = \hat{q} \!\cdot\! \hat{P}_+\;, \;\;  1 \pm \sigma_\pm = \sqrt{\frac{q_\pm^2}{q^2}} \;  ,  \;\; 1 + \tilde{\sigma} = \sqrt{\frac{P_+^2}{P^2}}\;.
            \end{align*}
            In the case where only the dominant diquark amplitude is retained and its dressing function $f_1$ (either scalar or axial-vector) depends solely on $q^2$,
            these ans{\"a}tze reduce to the forms $M^\mu = W^\mu \,i\gamma^5 C$ and $M^{\mu,\alpha} = W^\mu \,i\gamma^\alpha C$ (c.f. Refs.\,\cite{Oettel:1999gc,Oettel:2000jj,Oettel:2002wf,Alkofer:2004yf}),
            where
            \begin{equation*}
                W^\mu = e_- X_+^\mu \left( f_1(q_+^2)-f_1(q^2) \right) + e_+ X_-^\mu \left( f_1(q_-^2)-f_1(q^2) \right)\;.
            \end{equation*}
            The seagulls vanish if the diquark amplitudes are taken to be pointlike, i.e., $f_1(q^2)=const.$


\subsection{Color, flavor and charge coefficients} \label{app:color-flavor-charge}

             The current matrix diagrams still have to be endued with color and flavor-charge coefficients.
             The color traces for the impulse approximation and exchange/seagull diagrams are given by
             \begin{equation}
                 \frac{\delta_{BA}}{\sqrt{3}}\,\frac{\delta_{AB}}{\sqrt{3}}=1\;, \quad
                 \frac{\delta_{BA}}{\sqrt{3}} \, \frac{\varepsilon_{AED}}{\sqrt{2}}\,\frac{\varepsilon_{CEB}}{\sqrt{2}}\,\frac{\delta_{CD}}{\sqrt{3}}=-1\;,
             \end{equation}
             respectively.
             With the diquark and quark-diquark isospin matrices of Secs.\,\ref{app:mesondiquark-general} and \ref{app:qdqamp}:
             $\mathsf{s_i}$ and $\mathsf{t_i}$, $i=0\dots 3$,
             and the quark charge matrix $\mathsf{Q}=\text{diag}(q_u,q_d)$,
             the flavor-charge matrices for the quark-photon, diquark photon and exchange diagrams read:
             \begin{equation} \label{flavorcharge:1}
                 \sum_{ij} \delta_{ij} \,\mathsf{t_i^\dag} \,\mathsf{Q}\, \mathsf{t_j}\;, \quad
                 \sum_{ij} \mathsf{t_i^\dag} \, \mathsf{t_j} \,2\, \text{Tr}\left\{ \mathsf{s_i^\dag}\,\mathsf{s_j}\,\mathsf{Q} \right\}\; , \quad
                 \sum_{ij} \mathsf{t_i^\dag} \,\mathsf{s_j}\,\mathsf{Q}\, \mathsf{s_i^\dag}\,\mathsf{t_j}\;.
             \end{equation}
             The traces for proton and neutron are obtained by sandwiching these expressions between the isospinors $\mathsf{u}=(1,0)$ or $\mathsf{d}=(0,1)$, respectively.
             The index range of the sums in \eqref{flavorcharge:1} depends on the type of the quark-diquark amplitude in the initial and final state: e.\,g.,
             for a scalar quark-diquark amplitude in the final state (i.\,e., on the left-hand side): $i=0$, for an axial-vector amplitude: $i=1\dots 3$.
             For instance, with the four contributions:
             \begin{align*}
                 &\text{S}\leftarrow\text{S}: &\mathsf{u^\dag}\left(\mathsf{t_0^\dag}\,\mathsf{s_0}\,\mathsf{Q}\, \mathsf{s_0^\dag}\,\mathsf{t_0}\right)\mathsf{u} &= \frac{q_d}{2}\;, \qquad\\
                 &\text{S}\leftarrow\text{A}: &\mathsf{u^\dag}\left(\sum_{j=1}^3\mathsf{t_0^\dag}\,\mathsf{s_j}\,\mathsf{Q}\, \mathsf{s_0^\dag}\,\mathsf{t_j}\right)\mathsf{u} &= -\frac{2q_u+q_d}{2\sqrt{3}}\;, \qquad\\
                 &\text{A}\leftarrow\text{S}: &\mathsf{u^\dag}\left(\sum_{i=1}^3,\mathsf{t_i^\dag}\,\mathsf{s_0}\,\mathsf{Q}\, \mathsf{s_i^\dag}\,\mathsf{t_0}\right)\mathsf{u} &= -\frac{2q_u+q_d}{2\sqrt{3}}\;, \qquad\\
                 &\text{A}\leftarrow\text{A}: &\mathsf{u^\dag}\left(\sum_{i=1}^3\sum_{j=1}^3\mathsf{t_i^\dag}\,\mathsf{s_j}\,\mathsf{Q}\, \mathsf{s_i^\dag}\mathsf{t_j}\right)\mathsf{u} &= -\frac{4q_u-q_d}{6}\;, \qquad
             \end{align*}
             the full exchange contribution to the proton's electromagnetic current (including the color factor) is given by
             \begin{equation}
                 J^\text{EX}_p = -\frac{q_d}{2}\, J^\text{EX}_{SS} + \frac{2q_u+q_d}{2\sqrt{3}}\,
                                  \left( J^\text{EX}_{SA}+J^\text{EX}_{AS}\right) + \frac{4q_u-q_d}{6}\, J^\text{EX}_{AA}\;,
             \end{equation}
             where $J^\text{EX}_{SS}$, $J^\text{EX}_{SA}$, $J^\text{EX}_{AS}$ and $J^\text{EX}_{AA}$ denote the Dirac parts of
             the exchange contributions in Eq.\,\eqref{ff:current2}.
             For the seagull contributions, the coupling of the photon to both quark lines and the diquark line in the seagull
             amplitudes (cf. Fig.\,\ref{fig:seagulls}) have to be taken into account. Therefore all
             occurences of $e_-$, $e_+$ and $e_{dq}$ have to be replaced by the combined flavor-charge factors
             \begin{align*}
                  \text{SG:} \qquad &
                               e_- \rightarrow    \sum_{ij} \mathsf{t_i^\dag} \,\mathsf{s_j}\,\mathsf{Q}\, \mathsf{s_i^\dag}\,\mathsf{t_j}\;, \quad
                               e_+ \rightarrow    \sum_{ij} \mathsf{t_i^\dag} \,\mathsf{Q}\,\mathsf{s_j}\, \mathsf{s_i^\dag}\,\mathsf{t_j}\;, \quad \\
                             & e_{dq} \rightarrow \sum_{ij} \mathsf{t_i^\dag} \,\mathsf{s_j}\, \mathsf{s_i^\dag}\,\mathsf{t_j} \;2\, \text{Tr}\left\{ \mathsf{s_j^\dag}\,\mathsf{s_j}\,\mathsf{Q} \right\} \\
                  \overline{\text{SG:}} \qquad &
                               e_- \rightarrow    \sum_{ij} \mathsf{t_i^\dag} \,\mathsf{s_j}\,\mathsf{Q}\, \mathsf{s_i^\dag}\,\mathsf{t_j}\;, \quad
                               e_+ \rightarrow    \sum_{ij} \mathsf{t_i^\dag} \,\mathsf{s_j}\, \mathsf{s_i^\dag}\,\mathsf{Q}\,\mathsf{t_j}\;, \quad \\
                             & e_{dq} \rightarrow \sum_{ij} \mathsf{t_i^\dag} \,\mathsf{s_j}\, \mathsf{s_i^\dag}\,\mathsf{t_j} \;2\, \text{Tr}\left\{ \mathsf{s_i^\dag}\,\mathsf{s_i}\,\mathsf{Q} \right\} \nonumber
             \end{align*}
             and equipped with the overall color factor $-1$.

\section{Singularities and resulting boundaries}\label{app:sing}

             Each of the momenta \eqref{nuc:momenta} that enter the quark-diquark BSE can generically be written as
             \begin{equation} \label{sing:mom}
                 l_X = \sum_i\alpha_X^{(i)}(\eta)\,q_i + \beta_X(\eta)\,P\;,
             \end{equation}
             with loop momenta $q_i$ ($q_i^2 \in \mathds{R}_+$) and the nucleon's total momentum $P$ ($P^2=-M^2$).
             The real coefficients $\alpha_X^{(i)},\,\beta_X$ depend on the momentum partitioning parameter $\eta$.
             The complex argument $l_X^2$ of each quantity $X$ (quark propagator, diquark propagator and amplitudes)
             that is sampled by the BSE is therefore bounded by the parabola $(t \pm i \beta_X(\eta)\,M)^2$, with $t \in \mathds{R}_+$.
             Since we do not explicitly include residue contributions in our calculations,
             the largest possible parabola is the one which contains the singularity (if any) in the quantity $X$ closest to the origin at $l_X^2=p_X^2$:
             $(t \pm i m_X)^2$, with $m_X=|\text{Im}\sqrt{p_X^2}|$. For instance, the quark propagator could have complex conjugate poles,
             the diquark propagators by construction exhibit timelike poles (e.\,g., for the scalar diquark propagator: $p_X^2=-m_\text{sc}^2$) etc.
             For each momentum $l_X$, this leads  to a restriction $M<m_X/|\beta_X(\eta)|$, and in total to an upper limit
             for the nucleon mass:
             \begin{equation}\label{sing:cond}
                 M < f(\eta) := \text{min} \, \left\{ \, \frac{m_q}{\eta}\;, \,
                                                        \frac{m_q}{|1-2\eta|}\;, \,
                                                        \frac{m_d}{1-\eta}\;, \,
                                                        \frac{2\lambda}{|1-3\eta|} \,\right\}\;.
             \end{equation}

             Here $m_q$ is the quark "pole" mass (cf. Table \ref{tab:dse2}),
             $m_d=\text{min}(m_\text{sc},\,m_\text{av})$ is the smallest
             diquark mass (i.\,e., the scalar diquark mass) and $\lambda$ is the "pole" mass
             of the closest singularity in the relative momentum entering the diquark amplitudes.
             The dependence of the diquark amplitudes on the \textit{total} diquark momentum is modeled via Eq.\,\eqref{dq:offshellansatz2}
             to exhibit its first singularity at the diquark poles and therefore identical to the condition involving $m_d$.
             If any of the above quantities is free of singularities, its contribution can be removed from the bracket in \eqref{sing:cond}.
             As mentioned in Sec.\,\ref{sec:qdqbse}, the momentum partitioning parameter $\eta$ can be chosen to maximize the upper limit $f(\eta)$ for the nucleon mass.
             The analogous analysis for the quark momenta in the meson and diquark BSEs with momentum partitioning parameter $\sigma$ entails
             $f(\sigma) = \text{min}  \left\{ \, m_q/\sigma, \, m_q/(1-\sigma)\, \right\}$ with the maximum $f(\sigma=1/2)=2m_q$.
             This justifies the choice $\sigma=1/2$ and leads to the restrictions $m_\text{meson},\,m_\text{diquark} < 2 m_q$.

             With $q_1=Q$ and the replacement $P^2=-(M^2+Q^2/4)$, Eq.\,\eqref{sing:mom} can also be applied to the momenta appearing in the form factor diagrams.
             If the nucleon mass is already fixed, this leads to an upper limit for the potential photon momenta:
             $Q^2<4 (f(\eta)^2-M^2)$, which is typically of the size of $1 - 2\,\text{GeV}^2$ and explains the restrictions
             encountered in Sec.\,\ref{sec:results}. Possible additional singularities, e.\,g., in the inverse quark propagator's
             dressing functions $A$ and $M$ (which enter the Ball-Chiu vertex) have to be considered in this case;
             those may occur in a parametrization where $\sigma_v$ and $\sigma_s$ are entire functions \cite{Oettel:2002wf,Alkofer:2004yf,Maris:2003vk}.
             One can choose $\eta$ and $f(\eta)$ for each current diagram in Fig.\,\ref{fig:current} separately if for each value of $\eta$
             a corresponding quark-diquark BSE is solved, since the quark-diquark amplitudes depend on the
             momentum partitioning parameter but the single current contributions must be independent of $\eta$.

             The diquark amplitudes of Eqs.\,(\ref{dq:sc} -- \ref{dq:av}) depend on three complex variables $q^2$, $z$, and $P^2$,
             where possible singularities in $q^2$ and $P^2$ are avoided by virtue of Eq.\,\eqref{sing:cond}.
             The diquark BSEs \eqref{dq:bse} are explicitly solved for $z\in (-1,1)$ in terms of a Chebyshev expansion, Eq.\,\eqref{dq:chebyshev}.
             Within the convergence region $|z|<1$ of the Chebyshev polynomials this solution can be analytically continued into the complex $z$ plane.
             The quark-diquark BSE however samples the diquark amplitudes also in the region $|z|>1$ which is not accessible via a Chebyshev expansion.
             A BSE solution for all $z\in\mathds{C}$ would require refined methods due to the intricate singularity structure for $|z|>1$;
             furthermore, the residues of those singularities would have to be taken into account when inserting the diquark amplitudes into the quark-diquark BSE.
             We circumvent this problem by retaining only the zeroth Chebyshev moment in each diquark amplitude, i.e.
             assuming no $z$ dependence at all: $f_k(q^2,z)\approx f_k^0(q^2)$ in Eq.\,\eqref{sing:cond}.
             This is a reasonable approximation at least for $|z|<1$ where the full solution is known due to the strong suppression of higher Chebyshev moments.

\end{appendix}


\bibliographystyle{unsrt}
\bibliographystyle{elsart-num}
\bibliographystyle{elsart-num-mod}
\bibliography{had_nucl_graz}

\begin{thebibliography}{100}
\expandafter\ifx\csname url\endcsname\relax
  \def\url#1{\texttt{#1}}\fi
\expandafter\ifx\csname urlprefix\endcsname\relax\def\urlprefix{URL }\fi

\bibitem{Thomas:2001kw}
A.~W. Thomas, W.~Weise, {T}he {S}tructure of the {N}ucleon, Berlin, Germany:
  Wiley-VCH, 2001.

\bibitem{Arrington:2006zm}
J.~Arrington, C.~D. Roberts, J.~M. Zanotti, J. Phys. G 34 (2007) S23--S51,
  nucl-th/0611050.


\bibitem{Schladming06}
R.~Alkofer, A.~Krassnigg, W.~Schweiger (Eds.), {L}ecture notes of the 44th
  {S}chladming {W}inter {S}chool ``{H}adron {S}tructure and {N}onperturbative
  {QCD}'', {S}chladming, {A}ustria, {M}arch 11 - 18, 2006, Vol. 140 of Eur.
  Phys. J. ST, 2007, {L}ecture notes of the 44th {S}chladming {W}inter {S}chool
  ``{H}adron {S}tructure and {N}onperturbative {QCD}'', {S}chladming,
  {A}ustria, {M}arch 11 - 18, 2006.

\bibitem{Alkofer:2006jf}
R.~Alkofer, Braz. J. Phys. 37 (2007) 144--164, hep-ph/0611090.


\bibitem{Roberts:1994dr}
C.~D. Roberts, A.~G. Williams, Prog. Part. Nucl. Phys. 33 (1994) 477--575,
  hep-ph/9403224.


\bibitem{Alkofer:2000wg}
R.~Alkofer, L.~von Smekal, Phys. Rept. 353 (2001) 281, hep-ph/0007355.


\bibitem{Fischer:2006ub}
C.~S. Fischer, J. Phys. G 32 (2006) R253--R291, hep-ph/0605173.


\bibitem{Holl:2006ni}
A.~Holl, C.~D. Roberts, S.~V. Wright, nucl-th/0601071.


\bibitem{Nambu:1961fr}
Y.~Nambu, G.~Jona-Lasinio, Phys. Rev. 124 (1961) 246--254.


\bibitem{Nambu:1961tp}
Y.~Nambu, G.~Jona-Lasinio, Phys. Rev. 122 (1961) 345--358.


\bibitem{Buck:1992wz}
A.~Buck, R.~Alkofer, H.~Reinhardt, Phys. Lett. B 286 (1992) 29--35.


\bibitem{Weiss:1993kv}
C.~Weiss, A.~Buck, R.~Alkofer, H.~Reinhardt, Phys. Lett. B 312 (1993) 6--12,
  hep-ph/9305215.


\bibitem{Ishii:1995bu}
N.~Ishii, W.~Bentz, K.~Yazaki, Nucl. Phys. A 587 (1995) 617--656.


\bibitem{Bloch:1999ke}
J.~C.~R. Bloch, C.~D. Roberts, S.~M. Schmidt, A.~Bender, M.~R. Frank, Phys.
  Rev. C 60 (1999) 062201, nucl-th/9907120.


\bibitem{Bloch:1999rm}
J.~C.~R. Bloch, C.~D. Roberts, S.~M. Schmidt, Phys. Rev. C 61 (2000) 065207,
  nucl-th/9911068.


\bibitem{Oettel:1998bk}
M.~Oettel, G.~Hellstern, R.~Alkofer, H.~Reinhardt, Phys. Rev. C 58 (1998)
  2459--2477, nucl-th/9805054.


\bibitem{Oettel:1999gc}
M.~Oettel, M.~Pichowsky, L.~von Smekal, Eur. Phys. J. A 8 (2000) 251--281,
  nucl-th/9909082.


\bibitem{Hecht:2002ej}
M.~B. Hecht, et~al., Phys. Rev. C 65 (2002) 055204, nucl-th/0201084.


\bibitem{Oettel:2000jj}
M.~Oettel, R.~Alkofer, L.~von Smekal, Eur. Phys. J. A 8 (2000) 553--566,
  nucl-th/0006082.


\bibitem{Oettel:2002wf}
M.~Oettel, R.~Alkofer, Eur. Phys. J. A 16 (2003) 95--109, hep-ph/0204178.


\bibitem{Alkofer:2004yf}
R.~Alkofer, A.~Holl, M.~Kloker, A.~Krassnigg, C.~D. Roberts, Few Body Syst. 37
  (2005) 1--31, nucl-th/0412046.


\bibitem{Holl:2005zi}
A.~H\"oll, R.~RAlkofer, M.~Kloker, A.~Krassnigg, C.~D. Roberts, S.~V. Wright,
  Nucl. Phys. A 755 (2005) 298--302, nucl-th/0501033.


\bibitem{Maris:1997hd}
P.~Maris, C.~D. Roberts, P.~C. Tandy, Phys. Lett. B 420 (1998) 267--273,
  nucl-th/9707003.


\bibitem{Bender:2002as}
A.~Bender, W.~Detmold, C.~D. Roberts, A.~W. Thomas, Phys. Rev. C 65 (2002)
  065203, nucl-th/0202082.


\bibitem{Bender:1996bb}
A.~Bender, C.~D. Roberts, L.~Von~Smekal, Phys. Lett. B 380 (1996) 7--12,
  nucl-th/9602012.


\bibitem{Cahill:1988dx}
R.~T. Cahill, C.~D. Roberts, J.~Praschifka, Austral. J. Phys. 42 (1989)
  129--145.


\bibitem{Zwanziger:2001kw}
D.~Zwanziger, Phys. Rev. D 65 (2002) 094039, hep-th/0109224.


\bibitem{Lerche:2002ep}
C.~Lerche, L.~von Smekal, Phys. Rev. D 65 (2002) 125006, hep-ph/0202194.


\bibitem{Alkofer:2006gz}
R.~Alkofer, C.~S. Fischer, F.~J. Llanes-Estrada, hep-ph/0607293.


\bibitem{Fischer:2002hna}
C.~S. Fischer, R.~Alkofer, Phys. Lett. B 536 (2002) 177--184, hep-ph/0202202.


\bibitem{Fischer:2003rp}
C.~S. Fischer, R.~Alkofer, Phys. Rev. D 67 (2003) 094020, hep-ph/0301094.


\bibitem{Alkofer:2004it}
R.~Alkofer, C.~S. Fischer, F.~J. Llanes-Estrada, Phys. Lett. B611 (2005)
  279--288, hep-th/0412330.


\bibitem{Alkofer:2003jj}
R.~Alkofer, W.~Detmold, C.~S. Fischer, P.~Maris, Phys. Rev. D 70 (2004) 014014,
  hep-ph/0309077.


\bibitem{Jain:1991pk}
P.~Jain, H.~J. Munczek, Phys. Rev. D 44 (1991) 1873--1879.


\bibitem{Munczek:1991jb}
H.~J. Munczek, P.~Jain, Phys. Rev. D 46 (1992) 438--445.


\bibitem{Frank:1995uk}
M.~R. Frank, C.~D. Roberts, Phys. Rev. C 53 (1996) 390--398, hep-ph/9508225.


\bibitem{Alkofer:1995jx}
R.~Alkofer, C.~D. Roberts, Phys. Lett. B 369 (1996) 101--107, hep-ph/9510284.


\bibitem{Maris:1997tm}
P.~Maris, C.~D. Roberts, Phys. Rev. C 56 (1997) 3369--3383, nucl-th/9708029.


\bibitem{Maris:1999nt}
P.~Maris, P.~C. Tandy, Phys. Rev. C 60 (1999) 055214, nucl-th/9905056.


\bibitem{Alkofer:2002bp}
R.~Alkofer, P.~Watson, H.~Weigel, Phys. Rev. D 65 (2002) 094026,
  hep-ph/0202053.


\bibitem{Maris:1999bh}
P.~Maris, P.~C. Tandy, Phys. Rev. C 61 (2000) 045202, nucl-th/9910033.


\bibitem{Maris:2000sk}
P.~Maris, P.~C. Tandy, Phys. Rev. C 62 (2000) 055204, nucl-th/0005015.


\bibitem{Ji:2001pj}
C.-R. Ji, P.~Maris, Phys. Rev. D 64 (2001) 014032, nucl-th/0102057.


\bibitem{Maris:2002mz}
P.~Maris, P.~C. Tandy, Phys. Rev. C 65 (2002) 045211, nucl-th/0201017.


\bibitem{Jarecke:2002xd}
D.~Jarecke, P.~Maris, P.~C. Tandy, Phys. Rev. C 67 (2003) 035202,
  nucl-th/0208019.


\bibitem{Holl:2004fr}
A.~Holl, A.~Krassnigg, C.~D. Roberts, Phys. Rev. C 70 (2004) 042203,
  nucl-th/0406030.


\bibitem{Krassnigg:2004if}
A.~Krassnigg, P.~Maris, J. Phys. Conf. Ser. 9 (2005) 153--160, nucl-th/0412058.


\bibitem{Holl:2005vu}
A.~Holl, A.~Krassnigg, P.~Maris, C.~D. Roberts, S.~V. Wright, Phys. Rev. C 71
  (2005) 065204, nucl-th/0503043.


\bibitem{Maris:2005tt}
P.~Maris, P.~C. Tandy, Nucl. Phys. Proc. Suppl. 161 (2006) 136--152,
  nucl-th/0511017.


\bibitem{Bhagwat:2006pu}
M.~S. Bhagwat, P.~Maris, nucl-th/0612069.


\bibitem{Maris:2006ea}
P.~Maris, AIP Conf. Proc. 892 (2007) 65--71, nucl-th/0611057.


\bibitem{Bhagwat:2006xi}
M.~S. Bhagwat, A.~Krassnigg, P.~Maris, C.~D. Roberts, Eur. Phys. J. A31 (2007)
  630--637, nucl-th/0612027.


\bibitem{Bhagwat:2007rj}
M.~S. Bhagwat, A.~Hoell, A.~Krassnigg, C.~D. Roberts, S.~V. Wright, Few Body
  Syst. 40 (2007) 209--235, nucl-th/0701009.


\bibitem{Bowman:2002bm}
P.~O. Bowman, U.~M. Heller, A.~G. Williams, Phys. Rev. D 66 (2002) 014505,
  hep-lat/0203001.


\bibitem{Bowman:2005vx}
P.~O. Bowman, et~al., Phys. Rev. D 71 (2005) 054507, hep-lat/0501019.


\bibitem{Bhagwat:2003vw}
M.~S. Bhagwat, M.~A. Pichowsky, C.~D. Roberts, P.~C. Tandy, Phys. Rev. C 68
  (2003) 015203, nucl-th/0304003.


\bibitem{Fischer:2005nf}
C.~S. Fischer, M.~R. Pennington, Phys. Rev. D 73 (2006) 034029, hep-ph/0512233.


\bibitem{Fischer:2007ea}
C.~S. Fischer, M.~R. Pennington, Eur. Phys. J. A31 (2007) 746--749,
  hep-ph/0701123.


\bibitem{Fischer:2007ze}
C.~S. Fischer, D.~Nickel, J.~Wambach, Phys. Rev. D 76 (2007) 094009,
  arXiv:0705.4407 [hep-ph].


\bibitem{Capitani:1998mq}
S.~Capitani, M.~Luscher, R.~Sommer, H.~Wittig, Nucl. Phys. B 544 (1999)
  669--698, hep-lat/9810063.


\bibitem{Gattringer:2005ij}
C.~Gattringer, P.~Huber, C.~B. Lang, Phys. Rev. D 72 (2005) 094510,
  hep-lat/0509003.


\bibitem{Watson:2004kd}
P.~Watson, W.~Cassing, P.~C. Tandy, Few Body Syst. 35 (2004) 129--153,
  hep-ph/0406340.


\bibitem{Cloet:2007pi}
I.~C. Cloet, A.~Krassnigg, C.~D. Roberts, arXiv:0710.5746 [nucl-th].


\bibitem{Bhagwat:2004hn}
M.~S. Bhagwat, A.~Holl, A.~Krassnigg, C.~D. Roberts, P.~C. Tandy, Phys. Rev. C
  70 (2004) 035205, nucl-th/0403012.


\bibitem{Watson:2004jq}
P.~Watson, W.~Cassing, Few Body Syst. 35 (2004) 99--115, hep-ph/0405287.


\bibitem{Matevosyan:2006bk}
H.~H. Matevosyan, A.~W. Thomas, P.~C. Tandy, Phys. Rev. C 75 (2007) 045201,
  nucl-th/0605057.


\bibitem{Alkofer:2005ug}
R.~Alkofer, M.~Kloker, A.~Krassnigg, R.~F. Wagenbrunn, Phys. Rev. Lett 96
  (2006) 022001, hep-ph/0510028.


\bibitem{Maris:2002yu}
P.~Maris, Few Body Syst. 32 (2002) 41--52, nucl-th/0204020.


\bibitem{Maris:2004bp}
P.~Maris, Few Body Syst. 35 (2004) 117--127, nucl-th/0409008.


\bibitem{Cahill:1987qr}
R.~T. Cahill, C.~D. Roberts, J.~Praschifka, Phys. Rev. D 36 (1987) 2804.


\bibitem{Hecht:2000jh}
M.~B. Hecht, C.~D. Roberts, S.~M. Schmidt, Lect. Notes Phys. 578 (2001)
  218--234, nucl-th/0012023.


\bibitem{Alexandrou:2006cq}
C.~Alexandrou, P.~de~Forcrand, B.~Lucini, Phys. Rev. Lett. 97 (2006) 222002,
  hep-lat/0609004.


\bibitem{Osterwalder:1973dx}
K.~Osterwalder, R.~Schrader, Commun. Math. Phys. 31 (1973) 83--112.


\bibitem{Haag:1992hx}
R.~Haag, {L}ocal quantum physics: {F}ields, particles, algebras, Springer,
  1996, berlin, Germany: Springer (1992) 356 p. (Texts and monographs in
  physics).

\bibitem{Oehme:1994pv}
R.~Oehme, Int. J. Mod. Phys. A 10 (1995) 1995--2014, hep-th/9412040.


\bibitem{Roberts:2000aa}
C.~D. Roberts, S.~M. Schmidt, Prog. Part. Nucl. Phys. 45 (2000) S1--S103,
  nucl-th/0005064.


\bibitem{Hellstern:1997nv}
G.~Hellstern, R.~Alkofer, H.~Reinhardt, Nucl. Phys. A 625 (1997) 697--712,
  hep-ph/9706551.


\bibitem{Wetzorke:2000ez}
I.~Wetzorke, F.~Karsch, hep-lat/0008008.


\bibitem{Babich:2007ah}
R.~Babich, et~al., Phys. Rev. D 76 (2007) 074021, hep-lat/0701023.


\bibitem{Orginos:2005vr}
K.~Orginos, PoS LAT2005 (2006) 054, hep-lat/0510082.


\bibitem{Hess:1998sd}
M.~Hess, F.~Karsch, E.~Laermann, I.~Wetzorke, Phys. Rev. D 58 (1998) 111502,
  hep-lat/9804023.


\bibitem{Oettel:2001kd}
M.~Oettel, L.~Von~Smekal, R.~Alkofer, Comput. Phys. Commun. 144 (2002) 63,
  hep-ph/0109285.


\bibitem{Mandelstam:1955sd}
S.~Mandelstam, Proc. Roy. Soc. Lond. A 233 (1955) 248.


\bibitem{Haberzettl:1997jg}
H.~Haberzettl, Phys. Rev. C 56 (1997) 2041--2058, nucl-th/9704057.


\bibitem{Kvinikhidze:1998xn}
A.~N. Kvinikhidze, B.~Blankleider, Phys. Rev. C 60 (1999) 044003,
  nucl-th/9901001.


\bibitem{Kvinikhidze:1999xp}
A.~N. Kvinikhidze, B.~Blankleider, Phys. Rev. C 60 (1999) 044004,
  nucl-th/9901002.


\bibitem{Boinepalli:2006xd}
S.~Boinepalli, D.~B. Leinweber, A.~G. Williams, J.~M. Zanotti, J.~B. Zhang,
  Phys. Rev. D 74 (2006) 093005, hep-lat/0604022.


\bibitem{Frigori:2007wa}
R.~Frigori, et~al., arXiv:0709.4582 [hep-lat].


\bibitem{AliKhan:2003cu}
A.~Ali~Khan, et~al., Nucl. Phys. B 689 (2004) 175--194, hep-lat/0312030.


\bibitem{Leinweber:2003dg}
D.~B. Leinweber, A.~W. Thomas, R.~D. Young, Phys. Rev. Lett. 92 (2004) 242002,
  hep-lat/0302020.


\bibitem{Procura:2006bj}
M.~Procura, B.~U. Musch, T.~Wollenweber, T.~R. Hemmert, W.~Weise, Phys. Rev. D
  73 (2006) 114510, hep-lat/0603001.


\bibitem{Theberge:1980ye}
S.~Theberge, A.~W. Thomas, G.~A. Miller, Phys. Rev. D 22 (1980) 2838.


\bibitem{Thomas:1982kv}
A.~W. Thomas, Adv. Nucl. Phys. 13 (1984) 1--137.


\bibitem{Lu:1997sd}
D.-H. Lu, A.~W. Thomas, A.~G. Williams, Phys. Rev. C 57 (1998) 2628--2637,
  nucl-th/9706019.


\bibitem{Chodos:1974je}
A.~Chodos, R.~L. Jaffe, K.~Johnson, C.~B. Thorn, V.~F. Weisskopf, Phys. Rev. D
  9 (1974) 3471--3495.


\bibitem{Gasser:1983yg}
J.~Gasser, H.~Leutwyler, Ann. Phys. 158 (1984) 142.


\bibitem{Bernard:1995dp}
V.~Bernard, N.~Kaiser, U.-G. Meissner, Int. J. Mod. Phys. E4 (1995) 193--346,
  hep-ph/9501384.


\bibitem{Young:2002cj}
R.~D. Young, D.~B. Leinweber, A.~W. Thomas, S.~V. Wright, Phys. Rev. D 66
  (2002) 094507, hep-lat/0205017.


\bibitem{Hemmert:2002uh}
T.~R. Hemmert, W.~Weise, Eur. Phys. J. A15 (2002) 487--504, hep-lat/0204005.


\bibitem{Gockeler2005}
M.~Gockeler, et~al., Phys. Rev. D 71 (2005) 034508, hep-lat/0303019.


\bibitem{Young:2002ib}
R.~D. Young, D.~B. Leinweber, A.~W. Thomas, Prog. Part. Nucl. Phys. 50 (2003)
  399--417, hep-lat/0212031.


\bibitem{Donoghue:1998bs}
J.~F. Donoghue, B.~R. Holstein, B.~Borasoy, Phys. Rev. D 59 (1999) 036002,
  hep-ph/9804281.


\bibitem{Oettel:2002cw}
M.~Oettel, A.~W. Thomas, Phys. Rev. C 66 (2002) 065207, nucl-th/0203073.


\bibitem{Pearce:1986du}
B.~C. Pearce, I.~R. Afnan, Phys. Rev. C 34 (1986) 991.


\bibitem{AliKhan:2001tx}
A.~Ali~Khan, et~al., Phys. Rev. D 65 (2002) 054505, hep-lat/0105015.


\bibitem{Allton:2005fb}
C.~R. Allton, W.~Armour, D.~B. Leinweber, A.~W. Thomas, R.~D. Young, Phys.
  Lett. B 628 (2005) 125--130, hep-lat/0504022.


\bibitem{Fischer:2005en}
C.~S. Fischer, P.~Watson, W.~Cassing, Phys. Rev. D 72 (2005) 094025,
  hep-ph/0509213.


\bibitem{Young:2004tb}
R.~D. Young, D.~B. Leinweber, A.~W. Thomas, Phys. Rev. D 71 (2005) 014001,
  hep-lat/0406001.


\bibitem{Wang:2007iw}
P.~Wang, D.~B. Leinweber, A.~W. Thomas, R.~D. Young, Phys. Rev. D 75 (2007)
  073012, hep-ph/0701082.


\bibitem{Alexandrou:2006ru}
C.~Alexandrou, G.~Koutsou, J.~W. Negele, A.~Tsapalis, Phys. Rev. D 74 (2006)
  034508, hep-lat/0605017.


\bibitem{Gockeler:2007ir}
M.~Gockeler, et~al., arXiv:0709.3370 [hep-lat].


\bibitem{Friedrich:2003iz}
J.~Friedrich, T.~Walcher, Eur. Phys. J. A 17 (2003) 607--623, hep-ph/0303054.


\bibitem{Walker:1993vj}
R.~C. Walker, et~al., Phys. Rev. D 49 (1994) 5671--5689.


\bibitem{Jones:1999rz}
M.~K. Jones, et~al., Phys. Rev. Lett. 84 (2000) 1398--1402, nucl-ex/9910005.


\bibitem{Gayou:2001qd}
O.~Gayou, et~al., Phys. Rev. Lett. 88 (2002) 092301, nucl-ex/0111010.


\bibitem{Alkofer:2005jh}
R.~Alkofer, M.~Oettel, nucl-th/0507003.


\bibitem{Ball:1980ay}
J.~S. Ball, T.-W. Chiu, Phys. Rev. D 22 (1980) 2542.


\bibitem{Salam:1964zk}
A.~Salam, R.~Delbourgo, Phys. Rev. 135 (1964) B1398--B1427.


\bibitem{Wang:1996zu}
S.~Wang, M.~K. Banerjee, Phys. Rev. C 54 (1996) 2883--2893, nucl-th/9607022.


\bibitem{Maris:2003vk}
P.~Maris, C.~D. Roberts, Int. J. Mod. Phys. E12 (2003) 297--365,
  nucl-th/0301049.


\end{thebibliography}


\end{document}